\documentclass[%
 a4paper, twocolumn,11pt
amsmath,amssymb,
accepted=2024-09-29
]
{quantumarticle}
\pdfoutput = 1

\usepackage[utf8]{inputenc}
\usepackage[english]{babel}
\usepackage[T1]{fontenc}
\usepackage{xkeyval}
\usepackage{graphicx}% Include figure files
\usepackage[export]{adjustbox}% frame in figures
\usepackage{soul}
\usepackage{dcolumn}% Align table columns on decimal point
\usepackage{bm}
\usepackage{epstopdf}
 \usepackage[usenames,dvipsnames]{pstricks}
 \usepackage{epsfig}
 \usepackage{pst-grad} % For gradients
 \usepackage{pst-plot} % For axes
 \usepackage[space]{grffile} % For spaces in paths
 \usepackage{etoolbox} % For spaces in paths
 \makeatletter % For spaces in paths
 \patchcmd\Gread@eps{\@inputcheck#1 }{\@inputcheck"#1"\relax}{}{}
 \makeatother
\usepackage{booktabs}
\usepackage{boxhandler}
\usepackage{amsmath}
\usepackage{subcaption}
\usepackage{braket}

\usepackage{hyperref}
\hypersetup{
  colorlinks   = true,    % Colours links instead of ugly boxes
  urlcolor     = blue,    % Colour for external hyperlinks
  linkcolor    = blue,    % Colour of internal links
  citecolor    = blue      % Colour of citations
}

\usepackage{multirow}
\usepackage{enumitem}
\DeclareMathOperator*{\Argmax}{\arg\max}

\begin{document}

\title{Review on the decoding algorithms for surface codes}

\author{Antonio {deMarti iOlius} }
\email{toni.demarti@gmail.com}
\affiliation{Department of Basic Sciences, Tecnun - University of Navarra, 20018 San Sebastian, Spain.}
\author{Patricio Fuentes}
\email{pfuentes@photonic.com}
\affiliation{Photonic Inc., Vancouver, British Columbia, Canada.}
\author{Rom\'an Or\'us}
\email{roman.orus@multiversecomputing.com}
\affiliation{Multiverse Computing, Pio Baroja 37, 20008 San Sebastián, Spain}
\affiliation{Donostia International Physics Center, Paseo Manuel de Lardizabal 4, 20018 San Sebastián, Spain}
\affiliation{IKERBASQUE, Basque Foundation for Science, Plaza Euskadi 5, 48009 Bilbao, Spain}
\author{Pedro M. Crespo}
\email{pcrespo@unav.es}
\affiliation{Department of Basic Sciences, Tecnun - University of Navarra, 20018 San Sebastian, Spain.}
\author{Josu {Etxezarreta Martinez}}
\email{jetxezarreta@unav.es}
\affiliation{Department of Basic Sciences, Tecnun - University of Navarra, 20018 San Sebastian, Spain.}


\begin{abstract}
Quantum technologies have the potential to solve certain computationally hard problems with polynomial or super-polynomial speedups when compared to classical methods. Unfortunately, the unstable nature of quantum information makes it prone to errors. For this reason, quantum error correction is an invaluable tool to make quantum information reliable and enable the ultimate goal of fault-tolerant quantum computing. Surface codes currently stand as the most promising candidates to build near term error corrected qubits given their two-dimensional architecture, the requirement of only local operations, and high tolerance to quantum noise. Decoding algorithms are an integral component of any error correction scheme, as they are tasked with producing accurate estimates of the errors that affect quantum information, so that they can subsequently be corrected. A critical aspect of decoding algorithms is their speed, since the quantum state will suffer additional errors with the passage of time. This poses a connundrum, where decoding performance is improved at the expense of complexity and viceversa. In this review, a thorough discussion of state-of-the-art decoding algorithms for surface codes is provided. The target audience of this work are both readers with an introductory understanding of the field as well as those seeking to further their knowledge of the decoding paradigm of surface codes. We describe the core principles of these decoding methods as well as existing variants that show promise for improved results. In addition, both the decoding performance, in terms of error correction capability, and decoding complexity, are compared. A review of the existing software tools regarding surface codes decoding is also provided.

\end{abstract}

\maketitle

\tableofcontents

\section{Introduction}

Significant progress has been made in the field of quantum computing since Feynman first introduced the idea of computers that use the laws of quantum mechanics in 1982 \cite{feynman}. Quantum computers leverage the principles of quantum mechanics to accelerate computational processes known to be not efficiently solvable by means of classical machines\footnote{Efficient in this context means that the problem can be solved in a reasonable amount of time.}\cite{preskilSup}. Quantum algorithms that offer exponential speedups in terms of computational complexity have been proposed \cite{shorFact,exponentAdv}, implying that his technology has the unprecedented potential to solve computationally hard problems. In this sense, quantum processors are expected to revolutionize modern society by boosting fields like cryptography \cite{crypt1,crypt2}, optimization \cite{opt1}, macromolecule design \cite{macromolec1} or basic science \cite{basic1}, among others. Recent results have demonstrated experimental quantum supremacy or advantage \cite{sup1,sup2,sup3,sup4,sup5}, thus, proving that the potential promised by the theoretical quantum algorithms can be realized in physical machines.

There are many possible paths towards the construction of quantum processors such as qubit gate-based quantum computing \cite{nielsen}, qudit-based quantum computing \cite{quditComp}, measurement based quantum computing \cite{mbqc1,mbqc2}, quantum annealers \cite{anneal} or bosonic quantum computing \cite{boson}, to name a few. Many physical implementations of these types of technologies like superconducting circuits \cite{SCquantumcomp}, ion-traps \cite{ITquantumcomp} or photons \cite{photquantumcomp} are currently being investigated. In the present article, we will restrict our discussion for qubit gate-based quantum computation, in which the physical elements of a quantum computer are realized by discrete two-level systems. % tb para otros!

Despite recent advances, quantum noise still stands as a big obstacle to practical quantum computing \cite{shorQEC}. The qubits that make up current quantum computers are prone to suffer from errors, leading to inaccurate computations and unreliable results. There are numerous mechanisms by which qubits suffer from errors during computation. For example, quantum gates, which are the quantum analogue of classical logic gates, can cause unwanted errors to the qubits they are applied to due to imprecise implementation (this can be understood as an over- or under-rotation of the qubits targeted by the gate) \cite{fowlerReview}. Many of the errors arise due to similar imperfections in the equipment being used for preparing, controlling or measuring such quantum mechanical systems, and it may seem that dealing with errors suffered by qubits is an engineering problem \cite{fowlerReview,preskill}. However, qubits also suffer from errors whose origin is their inescapable interaction with their surrounding environment \cite{approximatingDec,decoherenceBench,TVQC}. These types of errors, which are grouped under the term decoherence, represent an unavoidable and hence, problematic error source. However, dealing with errors is not a new requirement in the field of computer science. Methods to prevent and mitigate the effects of noise in classical systems have been a thoroughly studied subject in the past century, and they have enabled commodities as important as WiFi. Unfortunately, phenomena particular to quantum mechanics, such as the no-cloning theorem or the effects of measurement on a quantum state, prevent the seamless implementation of classical methods in the quantum realm. Quantum error correction (QEC) was shown to be feasable (at least theoretically) when the 9-qubit Shor code was proposed in 1995 \cite{shorQEC}. The field was soon generalized following this demonstration with the theory of Quantum Stabilizer Codes, introduced by Gottesman in his PhD thesis \cite{gottesman}. The field of quantum error correction has advanced significantly since then and several families of quantum error correction codes (QECC) such as Topological codes \cite{fowlerReview,kitaev,xzzx,tuckettxy,heavyhexagon,colorcode}, Quantum Low-Density Parity Check (QLDPC) codes \cite{qldpc1,qldpc2,qldpc4,qldpc5} or Quantum Turbo Codes (QTC) \cite{qtc1} have been proposed.

The core principle of quantum error correction is embedding the information contained in a number of qubits (logical qubits) in a larger number of qubits (physical qubits) via a mapping known as encoding \cite{approximatingDec,gottesman,reviewpat}. To estimate the errors suffered by the logical qubits, non-destructive measurements (with this we mean that the actual qubits constituting the code are not directly measured) known as syndrome measurements, are performed \cite{approximatingDec}. These measurements yield data known as syndromes, which are used to produce an estimate of the error that took place, which, in turn, is used to return the quantum state to the previous undamaged state. This process, referred as syndrome decoding, depends on the specific code construction \cite{fowlerReview,xzzx,tuckettxy,heavyhexagon,colorcode,qldpc1,qldpc2,qldpc4,qldpc5,qtc1} and is critical for the QECCs to work. Decoding quantum error correction codes is further nuanced by a quantum unique effect known as code degeneracy \cite{qldpc2,reviewpat}, which refers to the existence of different errors that share the same syndrome (these errors cannot effectively be distinguished by measurements) but transform the quantum state in an indistinguishable manner. As a consequence of degeneracy, the optimal decoding of stabilizer codes falls within the $\#$P-complete complexity class, which is computationally harder than decoding classical linear codes \cite{reviewpat,PcompleteDec}. The complexity of the problem imposes a trade-off between decoding performance and decoding time. While decoders for quantum error correction codes need to be as accurate as possible, they should also be fast enough to correct the noisy quantum states before they suffer from more errors or decohere completely \cite{fastDec,sparse}. Thus, the design and implementation of efficient decoders for stabilizer codes, both in terms of speed and error correction capabilities, has become a hot topic within the QEC literature.

Surface codes are one of the most promising families of codes for constructing primitive fault-tolerant quantum computers in the near-term \cite{fowlerReview,kitaev,surfaceInt,surfaceTol,surfaceInt2,surfaceInt3,surfaceInt4}. This family of topological codes presents the benefit of being implementable in two-dimensional grids of qubits with local syndrome measurements, or check operators in the surface code terminology, while presenting a high tolerance to quantum noise \cite{surfaceTol}. Considering that many of the physical platforms being considered for constructing quantum processors such as quantum dots, cold atoms or even the mainstream superconducting qubits have architectural restrictions limiting qubit connectivity; surface codes represent a perfect candidate for implementing QEC using those technologies. Recently, major breakthroughs in the field of quantum error correction have been achieved with the first successful experimental implementations of surface codes over superconducting qubit and neutral atom processors \cite{wallraffSurf,googleSurf,neutralQEC}. The result by Google Quantum AI is specially relevant due to the fact that it experimentally proves that QEC improves its performance when the distance of the surface code increases \cite{googleSurf}. Importantly, the recent experiments by Harvard and QuEra also showed such scaling in the application of a transversal CNOT operation as well as other important results for QEC \cite{neutralQEC}.

In this sense, designing decoders for surface codes is a really important task for near-term quantum computers. At the time of writing, there are many potential methods in order to perform the inference of the error from the syndrome data, each of them with their own strong and weak points. Generally, the performance versus decoding complexity trade-off stands for those methods and, therefore, each of them is a potential candidate for being the one selected for experimental quantum error correction depending on the specific needs of the system to be error corrected. Due to the fact that surface code decoding is a relevant and timely topic, the main aim of this tutorial is twofold. The first is to provide a compilation and a comprehensive description of the main decoders for surface codes, while the second is to offer a comparison between those methods in terms of decoding complexity and performance. With such scope in mind, we first provide an introductory section describing the basic notions of stabilizer code theory in Section \ref{secstabcodes} so that surface codes can be introduced in Section \ref{secsurfcode}. We follow by discussing the noise sources in quantum computers and the way they are modelled in Section \ref{secnoisemod}. These sections serve as preliminary background in order to understand the decoding methods that are discussed in the core of the review, Section \ref{secdec}. We describe the most popular surface code decoding algorithms: the Minimum-Weight Perfect Matching (MWPM) decoder, the Union-Find (UF) decoder, the Belief Propagation (BP) decoder and the Tensor Network (TN) decoder. We present their functioning in a comprehensive manner and later discuss not only their performance under depolarizing and biased noise via simulations but also their computational complexity. In addition, we also discuss other decoding methods proposed through the literature for general surface (topological) codes. Also, we review existing and publicly available software implementations of the discussed decoding methods. We then provide a discussion section, Section \ref{secdiscussion}, where we compare the decoders for rotated planar codes and provide an overview of the challenges in decoding surface codes.

\section{Background}\label{secstabcodes}

Quantum computers leverage the principles of quantum mechanics to achieve computational capabilities beyond those of conventional machines, enabling them to process complex computations that would be infeasible for traditional computers. The basic unit of classical computation is the so called \emph{bit}, which represents a logical state with one of two possible values, i.e.

\begin{equation}
    x \in \{ 0,1 \}.
\end{equation}

In stark contrast, the building block of quantum computers are the elements referred as \emph{qubits}. Those quantum mechanical systems, are two-level quantum systems that allow coherent superpositions of them. In this sense, a qubit can be described as a vector in a two-dimensional complex Hilbert space, $\mathcal{H}_2$ \cite{nielsen}:

\begin{equation}
    \ket{\psi} = \alpha \ket{0} + \beta \ket{1},
\end{equation}\label{qubiteq}

where $\alpha,\beta\in\mathbb{C}$ and $\{\ket{0},\ket{1}\}\in\mathcal{H}_2$, usually referred as the computational basis \cite{nielsen}, form an orthonormal basis of such Hilbert space. In this sense, qubits allow to have linear combinations of the two orthonormal states.

This and other properties of quantum mechanics such as entanglement allow for a series of advantages in computation (speedups in algorithm complexity \cite{shorFact,exponentAdv}), or in communications  (supperadditivity of the quantum channel capacity \cite{super1}), among others. Nevertheless, such promises are put in question by the inherent noise present in these quantum mechanical systems. As good as it is to make use of quantum unique properties such as superposition or entanglement, quantum noise does also follow different laws and, thus, it is somewhat different to the noise in classical computers/communications. Classical bit noise can be summarized to flips, a logical operation which would turn the $0$ into $1$ and vice versa. We refer to Section \ref{secnoisemod} for understanding quantum noise in general, but as it will be seen there, for the case of a single qubit the noise can be described by means of Pauli channels. The Pauli channel refers to a stochastic model where a qubit can suffer from bit-flips, phase-flips or bit-and-phase-flips each with some probability. With this consideration in mind, these noise effects are described by the elements of the Pauli group \cite{nielsen}, $\mathcal{G}_1$, whose generators are the Pauli matrices $\langle X, Y, Z \rangle$. Those operators transform the state of an arbitrary qubit as in eq.\ref{qubiteq} in the following way:

\begin{equation}
\begin{split}
X\ket{\psi} = & \alpha \ket{1} + \beta \ket{0}, \\
Z \ket{\psi} = &\alpha \ket{0} - \beta \ket{1}, \\
Y \ket{\psi} = iXZ \ket{\psi}  =& i(\alpha\ket{1} - \beta \ket{0}).  
\end{split}  \label{eqop}
\end{equation}

Note that the the $Y$ operator not only does perform a bit-and-phase-flip operation on the arbitrary quantum state but also changes its overall phase. However, neglecting the global phase has no observable physical consequences and, thus, it is often completely ignored from the point of view of quantum error correction \cite{negglobphase}.

Dealing with these noise processes is a task of vital importance if complex quantum algorithms are meant to be executed reliably. In this context, there are two main approaches to deal with noisy quantum computations: quantum error mitigation (QEM) and quantum error correction (QEC). The two approaches have shown to be complementary with recent papers proposing schemes such as distance scaled zero-noise extrapolation (DSZNE) \cite{DSZNE}. QEM attempts to evaluate accurate expectation values of physical observables of interest by using noisy qubits and quantum circuits \cite{DSZNE,QEM4}, while the main objective of QEC is to obtain qubits and computations that experience error rates that are arbitrarily low. For example, it is considered that a logical error probability in the order of $10^{-14}-10^{-15}$ is required to run Shor's algorithm fault-tolerantly \cite{fowlerReview}. There are many approaches to quantum error correction, but the general idea behind QECCs is to protect the information of a number of qubits $k$, named \emph{logical qubits}, within a larger number of qubits $n$, named \emph{physical qubits} in a way that makes the whole system tolerant to a number of errors. Within these QECCs, many lay within the framework named Quantum Stabilizer Coding \cite{gottesman}. Stabilizer codes allow for a direct mapping of many classical methods into the quantum spectrum, making them very useful \cite{gottesman}. Since the surface code belongs to the family of quantum stabilizer codes, it is pertinent to cover the basics of such QEC constructions.

\subsection{Stabilizer Codes}

Quantum error correction is based on protecting the state of $k$ logical qubits by means of $n$ physical qubits so that the protected qubits operate as if they were noiseless. Note that the set of $n$-fold Pauli operators, $\mathcal{P}_n = \{\mathrm{I,X,Y,Z}\}^{\otimes n} $, together with the the overall factors $\{\pm 1, \pm i\}$ forms a group under multiplication, usually named as the $n$-fold Pauli group, $\mathcal{G}_n$ \cite{gottesman}. Generally, an unassisted\footnote{The so called entanglement assisted QECCs make use of Bell states as ancilla qubits for constructing the codes \cite{EAQEC}. Here we restrict our discussions to the regular stabilizer codes, where the ancilla qubits are usually defined by $\ket{0}$ states \cite{gottesman}.} $[[n,k,d]]$ stabilizer code is constructed using an abelian subgroup $\mathcal{S}\subset\mathcal{G}_n$ defined by $n-k$ independent generators\footnote{The stabilizer set will have $2^{n-k}$ elements up to an overall phase.} so that $k$ logical qubits are encoded into $n$ physical qubits with distance-$d$ \cite{gottesman, reviewpat, EAQEC}. The distance of a stabilizer code is defined as the minimum of the weights\footnote{The weight of an error is defined as the number of non-trivial elements of a Pauli operator in $\mathcal{G}_n$.} of the Pauli operators that belong to the normalizer of the stabilizer, which consists of the elements that commute with the generators but do not belong to the stabilizer \cite{josuPhD}. Thus, it is related to the maximum weight of the errors that can be corrected by the code. The codespace $\mathcal{T}(\mathcal{S})$ associated to the stabiliser set is defined as:

\begin{equation}
    \mathcal{T}(\mathcal{S}) = \{ \ket{\bar{\psi}}\in\mathcal{H}_2^{\otimes n} : M\ket{\bar{\psi}} = \ket{\bar{\psi}}, \  \forall M \in \mathcal{S} \},
    \label{Codespace}
\end{equation}
i.e. the simultaneous $+1$-eigenspace\footnote{Note that the subgroup should not contain any non-trivial phase times the identity so that the simultaneous $+1$-eigenspace spanned by the operators in $\mathcal{S}$ is non-trivial \cite{gottesman,EAQEC}.} defined by the elements of $\mathcal{S}$. We use the notation $\ket{\bar{\psi}}$ to refer that this state is within the codespace. Within that code, the physical qubits will experience errors that belong to the Pauli group\footnote{This comes from the so called \textit{error discretization} that arises from the Knill-Laflamme theorem \cite{nielsen,knillLaff}.} $\mathcal{G}_n$.

In stabilizer codes, the set of stabilizer generators of $\mathcal{S}$ are named checks and, thus, there will be $n-k$ checks. In order to perform quantum error correction, one must perform measurements of the checks in order to obtain information of the error that has occurred. The classical information obtained by measuring the checks of a stabilized code is named the \emph{syndrome} of the error, $\bar{s}$. Due to the fact that quantum measurements destroy superposition, these measurements must be done in an indirect way so that the codestate is not lost. This can usually be done by means of a Hadamard test that requires ancilla qubits that are usually referred as measurement qubits\footnote{Note that, for stabilizer codes, the measurement of the checks is responsible of the error discretization \cite{nielsen}.} \cite{approximatingDec}. Therefore, the error syndrome, $\bar{s}$, is defined as a binary vector of length $n-k$, $\bar{s} \in \mathbb{F}_2^{n-k}$. Given a set of checks, $\{ M_1, M_2, ..., M_{n-k}\}\in\mathcal{S}$, and a Pauli error, $E \in  \mathcal{G}_n$, the $i^{th}$ element of the syndrome will capture the commutation relationships of the error and the $i^{th}$ check. This comes from the common knowledge that any two elements of $\mathcal{G}_n$ commute or anticommute. Thus, this commutation relationship is captured by the syndrome as

\begin{equation}
    EM_i = (-1)^{s_i}M_iE,
    \label{syndromeformula}
\end{equation}

where $s_i$ represents the $i^{th}$ element of the syndrome vector.

One interesting thing to note from this construction is that, since the codespace is not altered by the application of stabilizers (recall eq.\ref{Codespace}), a channel error that coincides with those operators will have a trivial action on the codestate, i.e.

\begin{equation}
    E \ket{\bar{\psi}} =  \ket{\bar{\psi}},
    \label{coseet}
\end{equation}

if $E\in\mathcal{S}\subset\mathcal{G}_n$. In this sense, there will be different error operators that share the same error syndrome that affects the encoded quantum state in a similar manner. This phenomenon is usually termed as \emph{degeneracy}. The concept of error degeneracy has the consequence that the Pauli space that represents all possible error operators is not just partitioned into syndrome cosets, but also into degenerate error cosets\footnote{Note that degeneracy is somehow different in the entanglement-assisted paradigm \cite{EAQEC}} \cite{reviewpat}. Specifically, the Pauli group is partitioned in $2^{n-k}$ cosets that share error syndrome, and each of those cosets will be partitioned in to $2^{2k}$ cosets that contain $2^{n-k}$ errors that are degenerate among them \cite{super1,reviewpat}. How degenerate a quantum code is depends on the difference between the weight of its stabilizer generators and its distance. If $w(M_k) << d$, $\forall k \in {0,\ldots, n-k}$, where $M_k\in \mathcal{S}$ denotes a stabilizer generator and $w$ denotes the weight, then each logical coset (equivalence class) will contain many operators of the same weight and the code will be highly degenerate. In cases where $w(M_k) = d$, the code will be non-degenerate.

In summary, the checks give us a partial information of the error operator that corrupted the encoded information. Since the aim of quantum error correction is to recover the noiseless quantum state, an estimate of the channel error must be obtained so that the noisy state can be corrected. The process of estimating the quantum error from the measured syndrome is named decoding. Once a guess of the error, $\hat{E} \in \mathcal{G}_n$, is obtained by the decoder, it will be operated on the code. If the estimation turns out to be correct, the noisy quantum state will be successfully corrected since the elements of the Pauli group are self-inverse. Moreover, if the estimated error is not the exact element of the Pauli group but it belongs to the same degenerate coset, the correction will also be successful \cite{reviewpat}. Finally, whenever the estimated error does not fulfill any of those two cases, the correction operation will result in a non-trivial action on the logical qubits encoded in the state, implying that the error correction method has failed.

\subsection{The decoding problem}

The decoding problem in QEC is different from the decoding problem in classical error correction due to the existence of degeneracy. In this sense, the following classification can be done as a function of the decoding problem being solved \cite{reviewpat,PcompleteDec}:
\begin{itemize}
    \item \textbf{Quantum maximum likelihood decoding (QMLD):} those are an extrapolation of the classical decoding methods where the estimation problem is described as finding the most likely error pattern associated to the syndrome that has been measured \cite{reviewpat,PcompleteDec}. Mathematically, 
    \begin{equation}\label{eq:QMLD}
        \hat{E} = \Argmax_{E\in\mathcal{G}_n} P(E|\bar{s}),
    \end{equation}
    where $P$ refers to the probability distribution function of the errors. Due to the fact that degeneracy is ignored by this type of decoding, it is also referred as non-degenerate decoding. For an independent, identically distributed noise model, this decoding rule results in a minimum-weight decoding rule, i.e. looking for the error with minimum weight that satisfies the measured syndrome.
    \item\textbf{Degenerate quantum maximum likelihood decoding\footnote{Note that the literature sometimes names this decoding rule as ``maximum likelihood decoding" \cite{mps}. However, both rules are actually maximum likelihood estimation problems. Their difference is the likelihood function to maximize. Therefore, we stick to the terminology in \cite{PcompleteDec}.} (DQMLD):} due to the existence of degenerate errors that form cosets of errors that affect the coded state in a similar manner, it is possible that a the probability of occurrence of the coset containing the most probable error sequence (in the sense of QMLD) is smaller than other coset allowed by the measured syndrome. Thus, the QMLD decoder will be suboptimal as it ignores the degenerate structure of stabilizer codes. Therefore, DQMLD decoding can be described mathematically as \cite{reviewpat,PcompleteDec}:
    \begin{equation}\label{eq:DQMLD}
        \hat{L} = \Argmax_{L\in\mathcal{L}} P(L|\bar{s}),
    \end{equation}
    where by $\mathcal{L}$ we refer to the coset partition of $\mathcal{G}_n$ and $L$ a coset belonging to such partition. Note that once the coset is estimated, the decoding operation will be the application of any of the elements of such coset since the operation to the logical state is the same for all the elements of such coset.
\end{itemize}

Therefore, the optimal decoding rule for stabilizers is the DQMLD. However, it was proven that QMLD falls into the NP-complete complexity class (similar to the classical decoding problem), while DQMLD belongs to the $\#$P-complete class \cite{PcompleteDec}. The latter is computationally much harder than the other, implying that the optimal decoding rule may pose serious issues for the fast decoding needed in quantum error correction \cite{PcompleteDec}. Therefore, even if the optimal rule for decoding and, thus, code performance is obtained by using DQMLD, non-degenerate decoding is important and widely used as it is less expensive in terms of computational complexity.

\section{The surface Codes}\label{secsurfcode}

\begin{figure}
    \centering
    \includegraphics[width = 0.7\columnwidth]{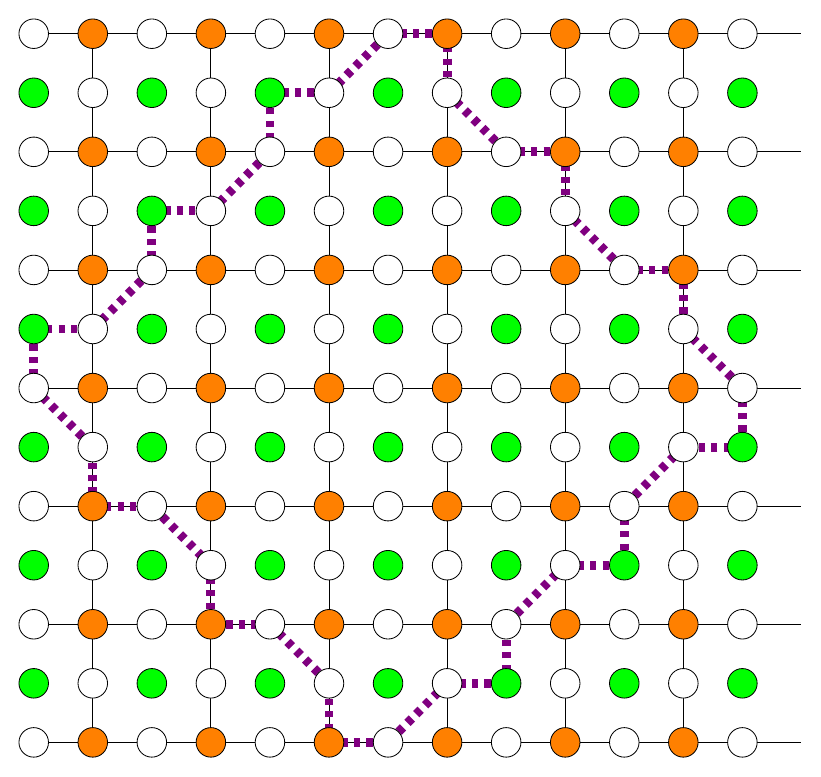}
    \caption{Distance-7 planar code. Data qubits are represented as white circles, $X$ and $Z$-checks are represented as green and orange circles respectively. }
    \label{planarcode}
\end{figure}

Surface codes are a family of quantum error correcting codes in which the information of logical qubits is mapped to a set of physical qubits, commonly named as data qubits, which are displayed in a lattice array. Moreover, the measurement qubits that are used to measure the checks are also displayed within the lattice. Alexei Kitaev first proposed the concept of surface codes in his prominent work \cite{kitaev}, where the qubits were displayed in a torus-shaped lattice. This toric code has periodic boundary conditions and it is able to protect two logical qubits. Nevertheless, such toric displacement of the qubits makes the hardware implementation and logical qubit connectivity complicated \cite{lattice} since many experimental implementations such as superconducting hardware, for example, may require the system to be placed in a two-dimensional lattice. Thus, the so called planar code encoding a single logical qubit is obtained by stripping the periodic boundary conditions from the toric code \cite{fowlerReview,kitPlanar}. Specifically, it will be a $[[d^2+(d-1)^2,1,d]]$ QECC. Furthermore, the number of data and measurement qubits used for protecting the single logical qubit can be lowered down, i.e. the rate of the code can be increased\footnote{The rate of a quantum error correction code is defined as the ratio between the number of logical qubits and the number of physical qubits, i.e. $R_\mathrm{Q}=k/n$}, by considering a specific set of data qubits and checks within the planar code. The obtained code is usually known as the rotated planar code, which will be a $[[d^2,1,d]]$ \cite{surfaceInt4}. FIG. \ref{planarcode} shows a distance-7 planar code where the dashed lines indicate the subset of qubits that form the rotated planar code with the same distance. In this tutorial we will consider the rotated planar code in a square lattice with Calderbank-Shor-Steane (CSS) structure \footnote{CSS codes refer to stabilizer codes admitting a set of generators that are either $X$ or $Z$-check operators. This means that the check operators will consist of either tensor products of identities with $X$ or with $Z$ operators exclusively \cite{CalSho,Steane}.} due to its practicality and relevance at the time of writing. Note that this code is the one that has been recently implemented experimentally by Wallraff's group at ETH Zurich \cite{wallraffSurf} and by the Google Quantum AI team \cite{googleSurf}. Nevertheless, the decoding methods discussed in this tutorial apply for all surface codes including the tailored versions proposed through the past years \cite{xzzx, tuckettxy, fragile, recMWPM} or the different lattices considered \cite{surfaceInt4,heavyhexagon,lattices}. From here through the end of the review, we will use the terms rotated planar code and rotated code interchangeably so as not to sound repetitive.

%  XZZX, tuckett, fragile, ton recursive, hexagonal, quantum error correction ZOO después de google Surf
% xzzx, tuckettxy,  fragile, ton recursive, recMWPM, heavyhexagon

\subsection{Stabilizer and check structure}
\begin{figure}
    \centering
    \includegraphics[width = 0.3\textwidth]{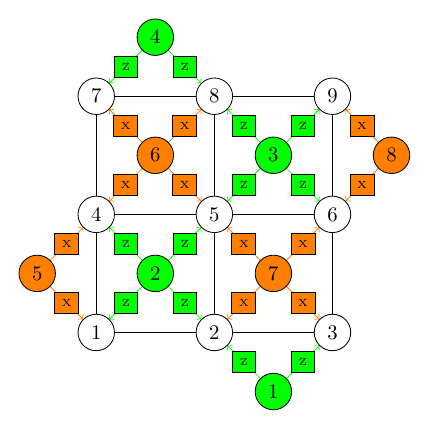}
    \caption{Distance-3 rotated planar code.  The stabilizing operators that the checks yield over their adjacent data qubits are denoted through green and orange squares, which act on their nearest data qubits.} 
    \label{rotatedplanardemo}
\end{figure}

Due to the structure of their stabilizers, CSS quantum surface codes have two types of checks: $X$-checks and $Z$-checks. The former detect $X$-errors, while the latter detect $Z$-errors. FIG. \ref{rotatedplanardemo} portrays the structure of the check operators in a distance-3 CSS rotated planar code. As it can be seen in the figure, this structure fulfills the condition that the stabilizer generators form an abelian group \cite{gottesman}. This can be seen by explicitly studying the stabilizer generators:

\begin{equation} \label{eq:stab_gens_d3}
    \begin{aligned}
        s_1 &= Z_2Z_3,\\
        s_2 &= Z_1Z_2Z_4Z_5,\\
        s_3 &= Z_5Z_6Z_8Z_9,\\
        s_4 &= Z_7Z_8,\\
        s_5 &= X_1X_4,\\
        s_6 &= X_4X_5X_7X_8,\\
        s_7 &= X_2X_3X_5X_6,\\
        s_8 &= X_6X_9,
    \end{aligned}
\end{equation}

where $s_i$ denotes the $i$th stabilizer generator or check. The locality of the check operations ensures that checks that are far apart commute with each other, while adjacent checks commute either because they are of the same type and, thus, apply the same operators to their adjacent data qubits; or because they anti-commute for two data qubits at the same time, making the whole operators to commute. For example,

\begin{equation}
\begin{split}
    M_\mathrm{x} M_\mathrm{z}\ket{\bar{\psi}} &= \mathrm{X}_1 \mathrm{X}_2 \mathrm{X}_3 \mathrm{X}_4 \mathrm{Z}_3 \mathrm{Z}_4 \mathrm{Z}_5 \mathrm{Z}_6\ket{\bar{\psi}} \\
    &= \mathrm{X}_1 \mathrm{X}_2 \mathrm{X}_3 \mathrm{Z}_3 \mathrm{X}_4 \mathrm{Z}_4 \mathrm{Z}_5 \mathrm{Z}_6\ket{\bar{\psi}} \\
    &= \mathrm{X}_1 \mathrm{X}_2 (-\mathrm{Z}_3 \mathrm{X}_3) (-\mathrm{Z}_4 \mathrm{X}_4) \mathrm{Z}_5 \mathrm{Z}_6\ket{\bar{\psi}} \\
    &=  \mathrm{Z}_3 \mathrm{Z}_4 \mathrm{Z}_5 \mathrm{Z}_6 \mathrm{X}_1 \mathrm{X}_2 \mathrm{X}_3 \mathrm{X}_4 \ket{\bar{\psi}} \\
    &= M_\mathrm{z} M_\mathrm{x}\ket{\bar{\psi}} = \ket{\bar{\psi}},
\end{split}
\label{commutationchecks}
\end{equation}
where $M_z$ and $M_x$ are two arbitrary adjacent $X$ and $Z$-checks in the bulk of the code, respectively, and the data qubits labeled with $3$ and $4$ are located in between both checks.

Surface codes, as do all CSS codes, exhibit a property known as \textit{transversal} $R_Z$, by which initializing all data qubits in the $\ket{0}$ state and measuring the stabilizers results in the fault-tolerant preparation of the logical zero state $\ket{\bar{0}}$ (the same is true for $\ket{\bar{1}}, \ket{\bar{+}},$ and $\ket{\bar{-}}$ logical state preparation). This property is also known as transversal initialization. For this reason, computations/experiments that employ surface codes all typically begin by initializing the data qubits in the $\ket{0}$ state \cite{wallraffSurf,googleSurf}.

\begin{figure}
\centering
    \includegraphics[width = 0.5\textwidth]{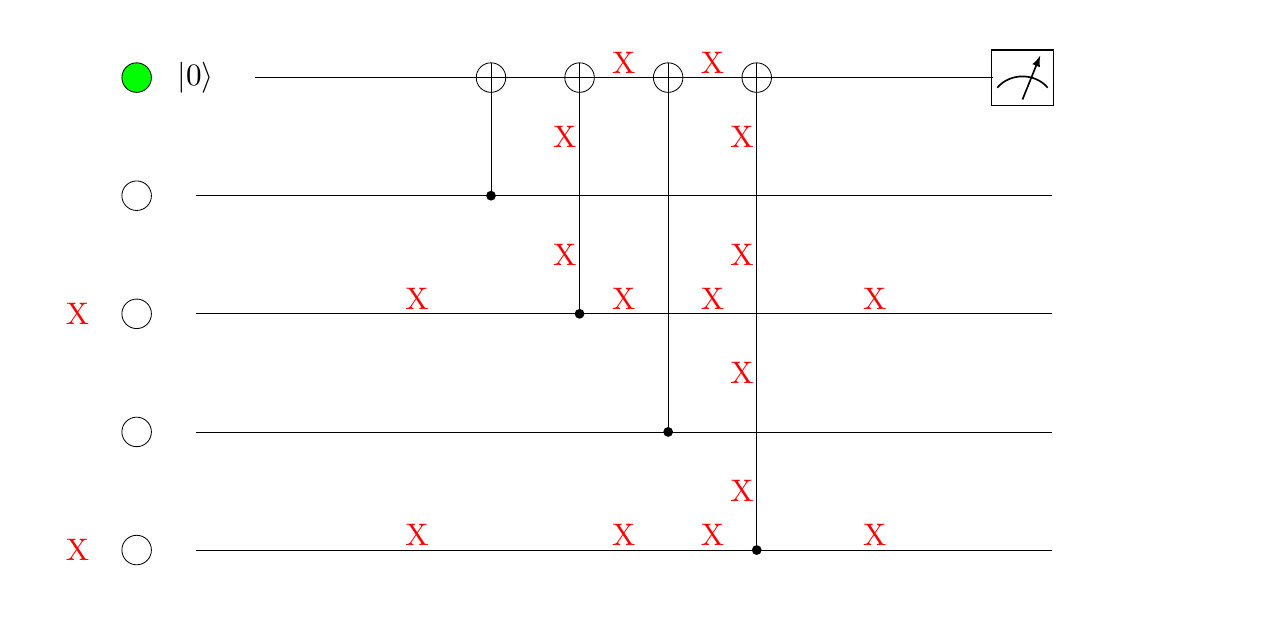}
    \includegraphics[width = 0.5\textwidth]{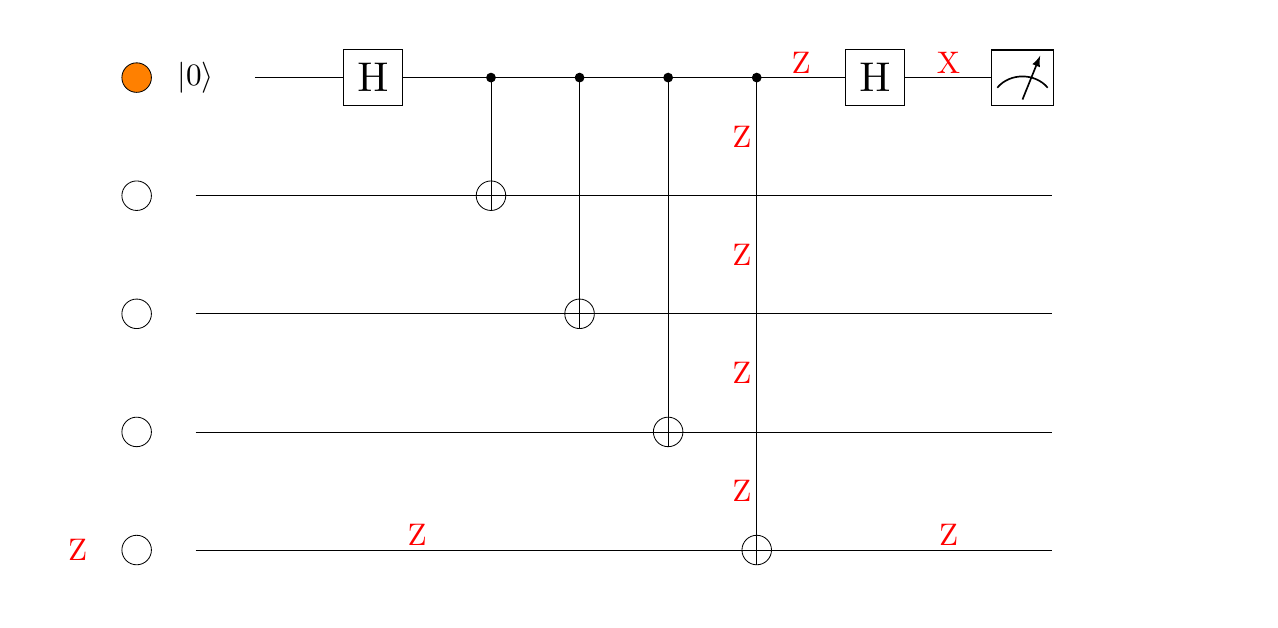}
    \caption{Stabilizing circuits of the $X$-checks (top) and the $Z$-checks (bottom). The green and yellow circles indicate the check qubits, and the grey qubits their 4 adjacent data qubits. In both cases an example of an error in the data qubits is introduced altogether with the path it follows within the circuit. }
    \label{stabcircuit}
\end{figure}

As explained before, the data qubits of surface codes may undergo a Pauli error. The syndrome of the associated error is measuring the check operators, which corresponds to the quantum circuits shown in FIG. \ref{stabcircuit}. The top circuit represents an $X$-check and the bottom circuit a $Z$-check. As seen in such figure, if an odd number of adjacent data qubits suffer from an $X$ or $Z$-error, the measurement of the respective $X$ or $Z$-checks will be triggered. However, as seen in the top image from FIG. \ref{stabcircuit}, in the event of an even number of errors, those will cancel, due to their unitary nature, and no error will be detected by the check operator measurement.

To sum up, whenever an error consisted of an odd number of $X$ or $Z$-errors affects the data qubits surrounding a check operator, the circuit from the picture will make said errors propagate to the measurement qubit associated to such check, changing the measurement and thus enunciating that an error has occurred in its vicinity \cite{fowlerReview}. 

\subsection{Types of errors and code threshold}
\begin{figure}
\centering
    \includegraphics[width = 0.4\textwidth]{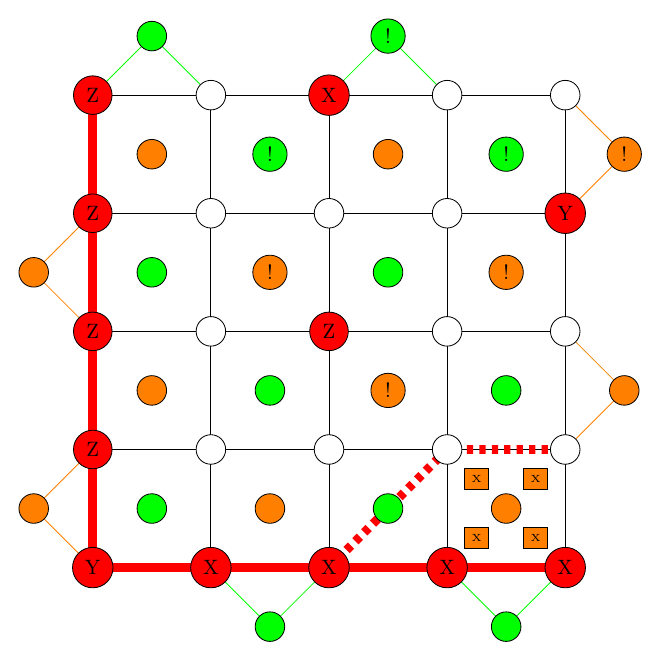}
    \includegraphics[width = 0.4\textwidth]{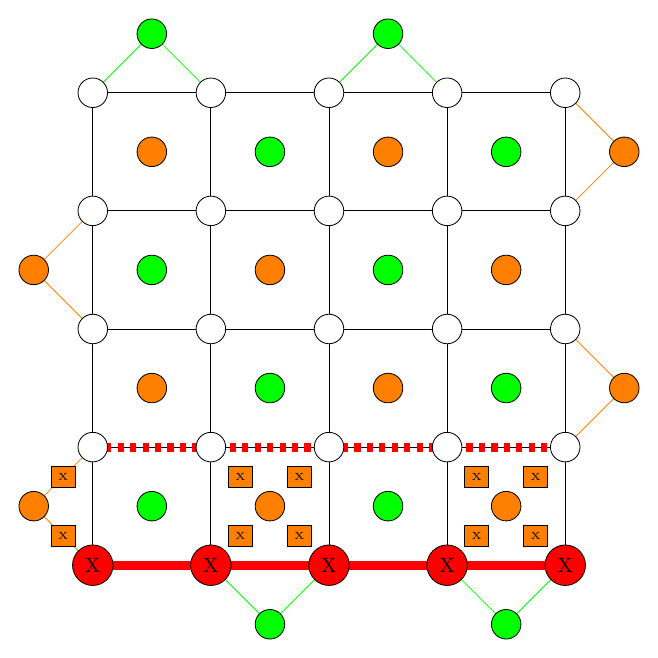}
    \includegraphics[width = 0.4\textwidth]{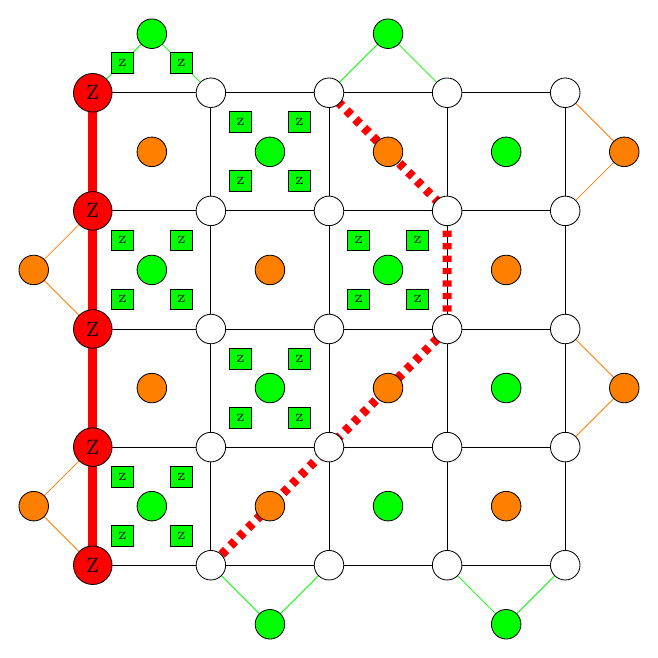}
    \caption{The figure illustrates three rotated codes undergoing a logical $Y_L, X_L$ and $Z_L$ operator from top to bottom. Physical non-trivial Pauli operators represented by red circles. Check measurements resulting in non-trivial syndrome elements are marked with an exclamation mark. The red thick line represent the shape of logical operators, while the red dashed line shows the changed path of a logical operator after interacting with a stabilizer the physical qubit operators of which are denoted as squares. }
    \label{rotateddegeneracy}
\end{figure}

In FIG. \ref{rotateddegeneracy} we show different types of errors that may arise in the rotated planar code. Beginning with the top figure, in the upper right section of the code, there are three isolated Pauli errors, namely an $X$, a $Y$ and a $Z$-error, which cause the adjacent susceptible checks to exhibit non-trivial syndrome elements upon measurement. Note that the $Y$-error triggers both the $X$ and $Z$-checks that are adjacent to such data qubit. This happens because $Y$-errors are a combination of $X$ and $Z$-errors, neglecting the global phase, as seen in eq.\eqref{eqop}. These isolated Pauli errors will be detected by the code, and using the syndrome information, the decoder will attempt to estimate them.

In addition, other errors forming vertical and horizontal chains along the boundaries are represented in all figures of FIG. \ref{rotateddegeneracy} and denoted with bold red lines. As seen in the figure, those error chains are not detected by the code since they do not trigger any of the surrounding measurement qubits. This is because each susceptible check is connected to two of the Pauli operators, i.e. it refers to the previously described case where there is an even number of operators acting on each of the checks. These error chains act non-trivially on the codestate without being part of the stabilizer group while presenting a trivial syndrome, i.e. they belong to the normalizer of the code. This type of errors receive the name of logical errors \cite{fowlerReview}. Specifically, in FIG. \ref{rotateddegeneracy}, both the vertical and horizontal chains from the top figure, the horizontal chain from the middle figure and the vertical chain from the bottom figure constitute the $Y_L, X_L$ and $Z_L$ operators respectively. Notice that, in the top figure, the anti-commutation relation $\mathrm{X}_L\mathrm{Z}_L = - \mathrm{Z}_L\mathrm{X}_L$ is preserved through the bottom left data qubit. Moreover, these are just three examples of the physical Pauli operators producing the logical errors. If a logical operator is applied, the resulting state will still be within the codespace, so eq.\eqref{Codespace} will still be preserved for the new sate and it will remain invariant upon the application of stabilizer operators. This can be reflected in all three figures within FIG. \ref{rotateddegeneracy}, where the application of three arbitrary stabilizers (one for each figure), produces another chain of Pauli operators spanning from one boundary to the other, commuting with the stabilizer set. The new operator, which is denoted by a dashed line and would act non-trivially on the data qubits it goes over, is different in terms of physical Pauli operators, but acts equally on the codespace.

When protecting against quantum noise, experiencing a logical error\footnote{Note that here we refer to the unwanted event that the codestate is altered by a logical error due to the noise. However, whenever the logical state is wanted to be manipulated with the scope of computation, the way of applying logical Pauli gates to the logical qubit is by means of these logical operators on the physical qubits.} is fatal since those alter the information stored in the code without any indication from the checks and, thus, cannot be detected nor corrected. One way of mitigating the impact of logical errors is to increase the size of the surface code. By doing so, the code distance increases and, therefore, the minimum number of Pauli operators needed to form a logical error will also be higher, making such event to be more improbable. This comes from the fact that logical errors belong to the normalizer of the code and the distance is defined as the minimum weight of those Pauli operators. Although it may seem intuitive that a larger surface code would perform better, this is not always the case. The momentous result in quantum computing named the threshold theorem 
states that, by means of quantum error correction, the logical error rate experienced by a quantum computer can be suppressed to an arbitrarily low level provided that the physical error rate of the data qubits is below a certain value \cite{shorQEC,kitaev,thresh1,thresh2}. This closely relates with the concept of code threshold ($p_{th}$), which refers to the maximum physical error rate for which increasing the distance of a code in consideration results in a better performance for a certain error model. Note that not all QECCs present a code threshold, e.g. the Bacon-Shor code \cite{BaconShor}, implying that increasing distance does not always result in lowering the logical error rate. Note that not having threshold is equivalent to saying that the threshold is zero. The considered rotated planar surface code presents non-trivial code thresholds for the noise models considered in theoretical QEC. Thus, as long as the physical error rate is below $p_{th}$, increasing the size of such code will lead to a better performance.

The probability threshold is a useful metric for benchmarking the performance of the surface code under a particular decoder. However, the value of the threshold does not only depend on the code and decoder in consideration but also on the structure of the underlying noise that affects the physical qubits of the code. In the following section, we will discuss the origin of quantum noise and the most relevant noise models considered in QEC.

\section{ Noise models }\label{secnoisemod}

The main obstacle to construct quantum computers is the proneness of quantum information to suffer from errors. There are many error sources that corrupt quantum information while being processed such as state preparation and measurement (SPAM) errors or errors introduced by imperfect implementations of quantum gates, for example. Many of the errors occur due to the fact that the technology used for manipulating qubits is imperfect \cite{fowlerReview,preskill}. However, qubits do also suffer from errors due to their undesired interaction with their surrounding environment. This last source of errors is named as environmental decoherence, and corrupts quantum information even if the quantum system is left to evolve freely \cite{nielsen,decoherenceBench,approximatingDec,TVQC}. Thus, decoherence poses a fundamental problem to the field of quantum information processing since its existence does not depend on imperfect implementations of qubit manipulations or measurements, which we may deem as engineering problems\footnote{It may seem unfair to state that decoherence is not related to engineering since the way qubits are constructed fundamentally determines how fast a qubit will decohere. However, even in the case that very long decoherence times are obtained, it would not be possible to apply an arbitrarily large amount of perfect quantum gates, since at some point the quantum information would be corrupted.}. Therefore, quantum error correction will be necessary if we want to run arbitrarily large quantum algorithms with enough precision. It is noteworthy to state that there are approaches to construct qubits that are naturally protected from decoherence by means of topological order or physical symmetries \cite{fastDec}. These methods are usually referred to as passive quantum error correction. However, in this review we are focusing on active error correction, in the sense that errors are corrected by actively detecting and correcting them.

The aim of this section is to describe the usual noise sources considered in theoretical quantum error correction as well as the motivation of certain error models, e.g. biased noise. In this sense, the usual error models are the code capacity noise, phenomenological noise and circuit-level noise. However, it is important to state that many important error sources are usually neglected by this models such as coherent errors, non-Markovian noise and correlated errors arising from crosstalk or cosmic rays (in superconducting qubits) \cite{surfaceInt,googleSurf,cosmic,nonMark}.

\subsection{Decoherence}
In general, decoherence comprises several physical processes that describe the qubit environment interaction, and the nature of those processes depends on the qubit technology\footnote{We consider conventional qubit-based quantum computation, i.e. discrete two-level coherent quantum systems.}, i.e. superconducting qubits, ion traps or NV centers, for example. However, most of those physical interactions can be grouped into three main decoherence mechanisms\footnote{Here we are considering noise sources that operate on the computational subspace of the qubit, i.e. transitions to other possible levels are neglected by now. This will be described later on.} since their operational effect on the two-level coherent system that is the qubit is the same \cite{approximatingDec,SchlorPhD}:
\begin{itemize}
    \item \textbf{Energy relaxation or dissipation:} this mechanism includes the physical processes in which a quantum mechanical system suffers from spontaneous energy losses. For example, atoms in excited states tend to return to the ground state by spontaneous photon emission. The amalgamation of relaxation processes is described by the so called relaxation time, $T_1$, which is the characteristic timescale of the decay process \cite{approximatingDec,TVQC,SchlorPhD}. 
    \item \textbf{Pure dephasing:} these physical processes involve the corruption of quantum information without energy loss. For example, this occurs a photon scatters in a random manner when going through a waveguide. Pure dephasing is also quantified by the characteristic timescale of the decay process \cite{TVQC,SchlorPhD}, which in this case is named as pure dephasing time, $T_\phi$.
    \item \textbf{Thermal excitation:} this refers to the undesired excitation of the qubit from the ground state to the excited state and the assisted relaxation from the excited state to the ground state caused by the finite temperature of the system \cite{nielsen,SchlorPhD,thermal}. Every qubit platform will be at a finite temperature in the real world, but the contribution of thermal excitation can be usually neglected when qubits are cooled down significantly\footnote{Note that superconducting qubits are cooled down to $T\approx 20$ $mK$, for example \cite{wallraffSurf,googleSurf,decoherenceBench,SchlorPhD}.}. Generally, the temperature of the system, $T$, and the energy levels of the ground state and the excited state quantify this effect on the quantum system.
\end{itemize}
For Markovian\footnote{Markovian noise assumes that the system-environment interaction is memoryless \cite{fastDec}.} noise, there are many ways in which this set of physical interactions can be mathematically described such as the Gorini-Kossakowski-Lindblad-Sudarshan master equation \cite{GKLS1,GKLS2} or the quantum channel formulation \cite{approximatingDec,choi}. In the context of QECC, quantum channels are used to describe noisy evolution. In general, quantum channels are linear, completely-positive, trace-preserving maps between spaces of operators. As a consequence of those properties, quantum channels fulfill the Choi-Kraus theorem, implying that the application of such maps on a density operator $\rho$ can be written as the following decomposition \cite{nielsen,approximatingDec,choi}:
\begin{equation}
    \mathcal{N}(\rho) =\sum_k E_k \rho E_k^\dagger,
\end{equation}
where the $E_k$ matrices are named Kraus or error operators, and should fulfill $\sum_k E_k^\dagger E_k = \mathrm{I}$ since the quantum channels must be trace-preserving. Thus, quantum channels are characterized by sets of Kraus operators that are associated to some physical interaction of the qubit with its surrounding environment. 

The generalized amplitude and phase damping channel describes the evolution of a quantum state when decoherence arises from the three qubit-to-environment interactions presented before \cite{approximatingDec,TVQC,thermal,gadc1}. Such channel consists of a generalized amplitude damping channel describing the thermal and relaxation interaction \cite{thermal,gadc1} and of a pure dephasing channel \cite{approximatingDec,TVQC}. The action of the generalized amplitude damping channel is described by the damping parameter, $\gamma$, and the probability that the ground state is excited by finite temperature, $N$. The damping parameter relates to the relaxation time of the qubit as, $\gamma(t) = 1-e^{-(2n_\mathrm{th}+1)t/T_1}$, with $t$ being the evolution time, while $n_\mathrm{th}$ and $N(n_\mathrm{th})$ depend on the temperature and the energy gap of the system \cite{thermal,gadc1}. The Kraus operators of this channel are:
\begin{align}
   & E_0  = \sqrt{1 - N(n_\mathrm{th})} \begin{pmatrix}
    1 & 0 \\ 0 & \sqrt{1- \gamma(t)}
    \end{pmatrix}, \\
    & E_1 = \sqrt{1 - N(n_\mathrm{th})}\begin{pmatrix}
        0 & \sqrt{\gamma(t)} \\ 0  &  0
    \end{pmatrix}, \\
    & E_2  = \sqrt{N(n_\mathrm{th})} \begin{pmatrix}
    \sqrt{1- \gamma(t)} & 0 \\ 0 & 1
    \end{pmatrix}, \text{ and} \\
    & E_3 = \sqrt{N(n_\mathrm{th})}\begin{pmatrix}
        0 & 0 \\ \sqrt{\gamma(t)}  &  0
    \end{pmatrix}.
\end{align}
Whenever the thermal excitation is considered negligible, $n_\mathrm{th}\approx 0$ and $N(n_\mathrm{th})=0$; and such channel reduces to an amplitude damping channel. The pure dephasing channel is described by the so called scattering probability, $\lambda$, which relates to the pure dephasing time as, $\lambda(t) = 1-e^{-2t/T_\phi}$, with $t$ the evolution time again \cite{TVQC,josuPhD}. The Kraus operators of this channel are:
\begin{align}
    &E_0 = \begin{pmatrix}
        1  & 0 \\ 0 & \sqrt{1 - \lambda(t)}
    \end{pmatrix}, \text{ and}\\
    &E_1 = \begin{pmatrix}
        0 & 0 \\ 0 & \sqrt{\lambda(t)}
    \end{pmatrix}.
\end{align}

In this sense, the complete channel is defined by those parameters its Kraus operators are given by the serial concatenation of the damping and dephasing channels \cite{josuPhD}.

\subsection{Twirled quantum channels}

Therefore, the introduced channel describes the evolution that a qubit undergoes when the considered decoherence processes are considered under the Markovian approximation. The problem with this quantum channel is the fact that it is not possible to simulate it efficiently by means of classical computer as the dimension of the Hilbert state increases exponentially with the number of qubits considered \cite{approximatingDec}. This makes it impossible to construct and simulate efficient error correction codes that will be used for protecting quantum information by using conventional methods. That is why a technique named twirling is usually employed in order to obtain channels that can be managed by classical computers and that capture the essence of more general channels \cite{approximatingDec,TVQC,twirl1,twirl2}. The significance of the twirling method comes after the fact that a correctable\footnote{Correctable codes refers to the fact that the Knill-Lafalamme conditions for quantum error correction are fulfilled \cite{knillLaff}.} code for the twirled channel will also be a correctable code for the original channel \cite{approximatingDec,lemmatwirl} and, thus, we can consider the simplified channels for designing codes that will eventually be successful for the actual noise. Following this logic, the most common twirling operations are the so called Pauli \cite{approximatingDec,twirl1,josuPhD} and Clifford twirl approximations \cite{approximatingDec,TVQC,josuPhD, twirl2}, where the quantum channel in consideration is averaged uniformly with the elements of the Pauli group, $\mathcal{P}$, and the elements of the Clifford group, $\mathcal{C}$, respectively. Twirling the generalized amplitude and phase damping channel with these two groups results in Pauli channels, i.e. those with Kraus operators $\{\sqrt{(1-p_x-p_y-p_z)}\mathrm{I},\sqrt{p_x} \mathrm{X},\sqrt{p_y} \mathrm{Y},\sqrt{p_z} \mathrm{Z} \}$ with probabilities:
\begin{itemize}
    \item \textbf{Pauli twirl:} $p_x=p_y=\frac{\gamma}{4}$ and $p_z = \frac{2-\gamma-2\sqrt{1-\gamma-(1-\gamma)\lambda}}{4}$.
    \item \textbf{Clifford twirl:} depolarizing channel, $p_x=p_y=p_z=\frac{2+\gamma-2\sqrt{1-\gamma-(1-\gamma)\lambda}}{12}$.
\end{itemize} 

The usefulness of Pauli channels resides in the fact that they can be efficiently simulated in classical computers since they fulfill the Gottesman-Knill theorem \cite{nielsen,GKthm}. Thus, we can use them in order to construct and simulate quantum error correction codes that will then be useful to protect qubits from the more general noise \cite{approximatingDec}. Note that the Clifford twirl approximation results in a depolarizing or symmetric Pauli channel where all the errors are equiprobable, while the Pauli twirl approximation presents a probability bias respect to the $\mathrm{Z}$ types of errors. Therefore, the latter is usually referred as the biased noise model, where the bias\footnote{The bias is usually defined using the so called Ramsey dephasing time, $T_2$, that includes the dephasing induced by relaxation \cite{TVQC}. In this sense, the parameters relate as $1/T_2 = 1/(2T_1) + 1/T_\phi$ \cite{TVQC,decoherenceBench}.} is defined as $\eta = p_z/(p_x+p_y)\approx T_1/T_2 - 1/2$ \cite{inid}. The bias of the channel varies significantly as a function of the technology or even as a function of the qubit of a processor being considered \cite{approximatingDec}.

\subsection{Noise models for multiple qubits}
There are several ways in order to construct the $n$-qubit quantum channel that is required to study the action of the quantum error correction code being designed \cite{approximatingDec,inid,corr1}. The literature on QEC usually assumes that each of the qubits of the system experience noise independently\footnote{Note that the independence assumption is not generally true since correlated noise has been considered in the literature. For surface codes, correlation between the nearest qubits of the code is considered whenever this scenario is studied, assuming that the other ones are far enough so that the correlations are negligible \cite{corr1,QEM2}. However, considering the channel to be memoryless is considered to be a reasonable assumption.}. In this sense, the following $n$-qubit twirl approximation channels will be considered:
\begin{itemize}
    \item \textbf{Independent and identically distributed (i.i.d.):} in this model each of the qubits will have the same experience of suffering a particular Pauli error \cite{approximatingDec}. Joining this with the fact that the noise is considered to be independent, the probability that a particular $n$-qubit Pauli error, $\mathrm{A}=\mathrm{A}_1\otimes \mathrm{A}_2\otimes\dots \otimes \mathrm{A}_n$ with $\mathrm{A}_j\in{\{\mathrm{I,X,Y,Z}\}}$, will be given by 
    \begin{equation}
        p_\mathrm{A}(\mu_{T_1},\mu_{T_2}) = \prod_{j=1}^n p_{\mathrm{A}_j}(\mu_{T_1},\mu_{T_2}),
    \end{equation}
    where $p_{\mathrm{A}_j}$ is described by the Pauli (biased) or Clifford (depolarizing) twirl approximations given before, and where $\mu_{T_1}$ and $\mu_{T_2}$ refer to the mean values of the relaxation and dephasing times. Taking the mean value is the usual approach.
    \item \textbf{Independent and non-identically distributed (i.ni.d.):} in this model every qubit experiences a different probability of suffering a Pauli error \cite{inid,roffeinid}. The motivation of this error model is the fact that state-of-the-art quantum processors are consisted of qubits whose relaxation and dephasing times differ significantly. Considering that the environment-to-qubit interaction is still independent from qubit-to-qubit, the probability of occurrence for a $n$-qubit Pauli error is given by
    \begin{equation}
        p_\mathrm{A}(\{T_1^j\}_{j=1}^n,\{T_2^j\}_{j=1}^n) = \prod_{j=1}^n p_{\mathrm{A}_j}(T_1^j,T_2^j),
    \end{equation}
    where $p_{\mathrm{A}_j}(T_1^j,T_2^j)$ is again given by the Pauli (biased) or Clifford (depolarizing) twirl approximations, but now each of the terms have particular values of relaxation and dephasing times.
\end{itemize}

\subsection{SPAM and gate errors}
These refer to errors arising from the imperfect implementation of the physical operations that are done whenever the qubits are prepared, measured or manipulated by means of quantum gates and the coupling to the environment while executing them. They are usually referred to as circuit-level noise \cite{nielsen,heavyhexagon,chamberland}. If only imperfect measurements are considered, the model is named phenomenological noise. These errors are usually classified and modelled in the following way:
\begin{itemize}
    \item \textbf{SPAM errors:} these refer to the errors that occur due to the imperfect preparation of the states that are needed to initialize the surface code and the fact that the measurement operations done to detect the syndrome are not always successful. Since it is usually considered that a surface code is initialized with all the physical qubits on the $\ket{0}$ state, state preparation errors are usually modelled so that a $\ket{1}$ state is prepared instead of the $\ket{0}$ with a probability of error $2p_\mathrm{prep}/3$ \cite{fowlerReview,heavyhexagon,chamberland}. This is the same as considering a depolarizing channel after state preparation. Imperfect measurements are usually modelled by considering that the single-qubit measurement is flipped with a probability $2p_\mathrm{meas}/3$ error \cite{fowlerReview,heavyhexagon,chamberland}.
    \item \textbf{Noisy single-qubit gates:} due to their imperfect implementation, single qubit quantum gates, $\mathcal{U}$, do not perform the desired operation in a perfect way and, thus, introduce noise to the qubit. In this sense, the noisy quantum gate, $\mathcal{\tilde{U}}$, can be seen as the operation of the quantum gate followed by a quantum channel, $\Lambda$, that describes the noise introduced by the gate, i.e. $\mathcal{\tilde{U}} = \Lambda \circ \mathcal{U}$ \cite{nielsen,QEM2,deconvolution}. In this sense, single-qubit gate errors are usually modelled by considering that they are followed by a depolarizing channel with probability of error $p_\mathrm{1Q}$ \cite{fowlerReview,heavyhexagon,chamberland}. This implies that an $\mathrm{X}$, $\mathrm{Y}$ or $\mathrm{Z}$-error will be applied to the physical qubit with probability $p_\mathrm{1Q}/3$.
    \item \textbf{Noisy two-qubit gates:} similar to the single-qubit gate, two-qubit gates are also modelled by a noisy channel being applied after the perfect operation. However, the usually considered error map is the two-qubit depolarizing channel with probability of error $p_\mathrm{2Q}$ \cite{fowlerReview,heavyhexagon,chamberland}. Therefore, a Pauli error of the set $\{\mathrm{I,X,Y,Z}\}^{\otimes 2} \setminus \mathrm{I}^{\otimes 2}$ will be randomly applied after the perfect two-qubit gate with probability $p_\mathrm{2Q}/15$.
\end{itemize}

It is important to state that a biased circuit-level noise model can also be considered if the depolarizing channels are changed by Pauli channels with a bias towards $\mathrm{Z}$-errors equal to $\eta$ \cite{chamberland}. % Has also been considered instead of can also be considered (?)

\subsection{Erasure errors}
To finish with this section, we will discuss another error type that can corrupt the qubits of a quantum computer, named erasure error, and that will be considered for the Union Find decoder \cite{UF,erasure}. Erasure errors come from two types of physical mechanisms that qubits may experience:
\begin{itemize}
    \item \textbf{Leakage:} qubits are defined as two-level coherent systems. However, when physically implemented, there exist other levels that can be populated.  This would imply that the qubit has left the computational subspace and, thus, it is not useful anymore \cite{UF,leak}. Leakage may arise due to decoherence processes that make the qubit to leave the computational space or due to leaky quantum gates.
    \item \textbf{Loss:} this refers to the scenario where the qubit is physically missing \cite{UF,loss}. For example, in a photonic system the qubit encoded in a photon may be lost. 
\end{itemize}
In this context, an erasure channel describes the fact that a qubit at a known location has been lost with probability $p_e$ \cite{erasure}. The fact that it is known which of the qubits is lost is important since it provides with useful information for treating those errors. The detection of leakage events in physical qubits can be done by means of the quantum jump technique or by means of ancillary qubits, for example \cite{erasure,leak,leakdetect}. Errors of this type with unknown locations are named deletion errors in the literature \cite{deletion}. The significant difference between deletion and erasure errors lies in the fact that a deletion error leads to a decrease in the number of qubits of the system, i.e. the qubit is effectively lost, while an erasure error occurrence does not decrease the number of qubits. In this sense, the erasure channel on a qubit, $\rho$, may be described as
\begin{equation}
    \mathcal{N}_\mathrm{er}(\rho) = (1-p_e)\rho + p_e |e\rangle\langle e|,
\end{equation}
where $|e\rangle$ refers to an erasure flag giving the information that such qubit has been erased. Since erasure errors are detected and their location is known, qubits subjected to such errors can be reinitialized, which results in those being subjected to a random Pauli error after the measurement of the stabilizers is performed \cite{UF}.

\section{Decoders for surface codes}\label{secdec}
As described in the previous sections, surface codes have the ability of detecting errors experienced by the data qubits, which can be accurately modelled by elements of the Pauli group $\mathcal{G}_n$. However, once the error syndrome is measured, an estimation of the channel error must be done using such information, $\hat{E}(\bar{s})$, so that active error correction can be performed on the noisy qubits. The methods used for performing this inference of the error are named decoders. Once the decoder makes a guess of the channel error, the recovery operation is performed by applying $\hat{E}^\dagger(\bar{s})$ since the elements of the Pauli group are unitary matrices.

Therefore, decoding methods for error correction codes are a critical element of the code itself since their efficiency on making correct guesses of channel error will be what will determine if the method is successful or not. In this sense, the threshold of a code is a function of the decoder in question, i.e. the code can perform better or worse as a function of the method used. Making the decoder to be more accurate usually comes with the drawback of increasing its computational complexity, which ultimately makes it to be slower in making guesses. Decoders must be fast enough since the action of decoherence will not stop while estimating the error after measurement, implying that the qubit may suffer from additional errors to which the decoder will be oblivious. Thus, a slow decoder will ultimately have a bad performance (See appendix \ref{appBacklog} for more details on why real time decoding is required). To sum up, the trade-off between the accuracy and complexity of the methods is vital for the field of quantum error correction \cite{fastDec,sparse}.

% Aqui quizas tb comentar algo de lo de QMLD y tal si al final lo comentamos en stab codes.

Surface codes can be decoded by using many methods \cite{fowlerReview,qldpc2,mps,UF,blossom1,fussionBlossom,pymatching,osd,renormalization,automata,neural}. In this section we will explain the operation and performance, in terms of correction ability and complexity, of the main decoders for surface codes: the minimum-weight perfect matching \cite{fowlerReview,blossom1,fussionBlossom,pymatching}, the Union-Find decoder \cite{UF}, the Belief Propagation decoder \cite{qldpc2, reviewpat,osd} and the Tensor Network or Matrix Product State decoder \cite{mps}. In addition, we will also discuss variants of those decoding methods that have proven to be more efficient in terms of error correction ability or complexity as the Belif-Propagation Ordered Statistics Decoder (BPOSD) \cite{qldpc2,osd}, for example. Furthermore, we discuss other decoding algorithms in the literature that can be used for decoding different types of surface codes. Many of those were proposed for decoding topological codes such as the toric or color codes, but could, in principle, be applied for the rotated planar code. Specifically, we discuss cellular-automaton \cite{automata}, renormalization group \cite{renormalization}, neural network or machine learning based \cite{neural} and MaxSAT \cite{maxsat} decoders. The end of the section includes an overview of the available software implementations available to the general public of all those decoding methods.

\subsection{The Minimum Weight Perfect Matching Decoder}

Before the operation of the Minimum Weight Perfect Matching decoder is described, some definitions of graph theory must be provided \cite{fowlerO1}. Consider a weighted graph $G$ composed by ($V_G$,$E_G$,$W_G$), where $V_G= \{ v_i \}$ are the vertices, $E_G = \{ e_ {ij} \}$ is the set of edges which satisfy $i \neq j$ and $e_{ij}=\{ v_i, v_j \}$, meaning that the edge $e_{ij}$ connects the nodes $v_i$ and $v_j$, and $W_G=\{w_e\}$, $e\in E_G$, which is the set of weights attributed to each edge. A matching of graph $G$ is a subset of edges, denoted as $M\subseteq E_G$, such that for any two edges $e$ and $f$ in $M$, $e$ and $f$ do not share any common vertices. In other words, $M$ is a set of edges without common endpoints. A perfect matching is a matching that additionally satisfies the condition that every vertex in $V$ is incident to exactly one edge in $M$. A minimum weight perfect matching is the perfect matching with the smallest possible weight among all possible perfect matchings \cite{blossom1,fowlerO1}, where the weight of a matching is defined as the sum of the weights of its edges: $\sum_{e\in M}w_e$. Additionally, a complete graph is a graph with the property that $\forall v_i, v_j \in V, i\neq j, \exists e_{ij} \in E_G$. The Minimum Weight Perfect Matching (MWPM) decoder makes use of these concepts from graph theory in order to develop an algorithm for decoding.
% La última frase es añadida.

\begin{figure}
\centering
    \includegraphics[width = 0.4\textwidth]{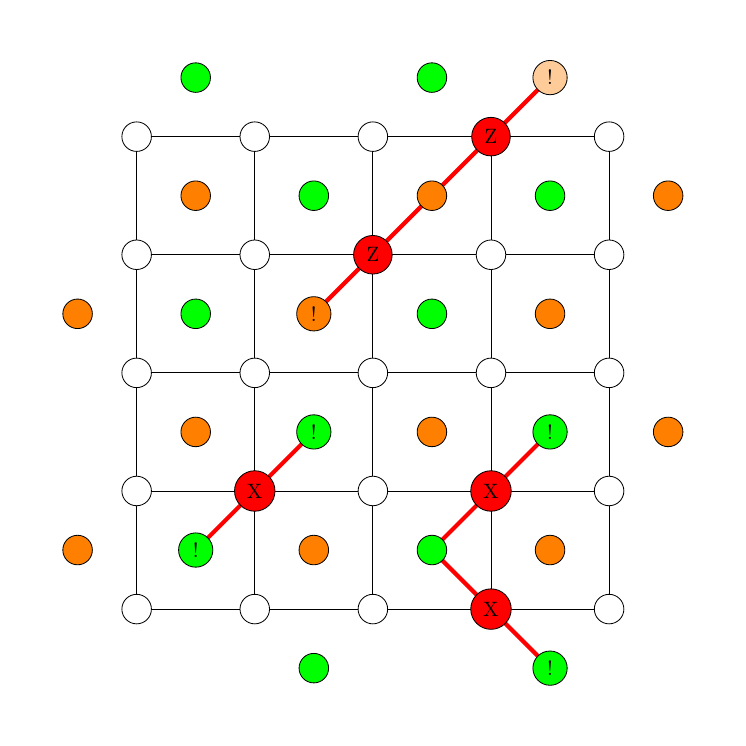}
    \caption{Graphical representation of the effect of a Pauli error in a 5x5 rotated surface code. $Z$-errors trigger orange $Z$-checks and $X$-errors trigger green $X$-checks. The red lines connecting non-trivial checks represent chains (when two data qubits trigger the same check it becomes trivial).}
    \label{chain}
\end{figure}

% Párrafo cambiado totalmente
%%%%%%%%%%%%%%%%%%%%%%%%%%%%%%%%%%%%%%%%%%%

% I explain chains & virtual checks.
As explained before, when the data qubits of the surface code experience a non-trivial Pauli error, the checks adjacent to an odd number of errors turn into non-trivial syndrome elements while checks adjacent to an even number of errors are not triggered by them, since the product of two errors of the same type will cancel. If we are to consider $X$ and $Z$-errors separately\footnote{Recall that $Y$-errors can be regarded as combinations of both $X$ and $Z$-errors on the same qubit.}, we can regard the error which affected the code as a set of chains of data qubits which have non-trivial checks on their endpoints. An example of this consideration is shown in FIG. \ref{chain}, notice how, adjacent $X$-errors on the bottom-left and bottom-right side of the figure form a chain denoted by a red line with non-trivial checks at its endpoints. There can also be the chance of errors ending in a boundary where there are no checks susceptible to them, this is the example of the top $Z$-error in FIG. \ref{chain}. For the sake of the MWPM decoding method, we are interested in chains having non-trivial checks on their endpoints, therefore, we consider the existence of \textit{virtual checks}, which are hypothetical $X$($Z$) check qubits on the left and right (top and bottom) boundaries that can be considered to be either trivial or non-trivial. Moving back to the top $Z$-error from FIG \ref{chain}, the top right non-trivial lighter color $Z$-check is a virtual check, since there is no physical check in that spatial position, we consider it so as to be able to have two non-trivial checks at the endpoints on the $Z$-chain.

% I explain the syndrome graph.
In this sense, due to the CSS structure of the rotated planar code, two separate graphs can be constructed: one for the $X$-checks and the other for the $Z$-checks, both combined are known as the \textit{syndrome graph} \cite{wu2}. Within the syndome graph, nodes represent the checks and virtual checks, either $X$-checks in the $X$-subgraph or $Z$-checks in the $Z$-subgraph. Also, edges represent data qubits, that is, if two checks are susceptible to a change in the parity of a specific data qubit ($X$-parity for the $X$-checks and $Z$-parity for the $Z$-checks), the two nodes which represent said checks will be connected by an edge representing that specific data qubit. This could be the case of the two non-trivial $X$-checks on the bottom-left side of FIG. \ref{chain}, which are both susceptible to the change in $X$-parity of the data qubit in between them, thus, within the constructed $X$-subgraph, there is an edge in  between them. All edges in the syndrome graph have the same weight (commonly 1), since we consider all data qubits to undergo non-trivial Pauli operators with the same probability.

% I explain the decoding graph & blossom algorithm.
Given a specific syndrome, we want to consider the error chains which may have produced it. As discussed earlier, chains have non-trivial checks on their endpoints, therefore, we can study this problem by looking for all possible same type non-trivial check pairs. Afterwards, we can create the \textit{decoding graph} \cite{wu2}. In the decoding graph, nodes correspond to non-trivial syndrome checks and some virtual checks. Moreover, it can also be separated into two independent subgraphs, the $X$-decoding subgraph and the $Z$-one. Nodes correspondent to non-trivial checks within each of the subgraphs will be connected with edges to all the other non-trivial checks of the same type, altogether with their nearest virtual-check. For example, in FIG. \ref{chain}, the nearest virtual check of the non-trivial $Z$-check in the bulk of the code will be the aforementioned top-right virtual check. Additionally, the weight of each edge in the decoding graph will be given by the weight of the minimum distance path in between both checks within the syndrome graph. For instance, going back to FIG. \ref{chain}, if we were to construct the $Z$-decoding subgraph, we would observe that is only composed by a node representing the non-trivial $Z$-check close to the center of the code and the virtual check from the top-right, connected via an edge of weight 2.

Once the two decoding subgraphs are obtained, one can compute their minimum weight perfect matching \cite{fowlerReview, fowlerO1}. The minimum weight perfect matching will be the chosen set of chains which is considered to have produced the syndrome. Afterwards, the edges belonging to said set of chains are mapped back to the syndrome graph, and the data qubits whose edges belong to the minimum weight perfect matching are considered to have undergone the non-trivial Pauli operators from the error. The decoding $X$-subgraph will return $X$-errors and the decoding $Z$-subgraph will return $Z$-errors.
%%%%%%%%%%%%%%%%%%%%%%%%%%%%%%%%%%%%%%%%%%%

\begin{figure*}
\centering
\begin{subfigure}[b]{.4\textwidth}
\centering
    \includegraphics[width = .65\textwidth, frame]{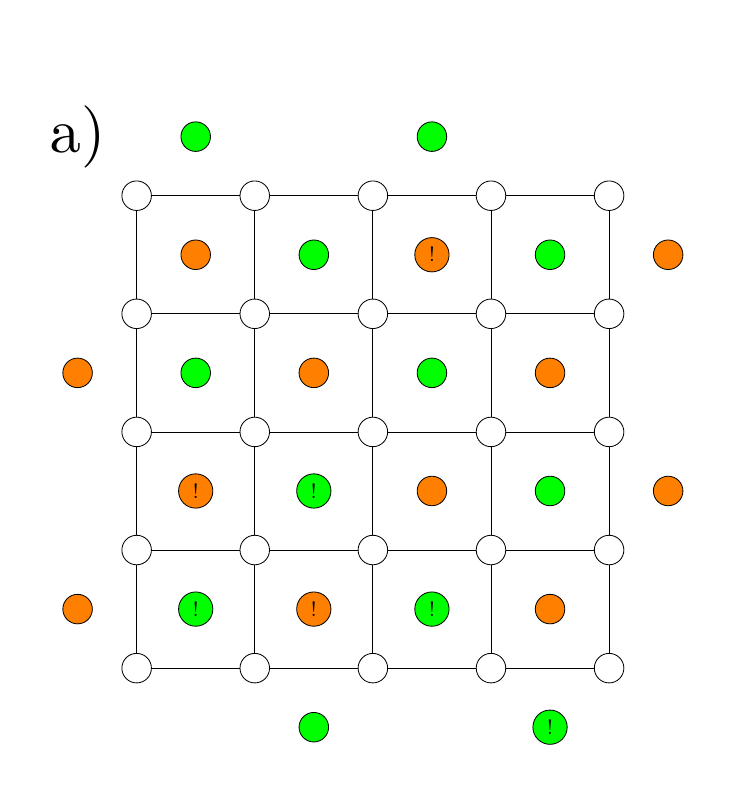} \\
    \includegraphics[width = .65\textwidth, frame]{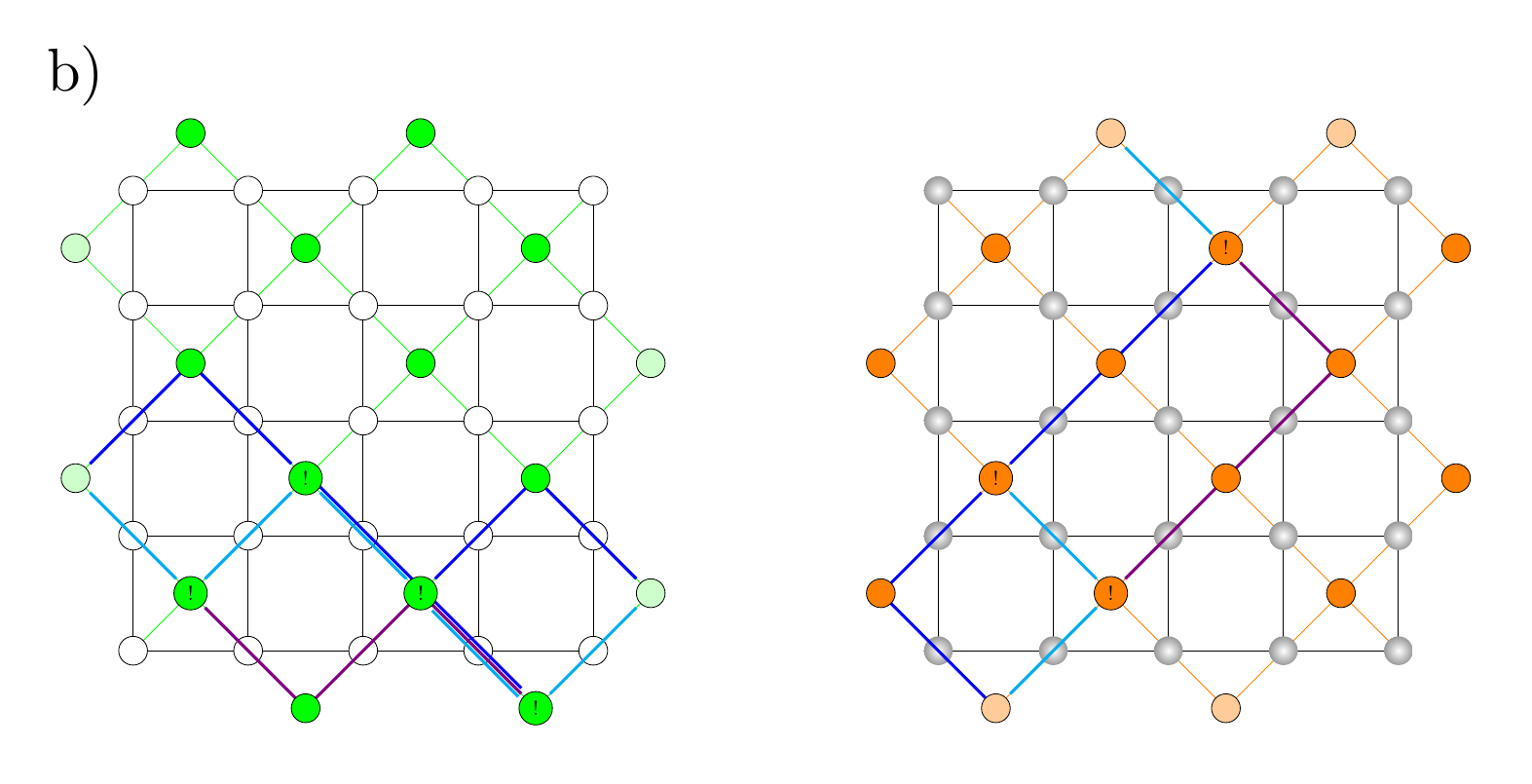} \\
    \includegraphics[width = .65\textwidth, frame]{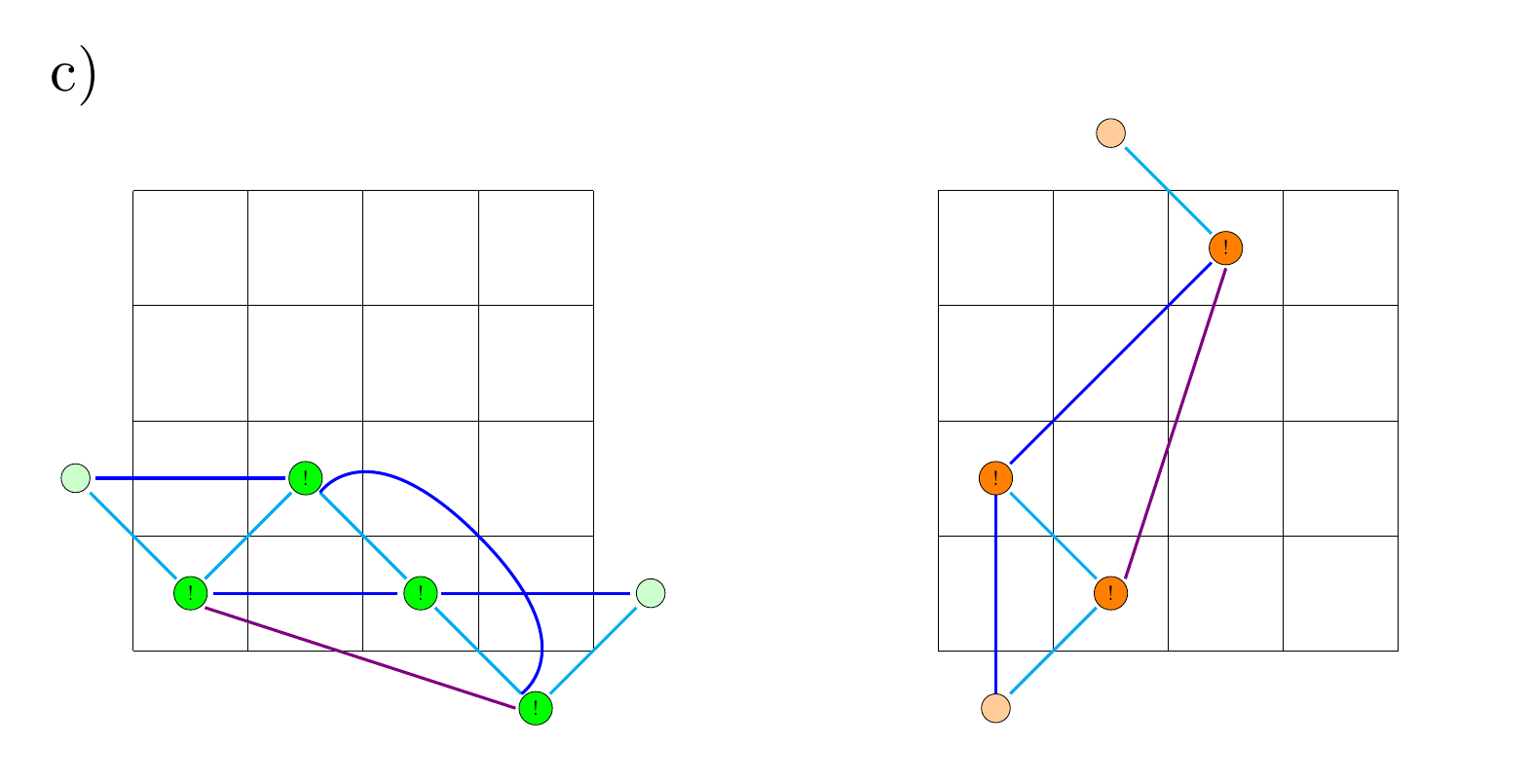} 
\end{subfigure}
\begin{subfigure}[b]{.4\textwidth}
\centering
    \includegraphics[width = .65\textwidth, frame]{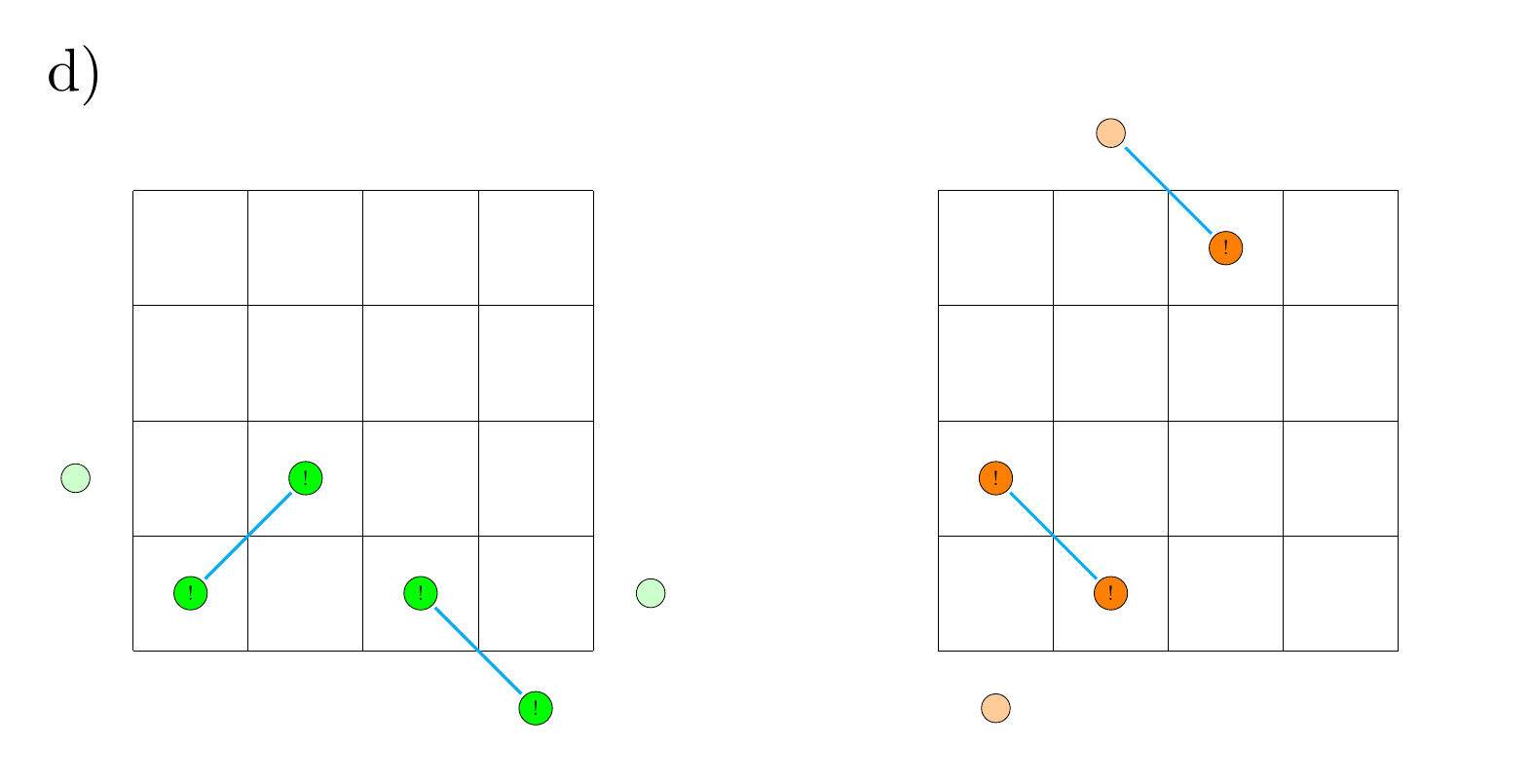} \\
    \includegraphics[width = .65\textwidth, frame]{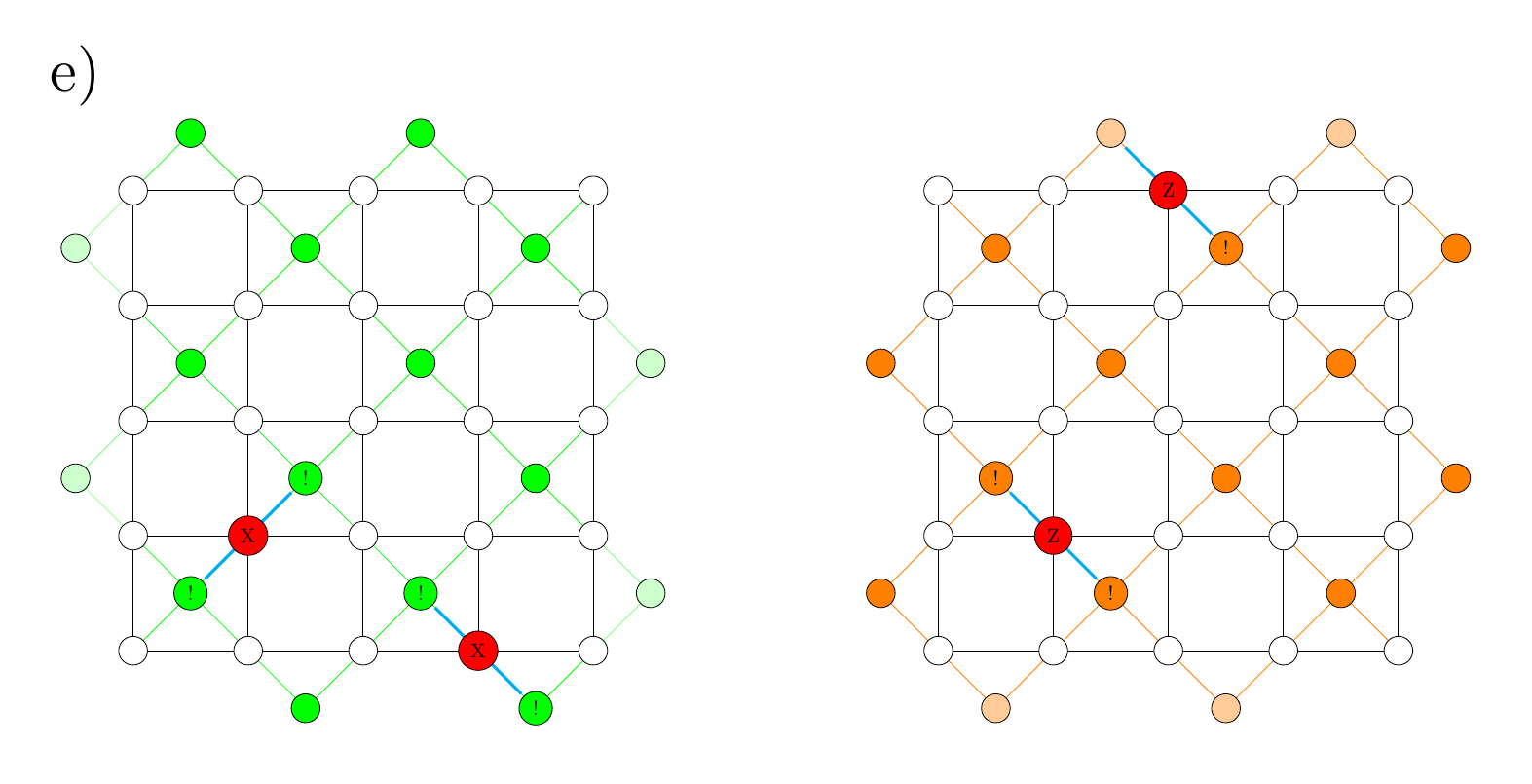} \\
    \includegraphics[width = .65\textwidth, frame]{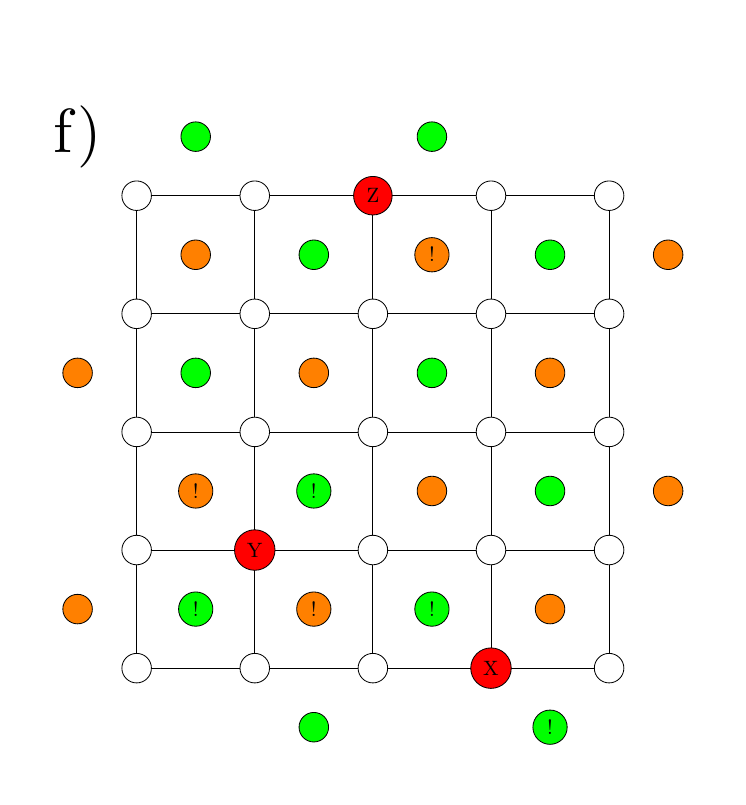}
\end{subfigure}
\caption{Graphical representation of a MWPM decoding process for a specific syndrome in a 5x5 surface  code. In \textbf{a)} there is a representation of a syndrome of a distance-5 rotated code. Images from \textbf{b)} represent the syndrome graphs of the distance-5 rotated code splitted in the $X$ and $Z-$check graphs and the illustrations from \textbf{c)} represent the decoding graphs for such syndrome for each check type. \textbf{d)} represents the obtained matchings from the decoding graphs and \textbf{e)} the recovered error for every subgraph within the syndrome graph. Finally \textbf{f)} shows the recovered error. }
\label{mwpmsketch}
\end{figure*}

In FIG. \ref{mwpmsketch}, we provide an example of the operation of the MWPM decoder for a specific detected error pattern syndrome. FIG. \ref{mwpmsketch} a) shows the considered syndrome, where the exclamation marks correspond to the checks that have measured a non-trivial syndrome element. FIG. \ref{mwpmsketch} b) depicts the two syndrome subgraphs, the $X$-subgraph (green nodes) and the $Z$-subgraph (orange nodes), edges in the $X$ and $Z$-subgraphs are denoted with green and orange lines respectively. Note that in both graphs, the previously discussed virtual checks are located in the boundaries: on the left and right boundaries for the $X$-checks and on the top and bottom for the $Z$-checks. The edges connecting nodes in the syndrome graph represent the data qubits which, when changing parity, trigger a change in the parity of the check nodes they are. Over these graphs, all the shortest paths connecting all non-trivial checks with each other and with their nearest virtual qubit are considered for the matching problem. These paths are represented with cyan, blue and violet colors for paths of weights $1$, $2$ or $3$ respectively. By means of those paths, the decoding subgraphs shown in FIG. \ref{mwpmsketch} c) are constructed, where the weights of the edges are given by the number of data qubits that are crossed when following the path from non-trivial check to non-trivial check, i.e. each of the edges of the graphs of the second row of the figure has weight 1. Then, in FIG. \ref{mwpmsketch} d), the minimum weight perfect matching of those subgraphs is computed. The result of said process can be seen in FIG. \ref{mwpmsketch} e), notice how virtual qubits can be left unmatched. The matchings of each of the subgraphs refer to $X$ or $Z$ operators applied to the data qubits that the matching crosses. Thus, for the example in consideration, the second row of the right columns presents the operators that will be applied on the data qubits of the rotated code. The $X$-checks recover $X$-errors and the $Z$-checks recover $Z$-errors. In FIG. \ref{mwpmsketch} f), we present the final recovery operator, where it can be seen that whenever an $X$ and $Z$-error are estimated from each of the graphs for the same data qubit, the resulting recovery operation is a $Y$ operator.

\begin{figure}
    \centering
    \includegraphics[width = 0.4\textwidth]{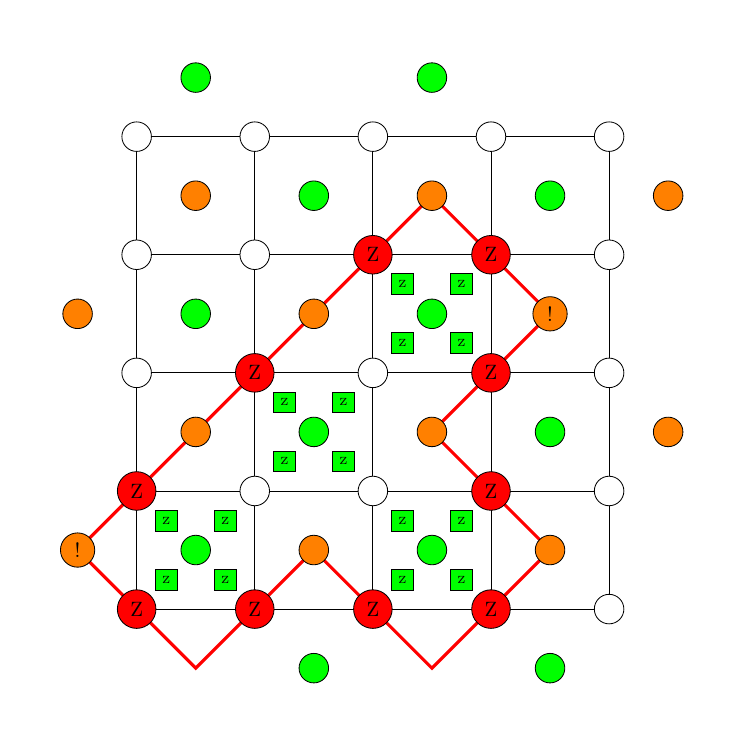}
    \caption{Two $Z$-error chains that share the same endpoints and the stabilizing operator checks which separate them. Since both chains are equal up to an stabilizer, they represent the same logical operation.}
    \label{mwpmdeg}
\end{figure}

The MWPM decoder estimates the Pauli error with minimum $X$  and $Z$ weight (since it considers said errors independently) that corresponds to a given syndrome. In this sense, this decoder always outputs an error whose syndrome is the same as the one measured\footnote{Other decoders such as the belief propagation decoder do not always estimate an error whose syndrome matches the true syndrome \cite{reviewpat,logicalsparse}.}. In addition to always matching the syndrome, if the recovered minimum weight perfect matching from the decoding graph succeeds, the recovered error will be correct independently of the path chosen within the syndrome graph. In other words, if the error can be represented as a set of chains which relates the same non-trivial checks as the minimum weight perfect matching, the recovered error will correct it even if they are not the same Pauli operator. The reason for this to happen is that the elements of the stabilizer in a 2D surface code correspond to Pauli sequences that form a closed loop in the surface code \cite{fowlerReview,inid}. Thus, if the estimated error forms a closed chain with the true error occurred on the surface code, the resulting Pauli element will belong to the stabilizer set of the code and, thus, the correction will have been successful (recall eq.\eqref{coseet} and degeneracy of errors). In FIG. \ref{mwpmdeg}, we pictorically present this scenario where two error chains with same endpoints are separated by a stabilizer element. Note also that whenever an error that forms an error chain that is a loop on the data qubits of the code, all the checks will have trivial values but the codestate will not be affected by it as it will be a stabilizer element. Thus, those types of chains will be non-detectable but unharmful for the code.

\subsubsection{Complexity}
As described before, the most critical part of the MWPM decoder consists in finding the perfect matching with minimum weight once the subgraphs are constructed from the syndrome information. An algorithm to efficiently solve such computational problem was proposed by Jack Edmonds back in the 60s, the so called blossom algorithm \cite{blossom1}. In general, the MWPM decoder is dominated by the blossom step of the algorithm whose worst-case complexity in the number of nodes $N$ is $\mathcal{O}(N^3 \log{(N)})$ \cite{sparse,pymatching}. However, the expected runtime of the decoder is roughly $\mathcal{O}(N^2)$ whenever the decoder is implemented such that all Dijktra searches are needed for computing the subgraphs where the matching of interest is needed \cite{sparse,pymatching}. Therefore, and due to the importance speed for real-time decoding, several implementations of the MWPM decoder have been proposed such as Fowler's implementation with $\mathcal{O}(1)$ parallel expected runtime \cite{fowlerO1} or the more recent sparse blossom by Higgot and Gidney with an observed complexity of $\mathcal{O}(N^{1.32})$ \cite{sparse} and Fusion Blossom by Wu with $\mathcal{O}(N)$, i.e. linear complexity \cite{fussionBlossom,wu2}. Each of the implementations have their own advantages and disadvantages, as for example, sparse blossom has a faster single-thread performance than Fusion Blossom, but the latter supports multi-thread execution, implying that it can be faster than the former if enough cores are available \cite{sparse}. Proposing faster MWPM implementations is an arduous but significant task, since large distances are precised in order to have a fault-tolerant quantum computer, and the decoding schemes also need to be scalable in the sense that they are fast enough when the distance of the code increases.

Following the previous discussions, it can be seen that the MWPM decoder follows the QMLD decoding rule as it aims to estimate the most probable error for the given syndrome. Note that, here, finding a perfect matching with minimum weight in the subgraphs formed with the measured syndrome implies that the Pauli element estimated will be the most probable to occur considering pure $X$ and $Z$ noise \footnote{This occurs because usually a i.i.d. model is assumed, implying that higher weigths imply less probability.}. Applying the suboptimal decoding rule has been observed to be an efficient approach for low physical error probabilities \cite {pymatching}. 
% esto ver como poner (sino sin lo del 10.3 y citar lo de pat o algo asi
%Nevertheless, starting at an error probability of $10.3\%$ for the planar code subgraph \cite{pymatching}, the most probable error no longer belongs to the most probable equivalence class probability, that is, the group of errors containing all possible stabilizing elements.

% Cambiar título del y-axis a P_L, arreglar puntos threshold bias.

\subsubsection{Performance and threshold}
\begin{figure}
    \centering
    \includegraphics[width = \columnwidth]{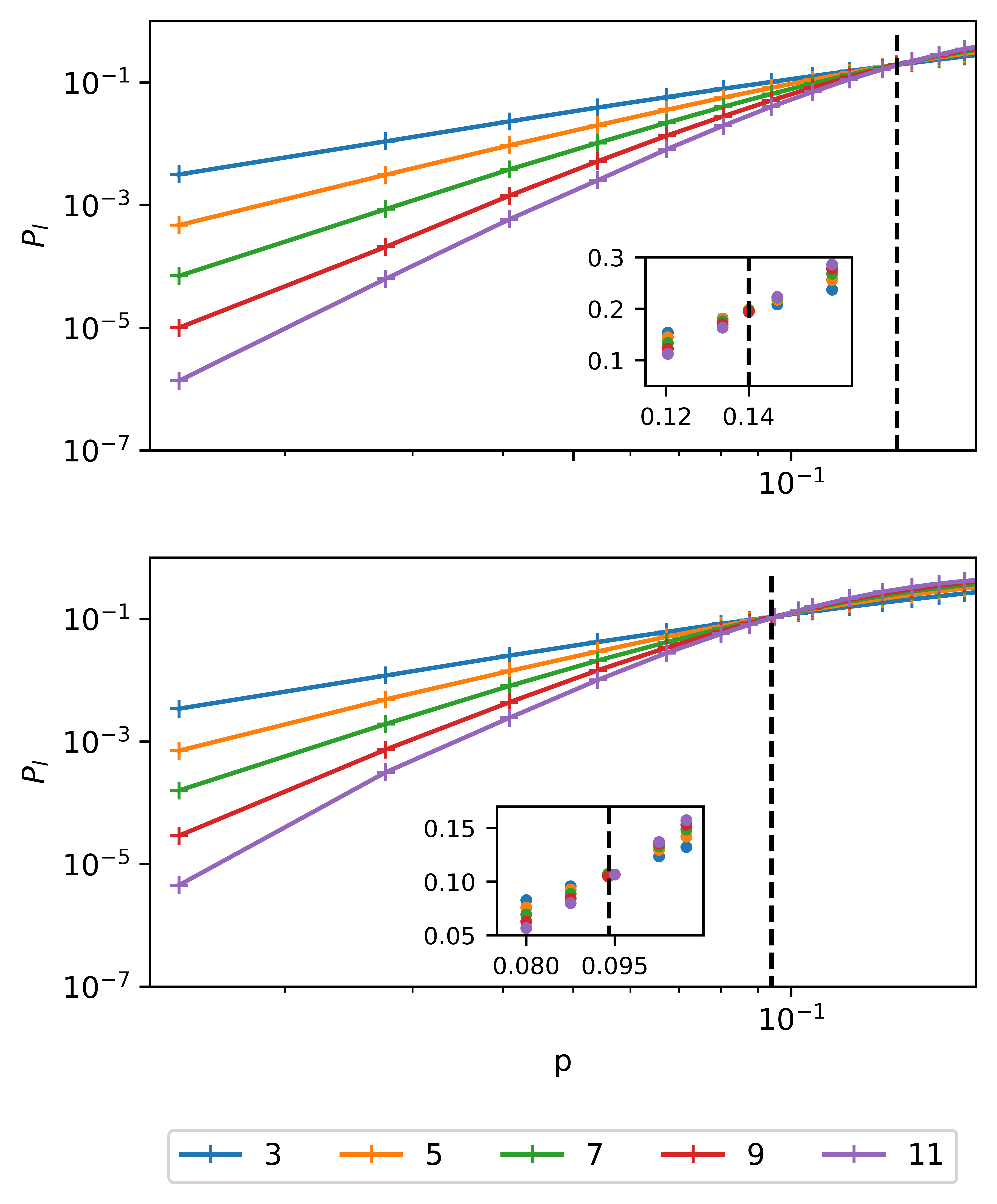}
    \caption{Logical error probability with dependence on the physical error probability under depolarizing (top) $\eta = 100$ (bottom) noise. The dashed line represents the $p_{th}$ case and the subplots are close ups to the points near the $p_{th}$. }
    \label{performancesmwpm}
\end{figure}

In FIG. \ref{performancesmwpm} we plot the performance of the rotated planar code in terms of the logical error probability ($P_L$), that is, the probability of the decoding process failing in predicting an error given its syndrome, as a function of the physical error probability ($p=p_x+p_y+p_z$) whenever it is decoded using a MWPM decoder. Two noise models are considered: the top figure considers an i.i.d. depolarizing error model ($p_x = p_y = p_z$), while in the bottom figure considers an i.i.d. biased Pauli channel with bias $\eta = 100$. The results show how the MWPM decoder performs better when considering noise channels closer to the depolarizing channel. Specifically, not only the logical error probabilities are significantly higher for the biased case when a physical error probability is fixed, but also the probability threshold $p_{th}$ is lower. This performance decrease when the channel is biased can be explained by the fact that both subgraphs are considered independently. Considering a bias towards $Z$-noise results in the $Z$ decoding subgraph being more dense, i.e. more non-trivial syndromes are triggered, as opposed to the $X$-checks one. Having more non-trivial checks within the $Z$ decoding subgraph makes it more probable for the recovered minimum weight perfect matching not to capture the error that the code underwent, as the performance of the MWPM decreases as the physical error rate $p$ increases. This results in the $Z$-subgraph reaching the probability threshold before the total physical error probability reaches the threshold of the depolarizing channel. Further increasing the bias of the channel will produce a decrease of the $p_{th}$ until the extreme value of $\eta\rightarrow \infty$, that is, a pure dephasing error model. At such point, all triggered syndromes will correspond to the same subgraph, i.e., the right column in FIG. \ref{mwpmsketch}. Table \ref{tablamwpm}, shows some $p_{th}$ for different biases when decoded using the standard MWPM decoder for the rotated planar code.

\subsubsection{Modifications and re-weighting for specific noise models}
Biased channels are important since for some experimentally implemented qubits, such as ion traps or spin qubits, have shown lower dephasing times, $T_2$, than relaxation times, $T_1$, which implies that those qubits are more prone to suffer from dephasing errors than bit-flip errors (recall Section \ref{secnoisemod}) \cite{approximatingDec,TVQC,inid}. Bias values in the range of $\eta\in [1,10^6]$ are typical depending on the technology used for constructing the qubits \cite{approximatingDec,josuPhD}. Therefore, some studies have been directed towards better adapting surface codes to $Z$ biased noise, either by introducing more $Z$-checks producing rectangular surface codes \cite{xzzx, fragile, rectangularbias}, tailoring the codes so as to have all checks susceptible to $Z$ noise \cite{xzzx, fragile} or by modifying the MWPM decoder by making it aware of the symmetries of the code \cite{tuckettmwpmbias}. In addition, the noise in experimentally constructed hardware is not identically distributed (recall the i.ni.d. error model) \cite{recMWPM, inid, roffeinid, fastfading}. The performance of surface codes is significantly affected by such non-uniformity of the noise in the data qubits of the lattice, as some of the data qubits will have a higher tendency to suffer from errors than other, and the standard MWPM decoder calculates the perfect matching considering that all the qubits are equiprobable. Thus, methods based on reweighting the syndrome subgraphs as a function of the probability of error of each of the qubits have been proposed so that the MWPM problem is solved over a weighted graph that takes such effects in consideration \cite{recMWPM,inid}. Another limitation of the standard MWPM decoder is the fact that since the $Z$ and $X$-subgraphs are decoded independently, $Y$-errors are underestimated. In fact, since $Y$-errors are combinations of bit- and phase-flips, it can be considered that there exists a correlation among them. Therefore, information can be passed from the $Z$-subgraph to the $X$-subgraph, and vice versa, so that the correlation between those events is taken into account by reweighting the other subgraph as a function of what has been estimated in the other \cite{recMWPM,xzcorrelation,fowlercorr}.

% Mirar a partir de quina probabilitat comença a fallar el subgraf de les X.
% Incluir tabla con thresholds

% XZZX, fragile, tuckettmwpmbias, rectangularbias

\subsubsection{Measurement errors}
% explicar el tema de measurement errors y propagacion y los multiple decoding round
Another important thing to consider when decoding surface codes is the fact that the measured syndromes might be erroneous \cite{fowlerReview,fragile,chamberland}. This means that the measured syndrome does not correspond to the Pauli operator affecting the data qubits after measurement. This may come due to the fact that the measurement operations are not perfect (recall SPAM errors) or due to the so called error propagation. Error propagation refers to the fact that Pauli errors on one qubit can propagate to other qubit when performing a two-qubit gate. For example, if we have an $X$ operator in the control qubit of a $\mathrm{CNOT}$ gate, such operator propagates to the target qubit, i.e. $\mathrm{CNOT}(\mathrm{X}\otimes \mathrm{I})=\mathrm{X}\otimes\mathrm{X}$. As a consequence of this, the circuit-level noise coming from the measurement qubits can propagate to the data qubits, changing the Pauli error due to propagation and, thus, making the measured error to be imperfect even with perfect SPAM. Those erroneous measurements have a big impact in code performance, lowering the code threshold in significant manner when circuit-level noise is considered \cite{fowlerReview,fragile,chamberland,stim}. In order to deal with this problem, several measurement rounds are done before decoding so that a space-time like graph is obtained in which the MWPM problem is performed to estimate the error. Usually, $d$ measurements are recorded for a distance-$d$ surface code \cite{fowlerReview,stim}. It is noteworthy to say that once the measurements are done, a non-trivial syndrome element for a measurement that follows another one will be a measurement that has flipped from such last round, refer to the Appendix \ref{appA} for a more detailed description. Then, the complete syndrome graph is constructed by connecting all those non-trivial elements both spatially and temporally. This consideration enlarges the size of the syndrome graph where the perfect matching with minimum weight must be computed and, therefore, the complexity of the algorithm increases considerably. Measurement errors highly compromise the decoding process under single-round syndrome extraction, due to the fact that they are not accounted for and will most likely cause a logical error. For example, consider an error consisting of a single measurement error in a check, which becomes non-trivial. As has been seen in this section, the MWPM decoder will consider this check altogether with its closest virtual check and return the minimum weight perfect matching of the decoding graph as the recovered error, which will be non-trivial. By considering this space-time decoding, the performance of the code will improve when compared to single-round decoding due to the fact that measurement errors will be efficiently considered. Nevertheless, the code threshold significantly decreases compared to the perfect measurement scenario \cite{fowlerReview, fragile, chamberland}. 

\begin{table}
\begin{center}
\begin{tabular}{|c | c|} 
 \hline
 $\eta$ &  $p_{th}$ \\ [0.5ex] 
 \hline\hline
 1/2 & 0.140  \\ 
 \hline
 1 & 0.138  \\
 \hline
 10 & 0.098  \\
 \hline
 100 & 0.095  \\
 \hline
 1000 & 0.088  \\ [1ex] 
 \hline
\end{tabular}
\caption{Probability threshold values for different biases in the rotated planar code under the MWPM decoding scheme.}
\label{tablamwpm}
\end{center}
\end{table}

\subsection{Union-Find decoder}

% Union-Find

The Union-Find decoder (UF) is a decoding scheme proposed by Nicolas Delfosse and Naomi Nickerson in 2017 which also consists in mapping the syndrome into a graph problem \cite{UF}. However, this decoder is based on clustering the non-trivial syndrome elements of the subgraphs by considering that Pauli errors at a known location can be treated as erasure errors. As mentioned in the error model section, an erasure error within a physical qubit can be treated as the qubit itself being in a mixed state (subjected to a random Pauli error). Therefore, having a uniform probability distribution for $I$-, $X$, $Y$ or $Z$ operators and, most importantly, a known location. Thus, for a pure erasure error model, all the qubits undergoing said erasure errors are localized. The surface codes under the erasure channel can be efficiently decoded in linear time, through a method named peeling decoding scheme \cite{peeling}. In light of this, the UF decoder is based on the idea of transforming the decoding problem of a surface code experiencing Pauli noise into an erasure error decoding problem. By doing this, the UF decoder achieves a decoding complexity of almost linear time $\mathcal{O}(n\alpha(n) )$ \cite{UF}, where $\alpha$ is the inverse of the Ackerman function and for all practical purposes $\alpha(n) \leq 3$ \cite{ackermann}. In order to do so, the UF decoding process consists of two different steps: \textit{syndrome validation} and \textit{erasure decoder}. The syndrome validation step consists on mapping the set of Pauli errors or a mixture of Pauli and erasure errors into clusters of overall erasure errors \cite{UF}. Note that mixtures of Pauli and erasure errors can be considered by the UF decoder since the idea is to only have erasure errors. Once the step is completed, the erasure decoder is based on the peeling decoder \cite{peeling}.

\subsubsection{Syndrome validation}
Similar to the MWPM decoder, the UF decoder works on two separate graphs, one for the $X$-checks and one for the $Z$-checks, and they include the boundary virtual checks discussed before. In the syndrome validation step, the checks within each of the graphs are considered to be even parity nodes if they correspond to a trivial measured syndrome element and odd parity nodes otherwise \cite{UF}. All odd parity nodes are considered clusters (at the beginning every cluster will have just one element). Then, every cluster will grow, encompassing its nearest check neighbours. When a cluster grows, its parity is updated to the one of the combined constituent checks. Checks with zero parity will contribute trivially to the overall parity of the cluster. When two different clusters come into contact, they merge into a single cluster the parity of which is the resulting one of combining the two previous ones. A cluster is frozen, i.e. it stops to grow if:
\begin{itemize}
    \item The updated parity of the cluster results in an even parity.
    \item The cluster reaches a virtual qubit.
    \item The growing cluster merges with another cluster that is frozen as a result of reaching a virtual qubit.
\end{itemize}
% frozen ? preguntar ton

\begin{figure}
    \centering
    \includegraphics[width = 0.25\textwidth]{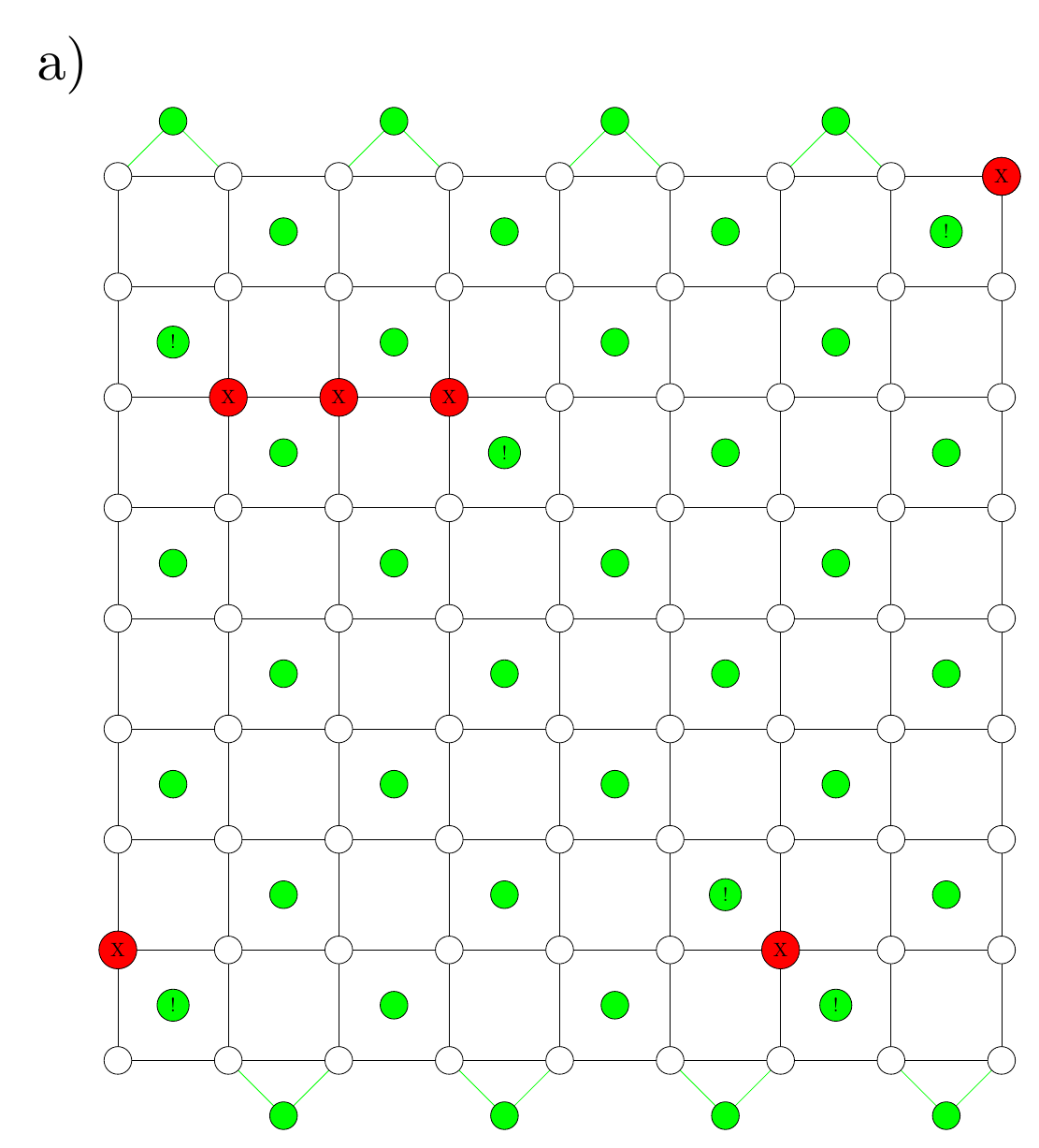} \\
    \includegraphics[width = 0.25\textwidth]{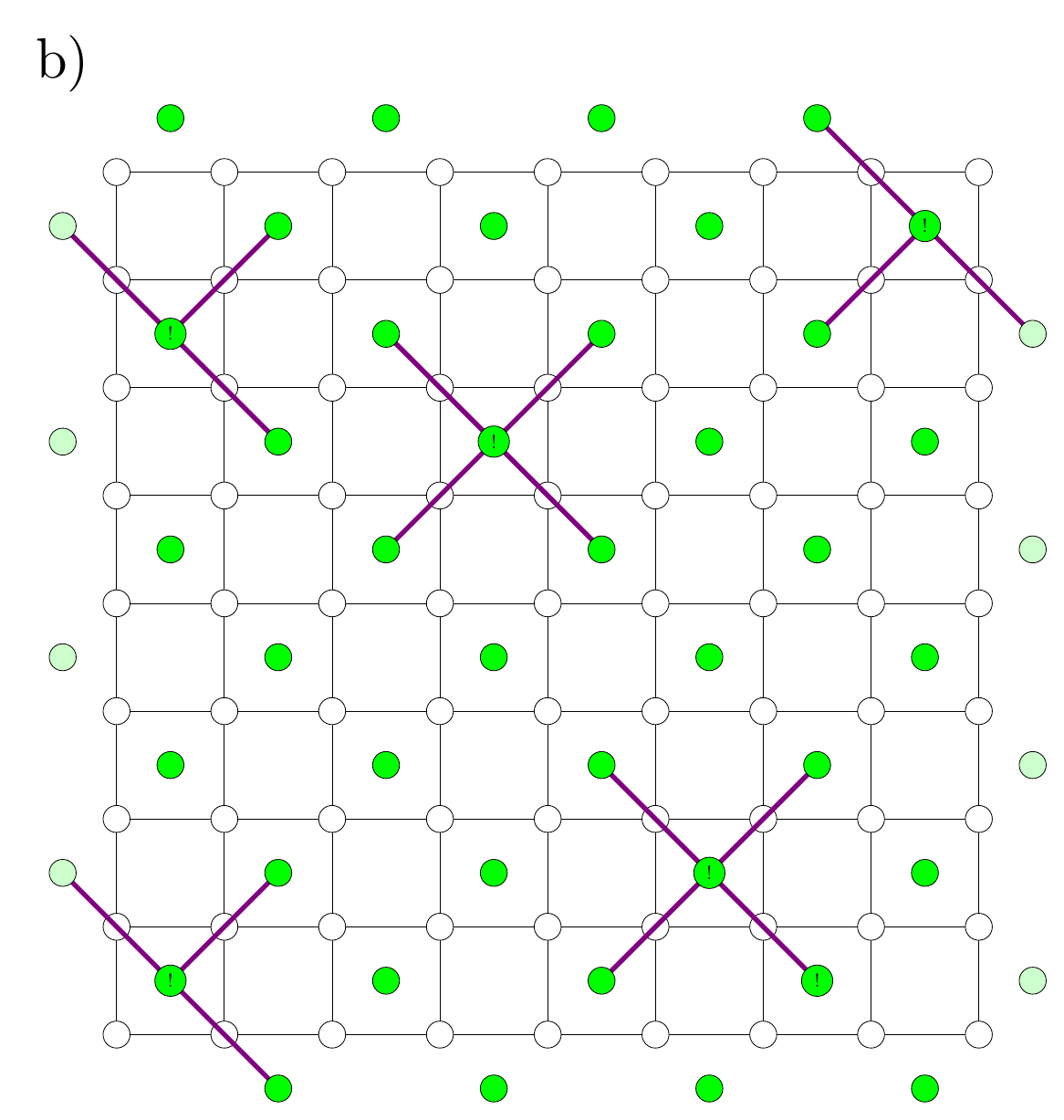} \\
    \includegraphics[width = 0.25\textwidth]{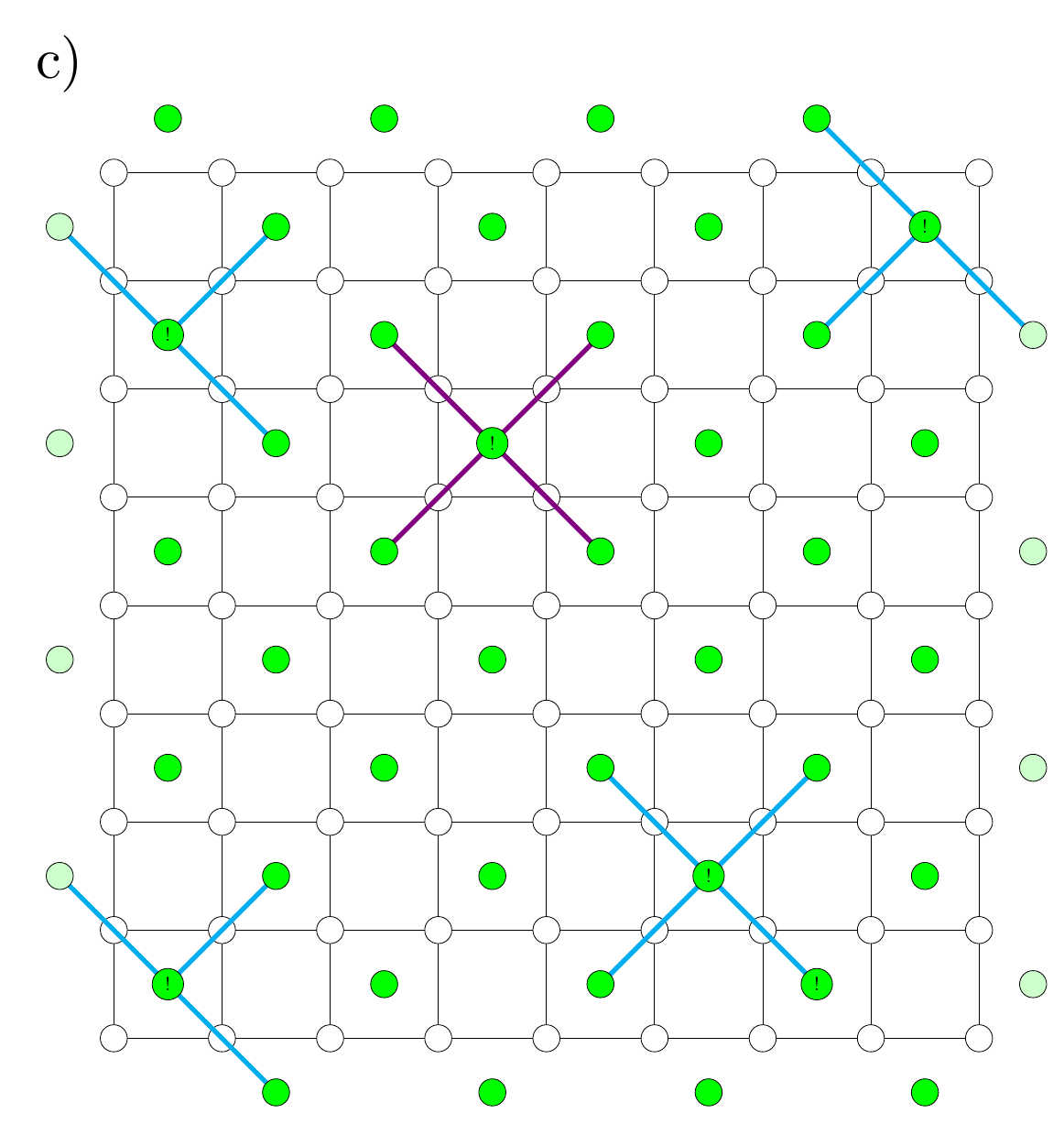} \\
    \includegraphics[width = 0.25\textwidth]{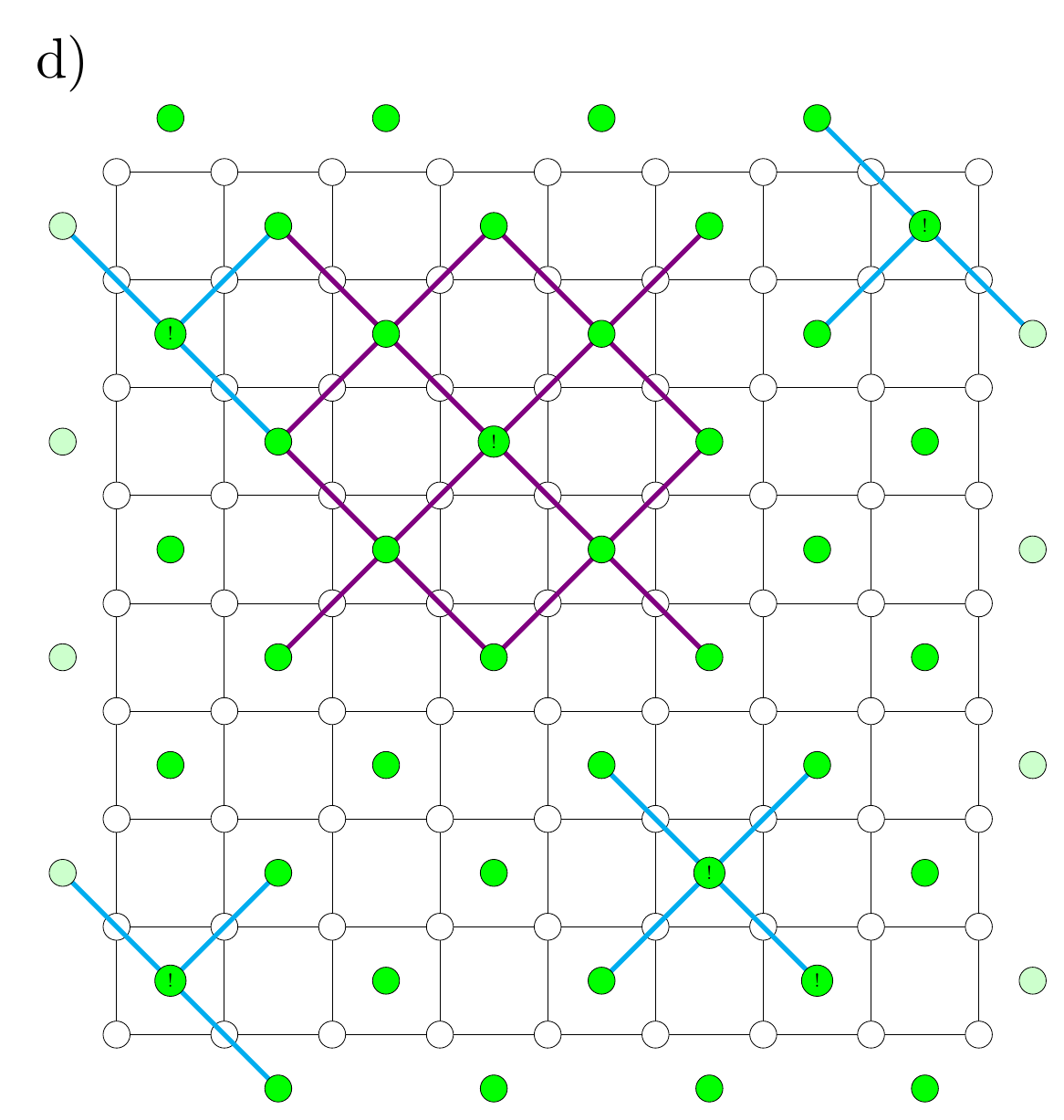} \\
    \includegraphics[width = 0.25\textwidth]{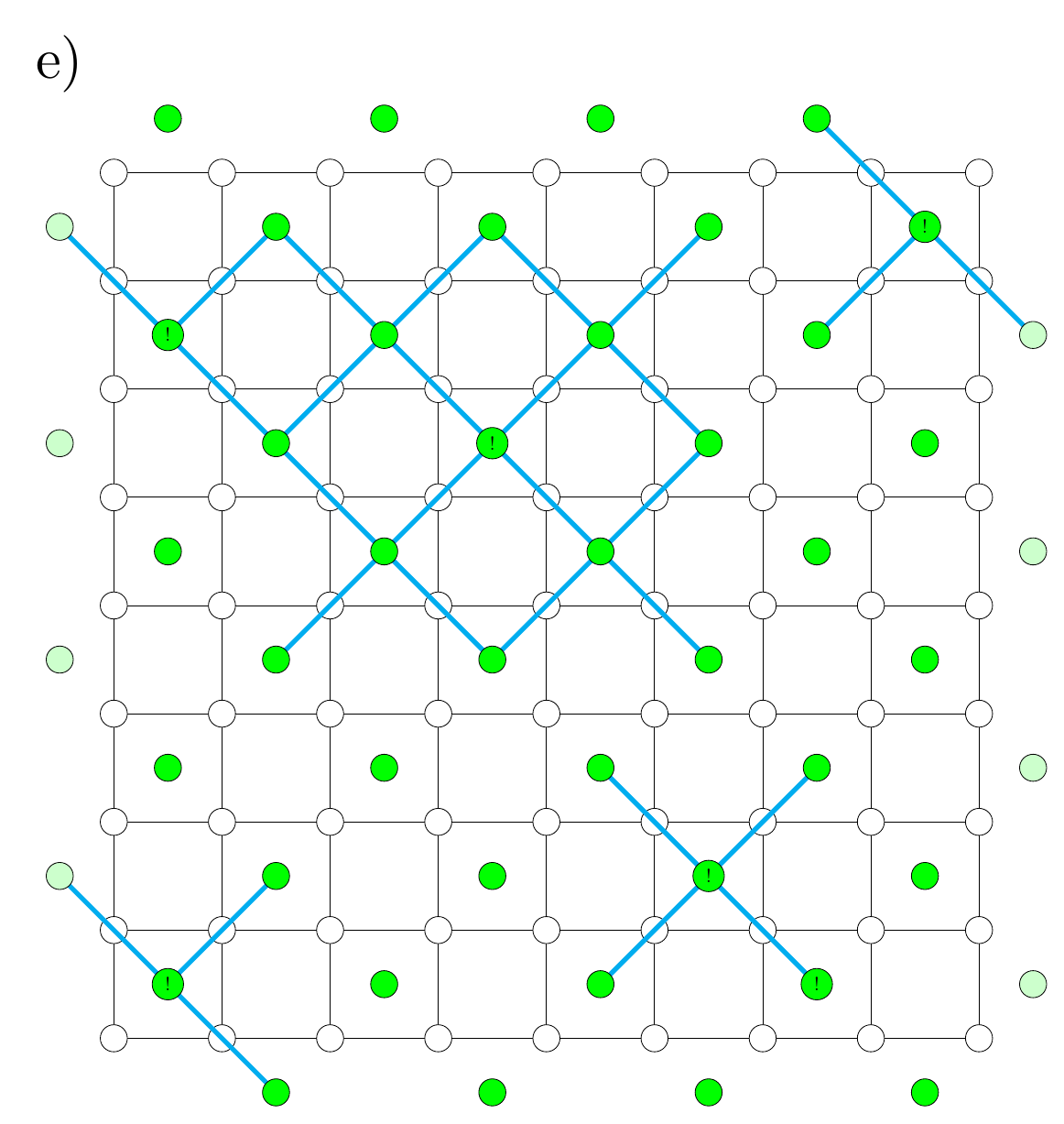} \\
    \caption{Graphical representation of the cluster growth in a distance-9 rotated planar code under a specific error. In a), an error and its syndrome is presented. From b) to e) illustrate the syndrome validation process for that syndrome. Violet lines indicate growing clusters and cyan lines indicate parity even and frozen clusters. }
    \label{clustergrowth}
\end{figure}

FIG. \ref{clustergrowth} presents an example of syndrome validation for the $X$-check graph (note that the execution of the Z-check graph will be performed in the same manner). FIG. \ref{clustergrowth} a) represents the error considered and the triggered checks. On FIG. \ref{clustergrowth} b), the clusters increase reaching the adjacent checks from the adjacent data qubits from the initial non-trivial checks. The leftmost and rightmost non-trivial checks reach the boundary and, thus, freeze, as shown in the third row. We depict frozen clusters with the cyan color. Moreover, the two triggered checks on the bottom right of the rotated code also freeze as a result of the even parity of the cluster. Therefore, by FIG. \ref{clustergrowth} c), only one cluster will continue to grow. For that reason, as it can be seen in FIG. \ref{clustergrowth} d), the cluster grows again making contact with one of the leftmost frozen clusters. Consequently, they merge into a single cluster which is frozen because the new cluster has reached a virtual qubit. This makes the syndrome validation step to conclude, as it can be seen in the FIG. \ref{clustergrowth} e).

\subsubsection{Erasure decoder}
Once the syndrome validation has been computed, the Pauli error within a surface code can be treated as an erasure error (due to its known location) and, thus, can be decoded with an erasure decoder. Due to its good performance and linear complexity, the erasure decoder of choice for this step is the peeling decoder \cite{peeling}, which will be covered in continuation. First of all, the structure of the clusters must be that of a spanning tree in order to execute such method. In graph theory, a tree is an undirected graph in which two vertices are connected by exactly one path, and so there are no cycles. A spanning tree is a tree which contains all vertices within a graph \cite{blossom1}. Consequently, since the clusters after syndrome validation may have cycles, one of the associated spanning trees must be chosen. If a cluster spans from one open boundary, i.e. a boundary with virtual checks, of the surface code to the other one, this is also considered a cycle, and so it must be split in two spanning trees, one adjacent to each side of the open boundary. The vertices of degree 1 within the spanning tree, that is, the ones that are adjacent to a single edge, are named leaves \cite{blossom1}. For the peeling decoder, one of the leaves of each spanning tree is selected as the root of the tree. If a spanning tree contains a number of virtual qubit leaves, one of them will be considered the root. The decoding process commences by selecting a non-root leaf for each cluster and applying the following rule \cite{UF}:

\begin{itemize}
    \item \textbf{If the leaf vertex is a non-trivial check: }the edge adjacent to it is stored as a matching (decoded non-trivial Pauli error), the vertex adjacent to it is flipped (if it is trivial it becomes non-trivial and vice versa), and both the leaf vertex and the edge connecting it to the rest of the spanning tree are erased from the spanning tree.
    \item \textbf{If the leaf is a trivial check: } the leaf and the edge adjacent to the leaf are removed from the spanning tree.
\end{itemize}

The peeling is directed from the leaves all the way to the root of the spanning trees. When the spanning tree is composed by only a single vertex the decoding process has been completed. It is worth mentioning that removing leaves changes the structure of the tree and may produce more leaves which must be later decoded. Moreover, virtual qubits play a somewhat ambiguous role in this decoding scheme because when they are considered as leaves they act as trivial checks, but when they are roots they are the last vertex to appear implying that it does not really matter how they are considered \cite{UF}. 

\begin{figure}
\centering
    \includegraphics[width = 0.45\columnwidth]{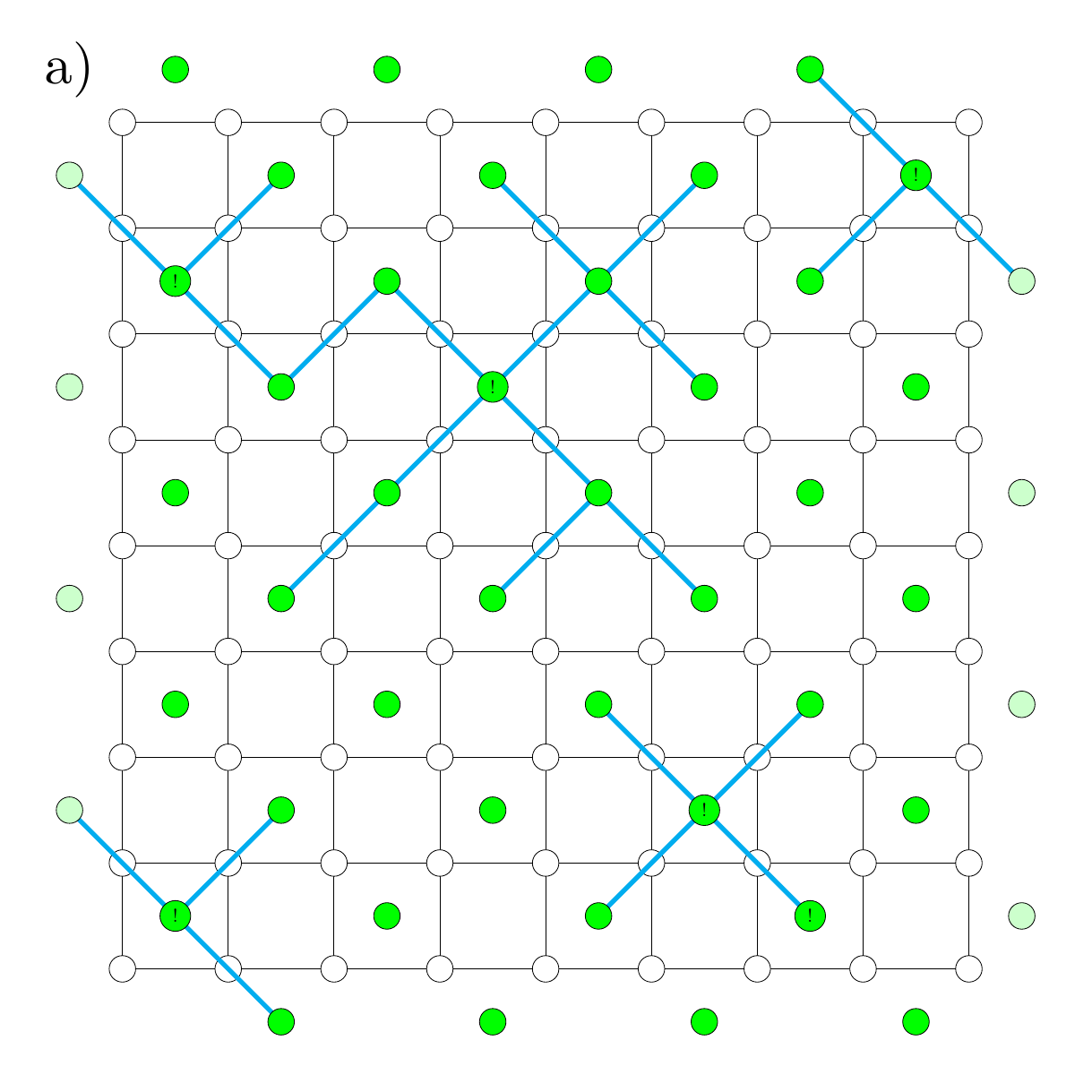} 
    \includegraphics[width = 0.45\columnwidth, frame]{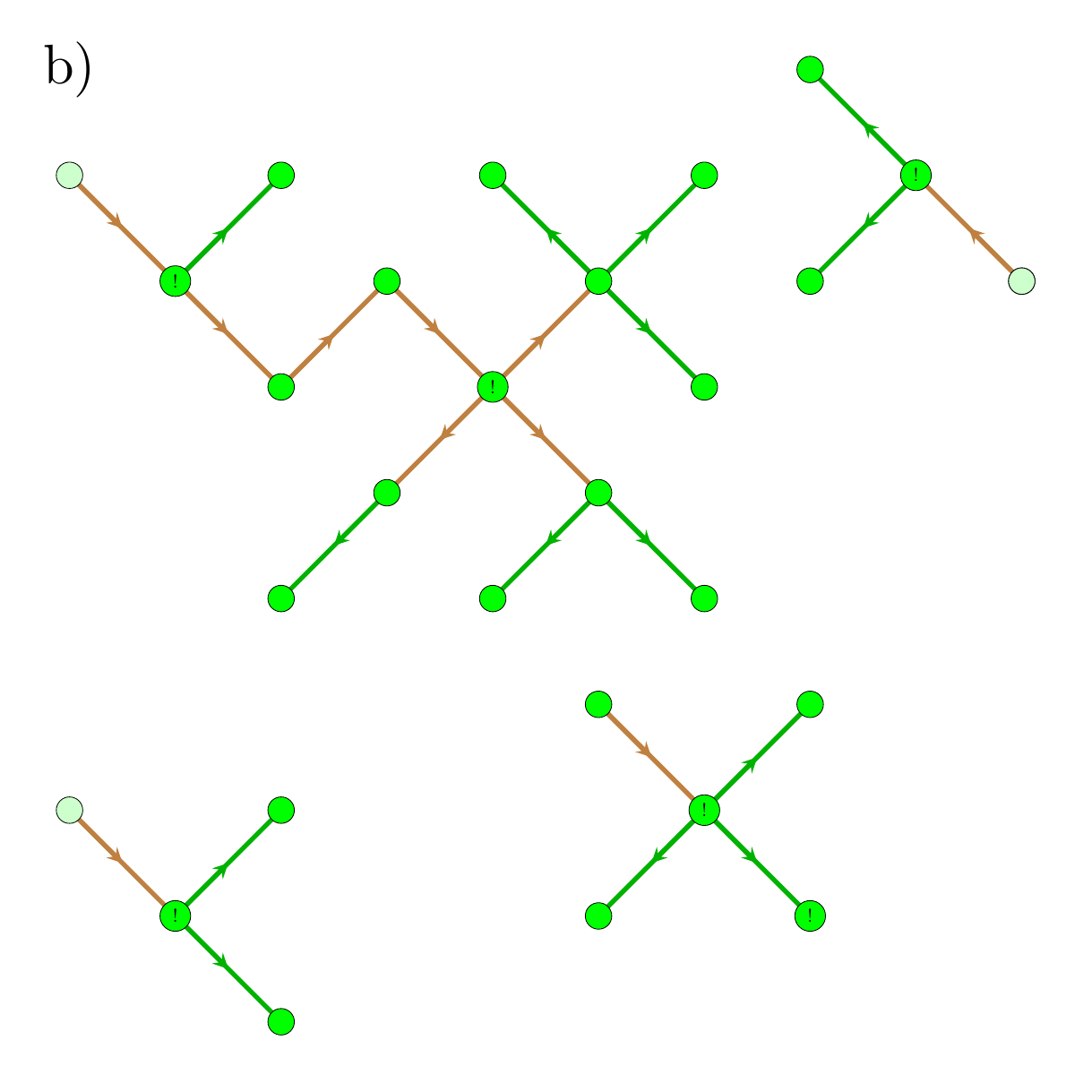} \\
    \includegraphics[width = 0.45\columnwidth, frame]{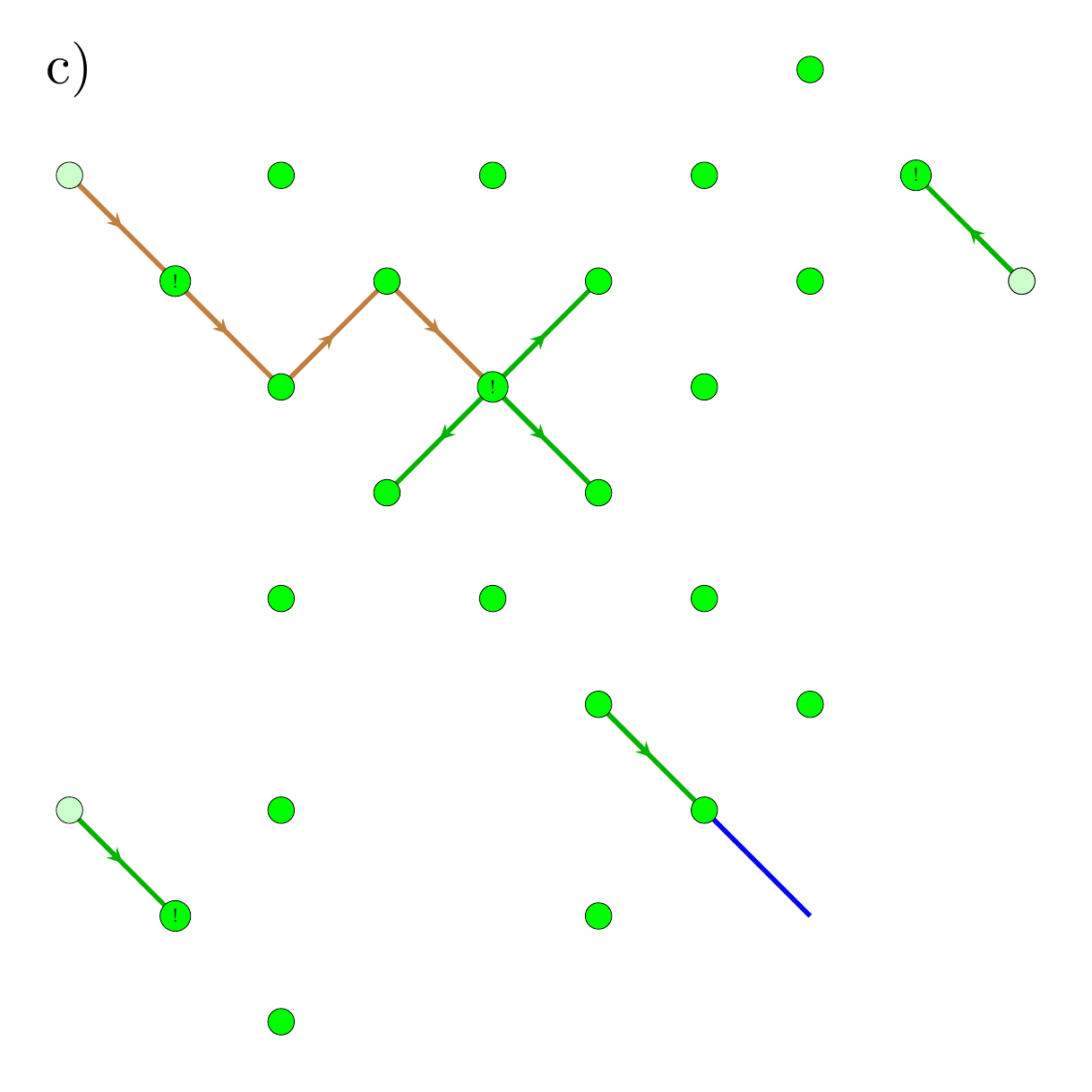} 
    \includegraphics[width = 0.45\columnwidth, frame]{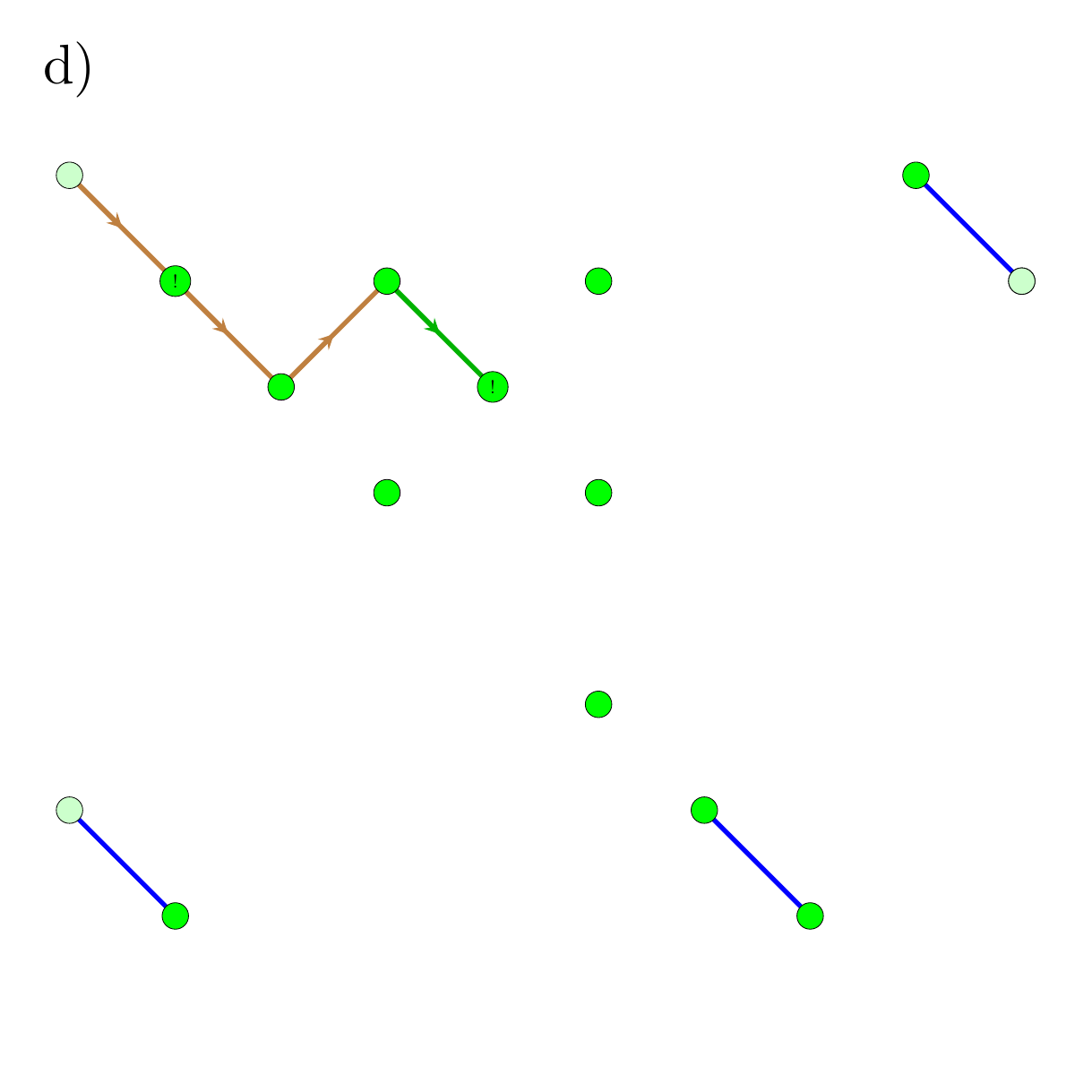} \\
    \includegraphics[width = 0.45\columnwidth, frame]{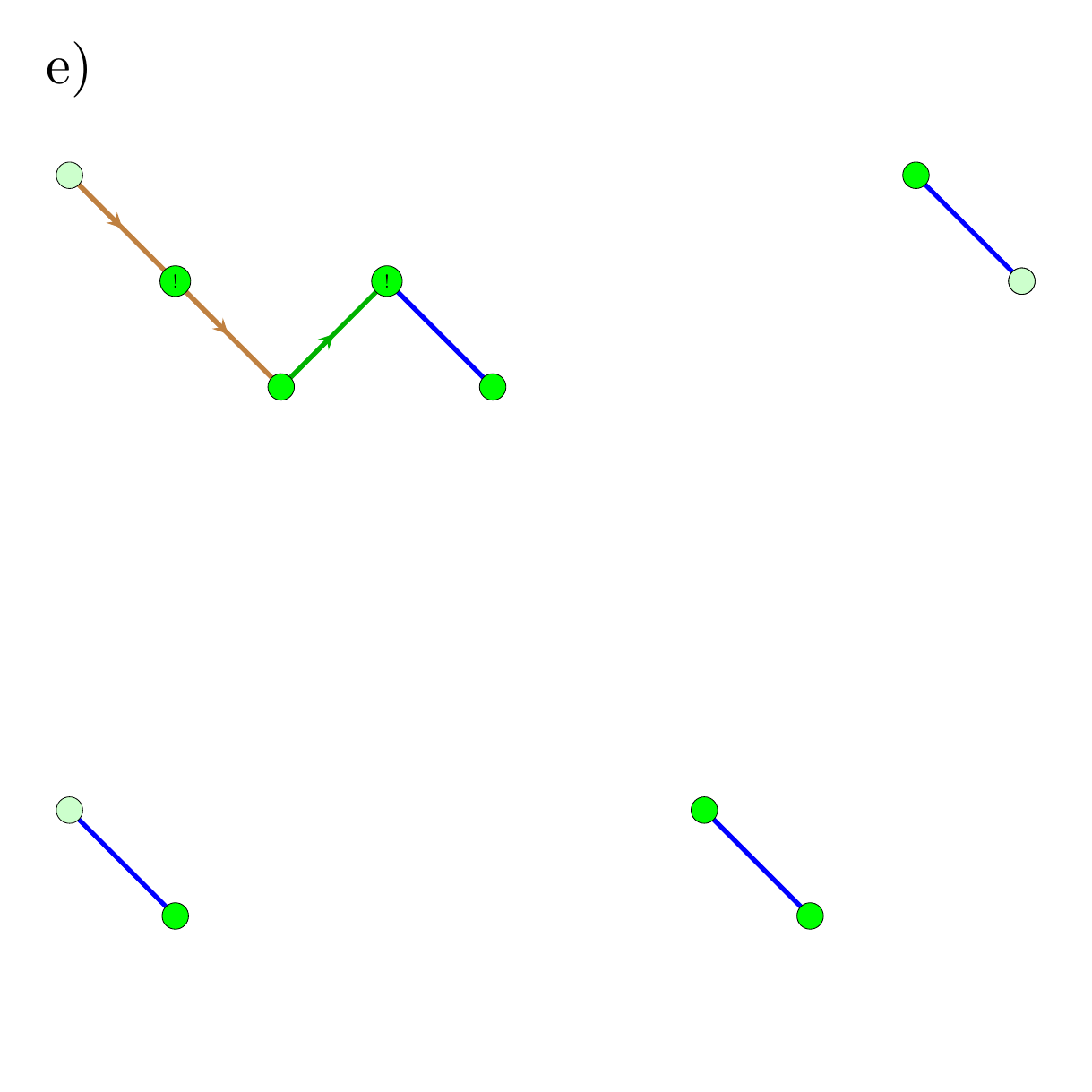} 
    \includegraphics[width = 0.45\columnwidth, frame]{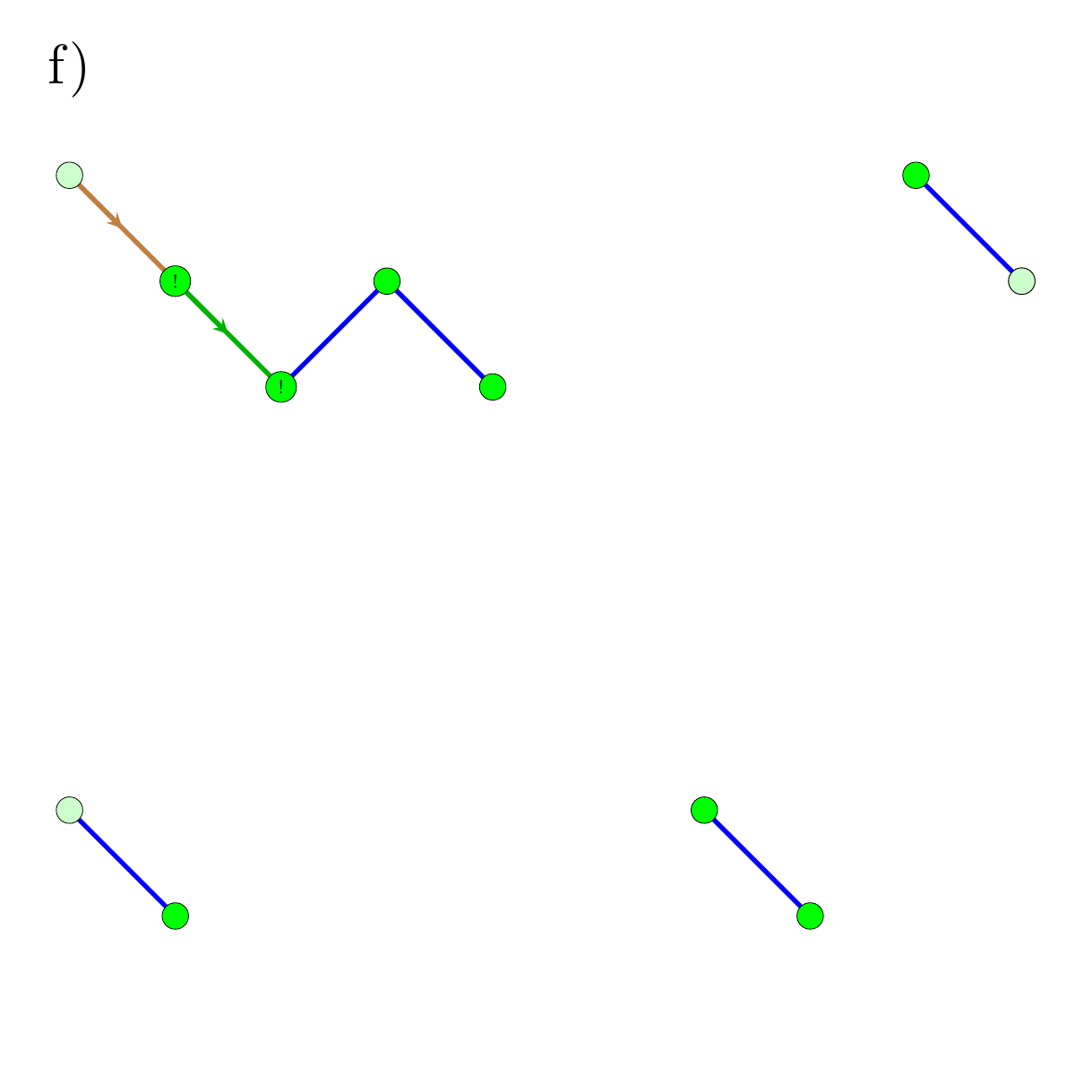} \\
    \includegraphics[width = 0.45\columnwidth, frame]{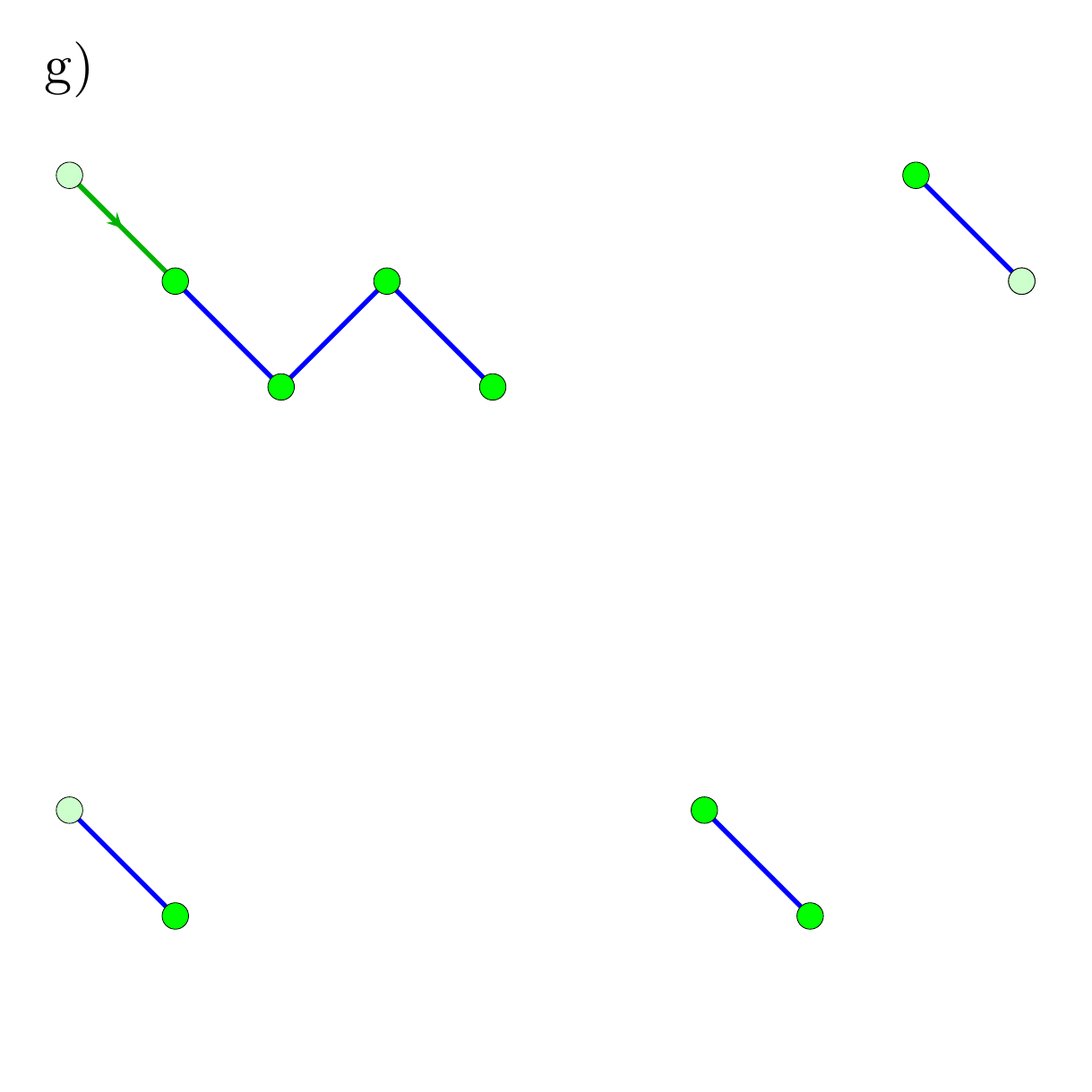} 
    \includegraphics[width = 0.45\columnwidth, frame]{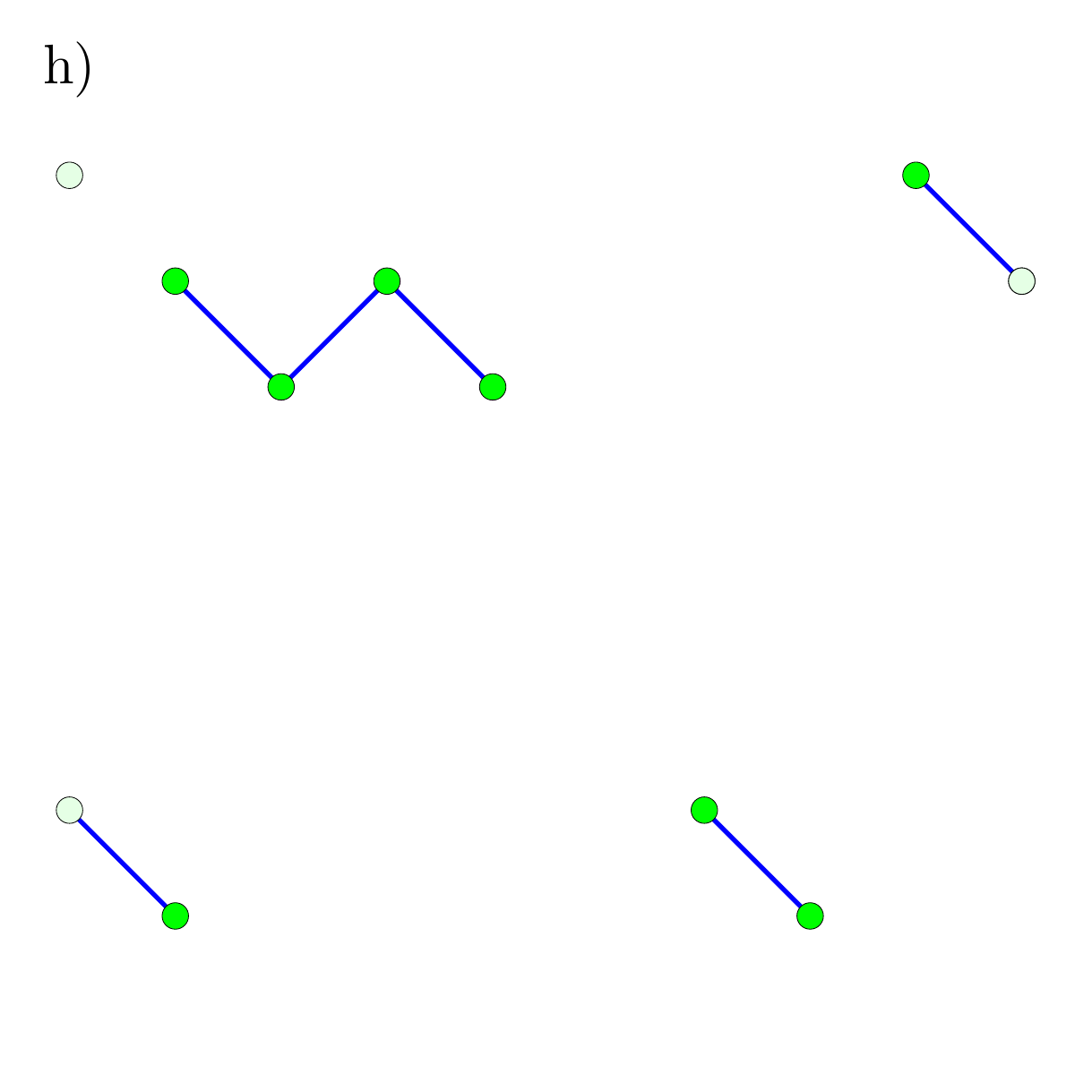} \\
    \includegraphics[width = 0.25\textwidth]{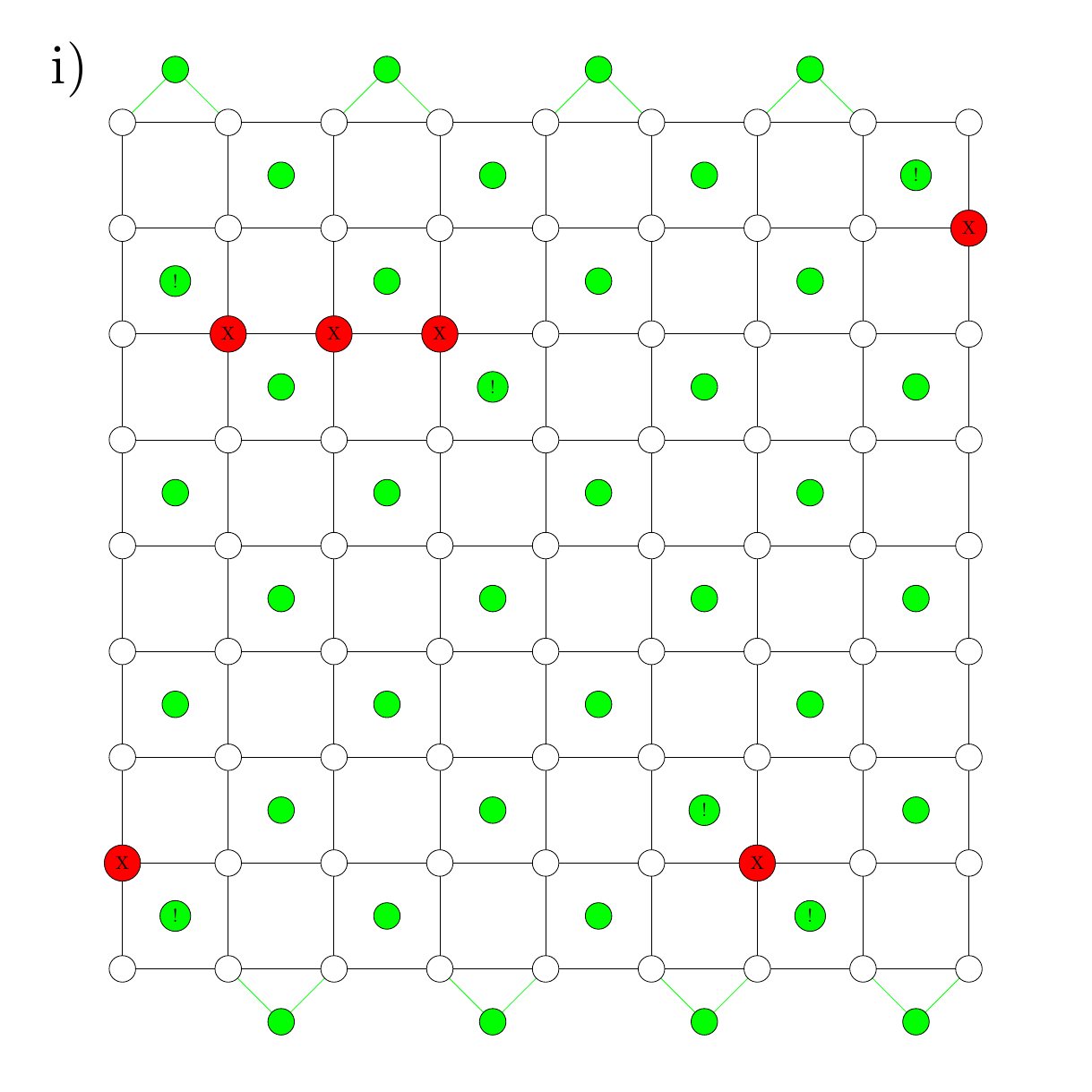} 
    \caption{Graphical representation of the forest peeling from a spanning forest chosen from the set of erasure errors from FIG. \ref{clustergrowth}. a) presents the chosen spanning forest and b) colors the spanning tree edges in green if they are leaves and brown otherwise. The peeling process lasts from c) to h), blue edges denote the non-trivial edges for the recovered error. In i), the recovered error is presented on the initial syndrome.} 
    \label{forestpeeling}
\end{figure}

In FIG. \ref{forestpeeling}, a possible peeling process for the error after syndrome validation in FIG. \ref{clustergrowth} is presented. Given the even parity clusters from FIG. \ref{clustergrowth}, a set of four spanning trees are chosen as portrayed in FIG. \ref{forestpeeling} a). In FIG. \ref{forestpeeling} b), these spanning trees are shown with identifying colors, green edges are edges incident to leaves and brown edges represent the trunks of the trees. Moreover, the arrows indicate the growth direction from the tree root to the leaves, the peeling should be done in the opposed direction. In FIG. \ref{forestpeeling} c), the result of peeling all the leaves from the FIG. \ref{forestpeeling} b) is shown. Since most of the leaves were trivial checks, no matchings arise except for one in the bottom right spanning tree, which is denoted with a blue line. The remaining FIG. \ref{forestpeeling} d), e), f), g) and h) show the progress of the peeling decoder until reaching the bottom figure which shows the recovered error. Edges which are part of the recovered error by the peeling decoder are represented with blue lines. Notice that the recovered Pauli error in FIG. \ref{forestpeeling} i) is the same one as the one in FIG. \ref{clustergrowth} a) up to a stabilizer from the top-right $X$-Pauli error implying a successful decoding round.

% Mention virtual vertices, mention if virtual vertices in both sides of the graph.

\subsubsection{Performance and threshold}
\begin{figure}
    \centering
    \includegraphics[width = \columnwidth]{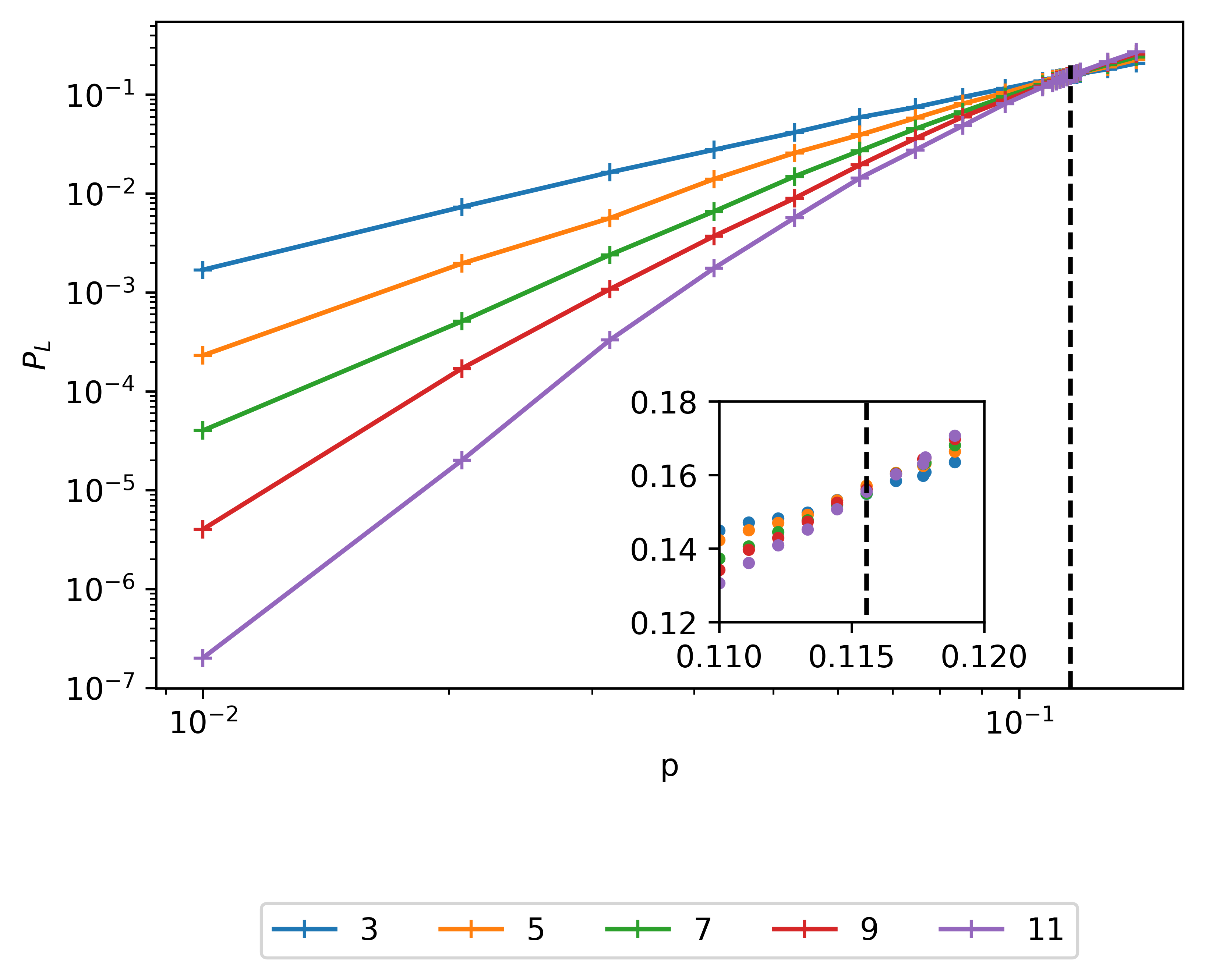}
    \caption{Logical error probability with dependence on the physical error probability from the UF decoding method under depolarizing noise.}
    \label{ufthreshold}
\end{figure}

In FIG. \ref{ufthreshold}, the performance of the rotated planar code over depolarizing noise when decoded with the UF decoder is presented. Inspecting the figure, and TABLE \ref{tablauf}, it can be seen that the code thresholds achieved by this decoding method are smaller than the ones obtained using the MWPM decoder for all biases considered (recall TABLE \ref{tablamwpm}). Interestingly, the UF decoder always returns an error suited for the syndrome facilitated, nevertheless, this error does not always result in the error of minimum weight. Thus, the decrease in performance when compared to the MWPM can be explained by the instances in which the peeling decoder misses to relate closest non-trivial syndrome elements. Several attempts have been done by the community in order to diminish the non-optimal choices made by the UF decoder while keeping its low complexity. At the time of writing, a popular approach towards this goal consists in reweighting the edges of the graph \cite{fragile, reweighteduf}. The reweighting is usually done by previously running some method in order to estimate which data qubits are more prone to have suffered from errors for such syndrome measurement. With such information, the edges representing data qubits that are more prone to errors will have lower weights, which implies that the vertices they connect are closer. Thus, when growing clusters in the syndrome validation phase, the radial growth is fixed and clusters are more prone to grow towards likely to fail data qubits. For example, the so-called belief-find decoder uses a belief propagation method to estimate such information and then continues to decode the error by the UF method \cite{fragile}.

%Crigerbeliefmatching

\begin{table}
\begin{center}
\begin{tabular}{|c | c|} 
 \hline
 $\eta$ &  $p_{th}$ \\ [0.5ex] 
 \hline\hline
 1/2 & 0.116  \\ 
 \hline
 1 & 0.114  \\
 \hline
 10 & 0.080  \\
 \hline
 100 & 0.078  \\
 \hline
 1000 & 0.077  \\ [1ex] 
 \hline
\end{tabular}
\caption{Probability threshold values for different biases in the rotated planar code under the UF decoding scheme.}
\label{tablauf}
\end{center}
\end{table}

\subsubsection{Measurement errors}
As explained before, the so-called measurement errors due to imperfect measurement operations and propagation of errors in stabilizer measurement stage have been considered for the MWPM decoder. In the case of the UF decoder, this type of effects have also been taken into account by considering multiple syndrome measurements for a single decoding round, refer to Appendix \ref{appA} for an extended description. In this sense, syndrome validation and peeling are realized over the space-time graph that is obtained. Reweighting methods for a better performance of the UF decoder over those space-time graphs have been discussed too \cite{fragile, reweighteduf}. In addition, and due to the fact that the complexity of the algorithm increases when the space-time graph is considered, truncated UF methods have also been proposed to maintain the fast decoding while not losing too much in terms of decoding success \cite{reweighteduf}.

Ultimately, UF proves to be a very efficient method for decoding the surface code and stands as a fair counterpart to the conventional MWPM decoding process. So much so, that the cluster growth process in the syndrome validation has inspired new methods for optimizing the computational complexity of the minimum-weight perfect matching decoder \cite{sparse, fussionBlossom}. Moreover, the UF-decoding method also yields the great advantage of successfully taking into account erasure errors. Were a surface code to undergo a mixture of Pauli and erasure errors, the only difference with the process explained before would be that on the syndrome validation step, there would also be erasure errors, frozen from the beginning, in the form of edges that would join to whichever cluster gets in contact with them. Lastly, there have also been studies trying to strictly relate the conditions under which UF will return the same error as the MWPM method \cite{wu2} and even more, it has been studied as a possible decoding method for QLDPCs \cite{ldpcuf}, although the complexity problem in this specific case has yet to be addressed. Due to its low complexity and high threshold, as of the moment of writing, the UF method seems to be a promising candidate for early experimental real-time surface code decoding.

\subsection{Belief Propagation}

Belief Propagation (BP) is a message-passing algorithm that can be used to solve inference problems on probabilistic graphical models \cite{BP}. It is also sometimes referred to as the Sum-Product Algorithm (SPA) \cite{SPA}, which is a more general-purpose algorithm that computes marginal functions associated with a global function. Although the terms BP and SPA are essentially interchangeable, throughout this paper we will use BP to refer to the algorithm applied to decode error correction codes.

Error correction codes, irrespective of being applied in a quantum or classical paradigm, can be represented by bipartite graphs known as factor graphs. A factor graph $G = (N_G, E_G)$ is defined by a set of nodes $N_G = V\cup C$, where $V, C$ represent two distinct types of nodes known as variable and check nodes, respectively, and $E_G$ is a set of edges. In the context of error correction codes, factor graphs are generally referred to as Tanner graphs. Throughout the rest of this paper, we will use the terms factor graph and Tanner graph interchangeably to refer to the graphical representation of an error correction code. A Tanner graph represents the structure of the error correcting code by means of variable nodes representing the physical bits and check nodes representing the checks. Every check will be connected with edges to the physical bits that can change its parity. Moreover, the Tanner graph can be represented through a binary matrix named the \textit{parity check matrix}. Within the parity check matrix, every column represents a physical bit and every row represents a check. Non-trivial elements within the parity check matrix define edges connecting the physical bit from the column to the check from the row. Then, given a binary error vector $\bar{e}$, where non-trivial elements indicate bit-flips, one can analytically obtain the syndrome as $\bar{s} = H\cdot \bar{e} \mod 2$, where $\bar{s}$ is the syndrome and $H$ the parity check matrix.

One can extend this logic to quantum error correcting codes, although taking into account some caveats\footnote{For a more detailed description to as to why parity checks are described like this in the quantum realm refer to \cite{negglobphase}.}. In general terms, quantum parity check matrices have $2n$ columns, due to the fact that each qubit is considered to have two parities, the $X$-error parity, which corresponds to the first $n$ columns, and the $Z$-error parity, which corresponds to the second $n$ columns. A $Y$-error is the result of a non-trivial $X$ and $Z$-error parity within the same qubit. Moreover, the checks are also represented through the rows. As an example, we can study the distance-$3$ planar code, which is defined by the set of stabilizer generators shown in \eqref{eq:stab_gens_d3} (relisted below to aid the reader):

\begin{equation*}
    \begin{aligned}
        s_1 &= Z_2Z_3,\\
        s_2 &= Z_1Z_2Z_4Z_5,\\
        s_3 &= Z_5Z_6Z_8Z_9,\\
        s_4 &= Z_7Z_8,\\
        s_5 &= X_1X_4,\\
        s_6 &= X_4X_5X_7X_8,\\
        s_7 &= X_2X_3X_5X_6,\\
        s_8 &= X_6X_9,
    \end{aligned}
\end{equation*}

As expressed in previous sections, the rotated code is CSS, which allows us to consider two separate parity check matrices, one for the checks susceptible to $X$-errors and another for the checks susceptible to $Z$-errors. Therefore, the parity check matrix susceptible to $X$-errors is given by stabilizers $s_1, s_2, s_3$ and $s_4$ and has the following expression:

\begin{equation}\label{eq:Hx_mat}
    H_X = \begin{pmatrix}
        0 & 1 & 1 & 0 & 0 & 0 & 0 & 0 & 0\\
        1 & 1 & 0 & 1 & 1 & 0 & 0 & 0 & 0\\
        0 & 0 & 0 & 0 & 1 & 1 & 0 & 1 & 1\\
        0 & 0 & 0 & 0 & 0 & 0 & 1 & 1 & 0\\
    \end{pmatrix},
\end{equation}

and the expression for the parity check matrix susceptible to $Z$-errors is given by stabilizers $s_1, s_2, s_3$ and $s_4$ and takes the form:

\begin{equation}\label{eq:Hz_mat}
    H_Z = \begin{pmatrix}
        1 & 0 & 0 & 1 & 0 & 0 & 0 & 0 & 0\\
        0 & 0 & 0 & 1 & 1 & 0 & 1 & 1 & 0\\
        0 & 1 & 1 & 0 & 1 & 1 & 0 & 0 & 0\\
        0 & 0 & 0 & 0 & 0 & 1 & 0 & 0 & 1\\
    \end{pmatrix}.
\end{equation}

Within both $H_X$ and $H_Z$, the columns represent the $X$ and $Z$-parity of the data qubits respectively and the rows represent the stabilizer generators.

% The set of generators $\{s_i\}_{i=1}^8$ can also be represented by the rows of the parity check matrix of the distance-3 rotated code}:

% \begin{equation} \label{eq:surfPCM}
% \setlength\arraycolsep{2pt} % Adjust column separation
% \left(
% \begin{array}{ccccccccc|ccccccccc}
% 0 & 0 & 0 & 0 & 0 & 0 & 0 & 0 & 0 & 1 & 1 & 0 & 0 & 0 & 0 & 0 & 0 & 0 \\
% 0 & 0 & 0 & 0 & 0 & 0 & 0 & 0 & 0 & 0 & 0 & 0 & 1 & 1 & 0 & 1 & 1 & 0\\
% 0 & 0 & 0 & 0 & 0 & 0 & 0 & 0 & 0 & 0 & 1 & 1 & 0 & 1 & 1 & 0 & 0 & 0\\
% 0 & 0 & 0 & 0 & 0 & 0 & 0 & 0 & 0 & 0 & 0 & 0 & 0 & 0 & 1 & 0 & 0 & 1\\
% 0 & 1 & 1 & 0 & 0 & 0 & 0 & 0 & 0 & 0 & 0 & 0 & 0 & 0 & 0 & 0 & 0 & 0\\
% 1 & 1 & 0 & 1 & 1 & 0 & 0 & 0 & 0 & 0 & 0 & 0 & 0 & 0 & 0 & 0 & 0 & 0\\
% 0 & 0 & 0 & 0 & 1 & 1 & 0 & 1 & 1 & 0 & 0 & 0 & 0 & 0 & 0 & 0 & 0 & 0\\
% 0 & 0 & 0 & 0 & 0 & 0 & 1 & 1 & 0 & 0 & 0 & 0 & 0 & 0 & 0 & 0 & 0 & 0\\
% \end{array}
% \right),
% \end{equation}

  Moreover, the CSS structure of the rotated planar code allows us to compute syndromes as: $\bar{s}_X = H_X\cdot \bar{e}_X \mod 2$ and $\bar{s}_Z = H_Z\cdot \bar{e}_Z \mod 2$, where , $\bar{e}_X$ ($\bar{e}_Z$) is a binary error vector correspondent to the bit-flips (phase-flips) and $\bar{s}_X$ ($\bar{s}_Z$) is a syndrome for the $X$ ($Z$) parity. Moreover, from $H_X$ and $H_Z$, we construct the Tanner graph as follows:
\begin{enumerate}
\item For every data qubit in the code, add two variable nodes to the graph. Place these nodes at the center of the graph. Our example will have $9$-pairs of variable nodes. Each node represents a different error parity, nodes to the left indicate $X$-parity and nodes to the right indicate $Z$-parity.
\item For every stabilizer generator in the code, add a check node to the graph. For checks correspondent to $H_X$, place the check nodes to the left of the variable nodes, checks correspondent to $H_Z$ should be placed to the right of the variable nodes.
\item Add an edge from every stabilizer generator to the parity it is susceptible to. This corresponds to the non-zero elements for their row in their parity check matrix. For example, $s_1$ corresponds to the first row in $H_X$, therefore it should be connected to the variable nodes $2$ and $3$.
\end{enumerate}

\begin{figure}
    \centering
    \includegraphics[width = 0.6\columnwidth]{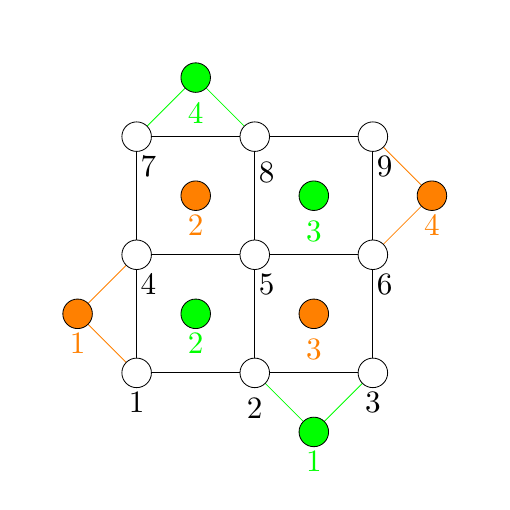}
    \includegraphics[width = 0.6\columnwidth]{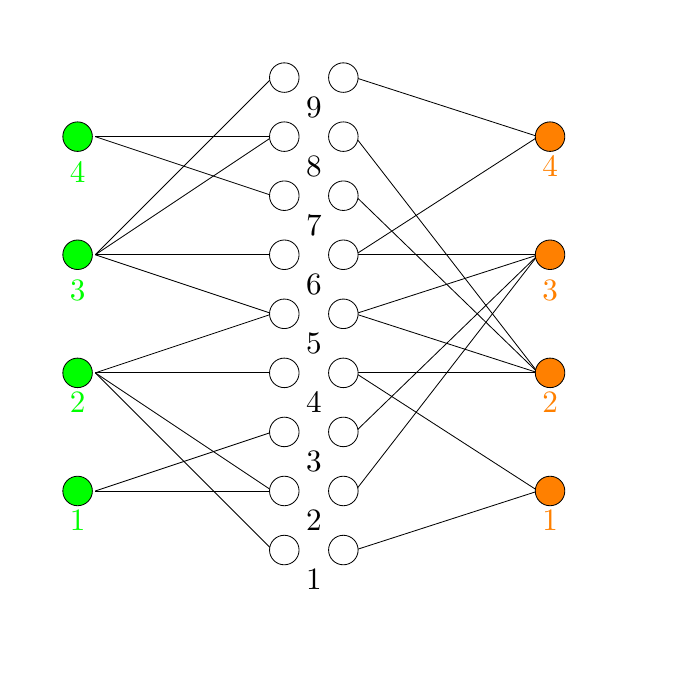}
    \caption{On the top, a representation of a distance-3 rotated planar code where both data and check qubits are labelled. On the bottom, the Tanner graph of said code. Note that the Tanner graph represents which data qubits interact with which checks.}
    \label{tannergraphplot}
\end{figure}

In FIG. \ref{tannergraphplot}, the distance-$3$ rotated planar code with parity check matrices \eqref{eq:Hx_mat} and \eqref{eq:Hz_mat} is illustrated altogether with its Tanner graph representation. Consider that, as has been seen, every data qubit is portrayed through two parities, the $X$-parity and the $Z$-parity.This is further represented in the Tanner graph of FIG. \ref{tannergraphplot}, where we can see each qubit represented by two variable nodes, one interacting with $X$-checks and another interacting with $Z$-checks.

Against this backdrop, BP can be understood as an algorithm that approximates the marginal probabilities of undergoing a parity change for every variable node in the Tanner graph through message exchanging between nodes. If BP runs over a graph that is a tree, it will converge to the exact marginal probability solution in a time bounded by the tree’s depth. That is, the marginal probabilities will provide an error which will be consistent to the syndrome. In scenarios where this does not hold, as is the case with most factor graphs that represent error correction codes, BP has proven to be a good heuristic decoding method, especially when the typical size of the loops in the graph is large.

\subsubsection{BP: Specifics}

In the classical context, BP decoders work by solving an approximation of the classical maximum likelihood decoding problem known as bit-wise maximum likelihood decoding \cite{reviewpat}. In the quantum context, an analogous principle is employed: instead of tackling QMLD, we use classical BP decoders to solve the problem of qubit-wise maximum likelihood decoding. 

% Pregunta de Ton: MWPM no es MLD decoder siendo complexity P? (Polinomial) - Pat: Creo que MWPM aproxima MLD, porque resolver QMLD es imposible de manera exacta (es un problema NP-hard)

Qubit-wise maximum likelihood decoding is different from QMLD in that, instead of looking for the most likely error given the syndrome, we will look for the qubit-wise most likely error, i.e., looking for the  {marginal probabilities of every column within the parity check matrix}. Qubit-wise maximum likelihood decoding can be written as:

 \begin{equation} \label{eq:bwmld}
        \hat{E}_i^{\text{bw}} = \Argmax_{E_i\in\mathcal{G}_1} \sum_{E_1,\ldots,E_{i-1},E_{i+1},\ldots,E_n}P(E_1\dots E_n|s),
    \end{equation}

where the qubit-wise most likely error is obtained by running through all values of $i$: $\hat{E}^{\text{bw}} = [\hat{E}^{\text{bw}}_1\ldots\hat{E}^{\text{bw}}_n]$. In general, $\hat{E}^{\text{bw}}$ need not coincide with the global optimum obtained by solving \eqref{eq:QMLD} \cite{degen3}.

Although computing solutions to \eqref{eq:bwmld} is also difficult, this task is amenable to BP. Given the Tanner graph of a code, BP can run over the aforementioned graph and obtain the qubit-wise most likely error in polynomial time. {As earlier seen, we can consider the code as two separate parity check matrices $H_X$ and $H_Z$ and, thus,}  the BP decoding problem can be understood as the execution of two classical BP decoders  {that are run simultaneously}, one for $X$-errors and the other for $Z$-errors.  {Thus, to understand BP-based decoding of CSS codes, we can limit ourselves to explaining how BP decoding works for classical codes} {To begin, the BP algorithm requires the following:}
\begin{itemize}
    \item A Tanner graph $G$ representing the parity check matrix $H$ of the $[n, k, d]$ classical code\footnote{ {The $[n, k, d]$ notation represents the fact that the code has $n$ physical bits, $k$ logical bits, and distance $d$.}} we are working with.
    \item An error syndrome $\bar{s}$ computed as  {$\bar{s} = H\cdot \bar{e}$}.
    \item The probability distribution of the error,  {$P(\hat{e})$.}
\end{itemize}

 Once $G$, $\bar{s}$, and  {$P(\hat{e})$} have been computed, the BP decoding algorithm operates in the following manner:

\begin{enumerate}

    \item  {\textbf{Initialization:} Each entry in the syndrome vector $\{s\}_{1}^{m}$ is provided to a check node $\{c\}_{1}^{m}$ of $G$, where $m$ indicates the number of checks. Additionally, each variable node $\{v\}_{1}^{n}$ is assigned a probability, known as the \textit{a priori bit error probability}, computed as $$\frac{p(\hat{e}_i=0)}{p(\hat{e}_i=1)},$$ where $i \in [1, \ldots, n]$. Once this information is provided to the check and variable nodes of $G$ the BP algorithm begins exchanging messages between them. These messages are understood in the context of \textit{decoding iterations}, which are defined as the bi-directional exchange of messages over all the edges of $G$. This means that an iteration completes whenever all connected variable and check nodes have both sent and received a message. The BP algorithm operates by executing numerous of these iterations to propagate information throughout $G$. In what follows we describe the message exchange procedure for an arbitrary iteration.}
    \item At any given decoding iteration $t$, each variable node, $v_i$, sends out a message, $\mu^{t}_{v_i\rightarrow c_j}$, to the check nodes, $c_j$, it is connected to.  {In the very first decoding iteration, $t=1$, this message is equal to the a priori bit error probability which was assigned to each variable node in the initialization stage. These messages are generally expressed in the log-likelihood domain. For $t=1$ the messages are given by $$\mu^1_{v_i\rightarrow c_j}=l_\text{ch}(\hat{e}_i) = \text{log}(\frac{p(\hat{e}_i=0)}{p(\hat{e}_i=1)}),$$ where $i \in [1,\ldots,n]$, $j \in [1,\ldots,n-k]$, and the term $l_\text{ch}$ represents the a priori channel \textit{log-likelihood ratio} (llr)}. In future iterations ($t > 1$), the message sent from a variable node to neighboring check nodes is given by  {$$ \mu^t_{v_i\rightarrow c_j} = l_\text{ch}(\hat{e}_i) + \sum_{k=1}^{\sigma-1}\mu^{t-1}_{c_k\rightarrow v_i}, $$}
    where $\mu^{t-1}_{c_k\rightarrow v_i}$ are messages received from check nodes in the previous iteration and $\sigma$ is the degree (number of connections) of the variable nodes. The sum goes to $\sigma - 1$ because the message received from check node $c_j$ in the previous iteration ($t-1$) is not considered.

    \item Once the check nodes have received a message from neighboring variable nodes, every check node will reply to all neighbouring variable nodes with the following message:

    $$
    \mu^t_{c_j\rightarrow v_i} = (-1)^{{s}_j}2\text{atanh}[\prod_{k=1}^{\psi-1}\text{tanh}(\mu^t_{v_k\rightarrow c_j})],
    $$

    where $\psi$ represents the degree of the check nodes, and $s_j$ denotes the syndrome bit associated to that particular check node. Notice how the product required to compute $\mu_{c_j\rightarrow v_i}$ only goes to $\psi - 1$. Once again, this is because the prior variable node message received from the node the current message will be sent to should not be considered. 

    \item  {Once all check node messages have been received by the variable nodes, the message exchange process for iteration $t$ has finished. At this point, variable nodes update their original probabilities (also referred to as beliefs or marginals) based on the information transmitted from the check nodes. Given this update, the a-priori llrs now become a-posteriori llrs. This update is done as $$l^t_\text{ap}(\hat{e}_i) =  l_\text{ch}(\hat{e}_i) + \sum_{k=1}^\sigma\mu^t_{c_k\rightarrow v_i}.$$
    Note how the a-posteriori llrs are a combination of the initial a-priori channel llrs and decoding information processing.}

    \item These updated a posteriori llrs is what the BP decoder uses to estimate \eqref{eq:bwmld}. First we obtain the estimate $\hat{e}$ by making hard decisions on the beliefs, $l^t_\text{ap}(\hat{e}_i)$ {, that is, we choose a threshold value for $l^t_\text{ap}(\hat{e}_i)$ at which we consider the parity to be non-trivial. The common threshold consists in considering $l^t_\text{ap}(\hat{e}_i) < 0$ to be flipped parities}. Then, we compute the syndrome associated to the decoding estimate as $$ \hat{s} = H\cdot\hat{e}.$$ If  {$\hat{s} = \bar{s}$} then decoding has been successful and the BP algorithm is halted. If not, then additional iterations will be run until either  {$\hat{s} = \bar{s}$} or a maximum number of iterations is reached.

\end{enumerate}

 {Therefore, one can see the BP algorithm as a method in which both variable and check nodes iterative send information about their state to their adjacent nodes regarding their marginal probability state. After sending said information in the form of a belief, their state is updated through the beliefs they receive. This local transmission of information which encompasses all the Tanner graph produces changes in the marginal probabilities of the variable nodes until it is halted. Afterwards, one can retrieve a recovered error $\hat{e}$ by considering that negative $l^t_\text{ap}(\hat{e}_i)$ correspond to flipped parities ($\hat{e}_i = 1$). This recovery is taken under the consideration that negative $l^t_\text{ap}(\hat{e}_i)$ mean resulting marginal probabilities indicating the variable node is more likely to be flipped than not to. }

\subsubsection{Aftermath of neglecting DQMLD}

The excellent performance of BP as a decoding algorithm for classical random-like codes, such as LDPC codes and turbo codes \cite{dBs}, is well-documented. In fact, classical LDPC codes essentially  {achieve theoretical performance limits} when decoded via BP \cite{dBs}. For this reason, along with their finite-rate guarantees, the design of so-called `good'\footnote{By `good' error correction codes we refer to codes whose number of encoded logical qubits and distance increase in a polynomial manner with the number of physical qubits. Essentially, $k, d \approx \mathcal{O}(n)$.} quantum LDPC codes has been a long-pursued topic in the field of QEC. Although the existence of sparse codes exhibiting such favorable parameter scaling remained unproven for the past two decades, groundbreaking results by Panteleev and Kalachev \cite{qldpc4} as well as \cite{fiber, balanced, QuantumTanner} have finally shown that quantum analogues of robust LDPC codes do actually exist.  {In light of these results and given the overhead advantages that quantum LDPC codes offer when compared to surface codes, industry leading companies have begun studying their commercial viability, as was shown in recently published results for circuit-level performance of quantum LDPC-based memories \cite{bravyildpc, QuEra}.}

\begin{figure}
    \centering
    \includegraphics[width = 1\columnwidth]{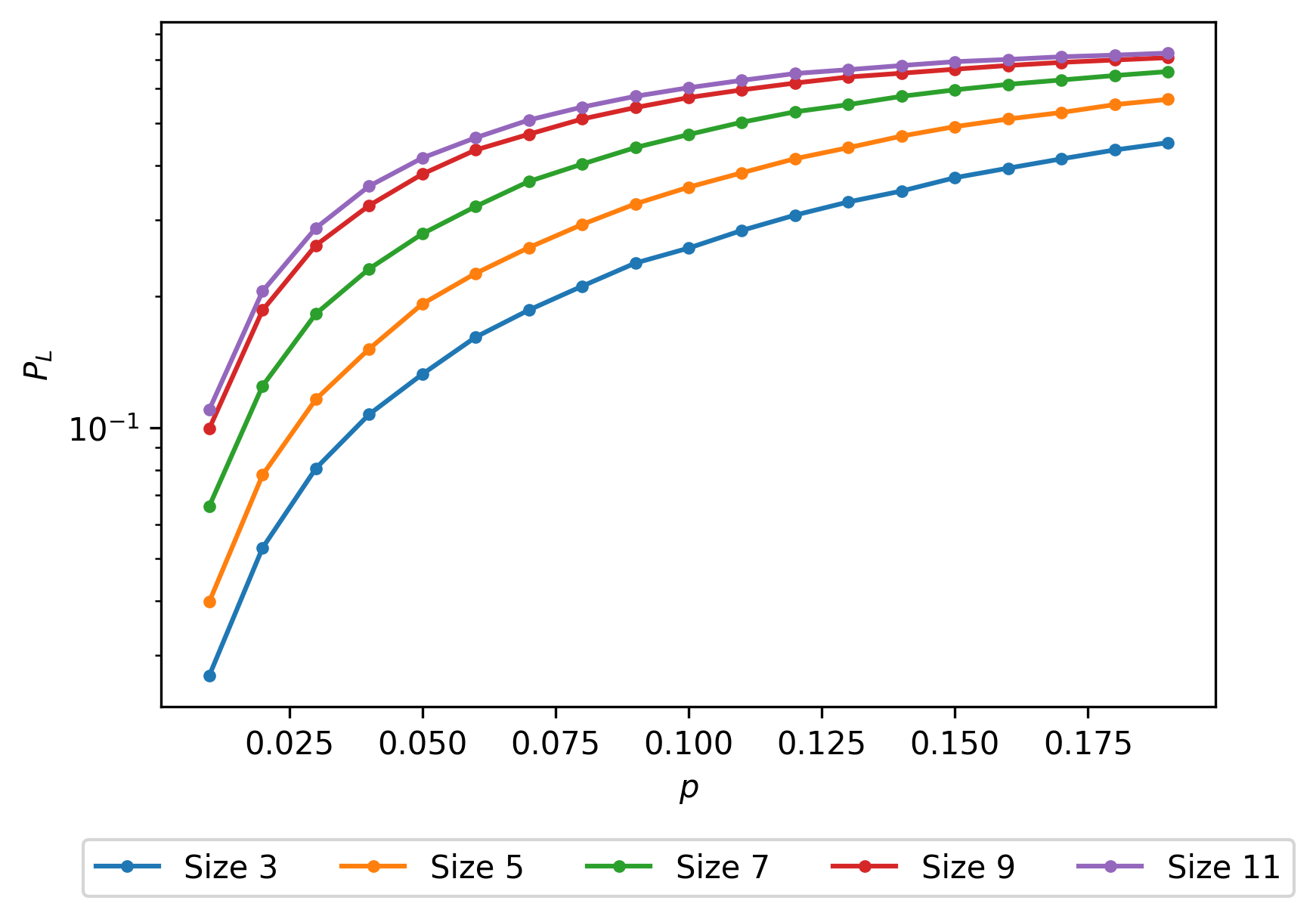}
    \caption{Logical error probability with dependence on the physical error probability under depolarizing noise with BP decoding.}
    \label{bpdepo}
\end{figure}

 {The design of good quantum LDPC codes is only half the problem, however, as decoding quantum codes with the classical BP algorithm leads to performance that is far from what would be expected based on the classical literature. Previously in this work we highlighted the differences between optimal decoding for degenerate quantum codes: DQMLD and non-degenerate quantum codes: QMLD. Applying the QMLD rule to a degenerate quantum code is suboptimal, as the operator with highest probability need not belong to the coset with highest probability. Quantum codes, especially quantum LDPC codes, are highly degenerate, so a performance penalty is already incurred when settling for the QMLD rule.} This performance difference is further aggravated if a BP decoder is chosen to solve QMLD\footnote{Recall that BP is an approximation of QMLD that relies on marginalization: it optimizes the error probability individually for each qubit rather than jointly as is called for by QMLD.}. This is due to the fact that degenerate quantum codes can simultaneously exhibit an error probability that is sharply peaked over a logical coset (which would lead to great performance under DQMLD) and that has a broad marginal distribution over individual qubits, which given the large number of low and similarly weighted operators of such codes, would severely hinder a BP decoder. The existence of the aforementioned operators and their equivalence under BP decoding lead to the peculiar \textit{symmetric degeneracy error} phenomenon \cite{degen3}, which is also referred to as \textit{split-belief} \cite{osd} or \textit{Quantum Trapping Set} (QTS) \cite{raven1}. Quite frankly, if a  code is degenerate enough (if the weight of its generators is substantially smaller than its distance), split beliefs can put the proverbial nail-in-the-coffin for a BP decoder in this context. This cannot be better exemplified than by the fact that the  {rotated} code exhibits no threshold under BP decoding \cite{osd}. {Rotated }codes are degenerate by nature, becoming even more so as their size is increased (the weight of their stabilizer generators remains the same but the distance grows). This makes the split-belief phenomenon more prevalent for larger {planar } codes, explaining the absence of a threshold for this family of codes under BP decoding, as can be seen in FIG. \ref{bpdepo}.

Split-beliefs have been analyzed in the literature \cite{degen3, raven1} and their impact has been successfully alleviated through myriads of post-processing routines added to general BP decoding. The most successful of these modified BP-based decoding techniques is known as BP-OSD \cite{qldpc2}. The performance increments this strategy provides have made it the front-runner in the conversation of a general purpose decoder for QLDPC codes. In fact, the {rotated} code actually exhibits a threshold when decoded via this more sophisticated algorithm. In consequence, this begs the question of whether BP-OSD can compete with the other decoding strategies for the planar code that we have seen thus far.

\subsubsection{Enhanced Belief Propagation: Quantum Ordered Statistics Decoding}

The post-processing algorithm known as Ordered Statistics Decoding (OSD) \cite{osdclassic1, osdclassic2} was originally designed to improve the performance of small classical codes, as well as to lower the error floors of certain LDPC codes. It was later adapted to the quantum paradigm by Panteleev and Kalachev in \cite{qldpc2}, where the authors successfully devised the so called qOSD routine via specific modifications to the classical OSD algorithm. In a similar fashion to other post-processing routines for sparse quantum codes, qOSD only works in conjunction with a BP decoder; i.e., it requires the {marginal probability} outputs of a BP decoder\footnote{Or any other decoder capable of yielding soft values as its output. Soft in this context means that the hard decisions on the most probable error have not been taken so essentially it is a probability distribution.} in order to function. For this reason, the decoding routine that combines both BP and qOSD is generally referred to as BP-OSD. In later work, BP-OSD was shown to work well for Toric codes and a novel class of semi-topological codes \cite{osd}, and recently, it has also been shown to be a valid decoding strategy for bias-tailored QLDPC codes \cite{bias}.

As is done in \cite{osd}, for the sake of notational simplicity, we describe OSD post-processing as applied to the classical decoding problem described by  {\begin{equation}\label{eq:synd}
\bar{s} = H\cdot \bar{e} .
\end{equation}}

 { As has previously seen, it} is obvious that this framework is equally applicable (albeit with minor programming modifications) to decoding the $X$ and $Z$ components of a CSS code.

The OSD post-processing algorithm is called upon whenever BP fails to produce an estimate $\hat{e}$ that matches the measured syndrome. Hence, our starting point is the following: after attempting to decode the measured syndrome  {$\bar{s}$} via BP we end up with the incorrect estimate of the error  {$\bar{s}$}. The estimate is known to be erroneous since its syndrome does not match the true syndrome. However, the fact that  {$\bar{s} \neq H\cdot \hat{e}$, where $H$ is the parity check matrix} does not imply that all components of $\hat{e}$ are wrong, a rationale that OSD will exploit to find a valid solution to \eqref{eq:synd}.

We begin by introducing a necessary set of concepts. We will call the set of indices $[I]$ for which  {$H\cdot \bar{e}_I = H\cdot \hat{e}_I$} the most reliable information set. The complement of this set $[\bar{I}]$, all those indices for which  {$H\cdot \bar{e}_{\bar{I}} \neq H\cdot \hat{e}_{\bar{I}}$}, will thus be referred to as the least reliable information set. As we explain in what follows, OSD post-processing exploits the concept of reliable information sets to solve the linear system described by \eqref{eq:synd}. The parity check matrix $H$ of a quantum error correction code is a rectangular matrix that does not have full column-rank, making it impossible to solve  {$\bar{s} = H\cdot \bar{e}$} via matrix inversion. However, the system described by an appropriate set of $n-k$ linearly independent columns $[S]$ of $H$ is actually solvable: 
 {
\begin{equation} \label{eq:synd_mod}
\bar{s} = H_{[S]}\hat{e}_{[S]},
\end{equation}}

can be solved as
 {
\begin{equation} \label{eq:synd_mod_sol}
H^{-1}_{[S]}\bar{s} = \hat{e}_{[S]},
\end{equation}}
(note that $H_{[S]}$ is a full-rank matrix). For every choice of $[S]$, the basis of linearly independent columns, we will obtain a unique solution  {$\hat{e}_{[S]}$}, that satisfies \eqref{eq:synd_mod}. Against this backdrop, we can understand what OSD post-processing is about: if a full-column rank subset of the parity check matrix can be found and used to solve \eqref{eq:synd_mod} via matrix inversion, we will always find a solution  {$\hat{e}_{[S]}$} that produces a matching syndrome  {$\bar{s}$}. Additionally, because this solution is unique, any symmetries that might hinder BP decoding (like those that cause split beliefs) are now broken.

Naturally, the next question becomes how do we pick the basis $[S]$ in a way that guarantees that this solution is actually `good', i.e., that it is the lowest weight operator associated to the measured syndrome. It is at this point that the previously introduced concept of reliable sets is applied. More precisely, we can use the soft-values (a posteriori llrs) produced in the final BP decoding iteration to rank the bits from most likely to least likely of being flipped (lowest to highest llr values). We can then apply this order to re-arrange the parity check matrix of the code into a new matrix $\Lambda$. It is clear that the basis $[J]$ defined by the columns of the first full column rank submatrix\footnote{This matrix is found by taking the first 
$\text{rank}(H)$ linearly independent columns of $\Lambda$.} $\Lambda_{[S]}$ of the rearranged matrix is the least reliable basis, as it is obtained from the indices of the linearly independent columns associated to the least reliable set of bits. By picking the OSD submatrix in this way, we are guaranteed to find a low weight solution to the syndrome equation. At this point, instances of the OSD algorithm with varying degrees of complexity, denoted as order-$w$ OSD or OSD-$w$, where $w \in [0,\ldots, K]$ and $K \in \mathbb{N}$, can be applied to search for the lowest weight solution to \eqref{eq:synd}. In what follows we detail the functioning of OSD post-processing and the differences between the lowest order version of OSD, OSD-$0$, and higher order intances, OSD-$w$.

\textbf{OSD-0}: Assume the following, after measuring a given syndrome  {$\bar{s}$} for a specific code with parity check matrix $H$, we have decoded via BP and obtained an estimate of the error $\hat{e}$, which unfortunately does not produce a matching syndrome:  {$\bar{s} \neq H\cdot \hat{e}$}. In this context, we would execute OSD, which would operate as follows:

\begin{enumerate}

\item Take the soft-outputs\footnote{The a-posteriori llrs  {are} estimated in the final decoding iteration.} of the BP decoder given by  {$l_{i,\text{ap}} = \frac{P(\hat{e}_i=0|\bar{s})}{P(\hat{e}_i=1|\bar{s})},\ \forall i \in {0,\ldots,n}$}, and order them from most-likely to least-likely to have been flipped (increasing order of magnitude). Store the list of bit-indices $[\text{OG}]$, as this defines the least reliable information set of bits.

\item Re-arrange the columns of the parity check matrix of the code according to the ranking defined by $[\text{OG}]$. We will denote this new matrix by $\Lambda$.

\item Select the first $n-k=\text{rank}(H)$ linearly independent columns of $\Lambda$ to obtain the submatrix $\Lambda_{[J]}$. The list of indices associated to these columns defines the least reliable information set of bits $[J]$. Note that these columns must be linearly independent, else $\Lambda_{[J]}$ will not have full-column rank.

\item Invert $\Lambda_{[J]}$ into $\Lambda^{-1}_{[J]}$. Calculate the solution  {$\hat{e}_{[J]}$} to the OSD syndrome equation as  {$\Lambda^{-1}_{[J]}\bar{s} = \hat{e}_{[J]}$}. 

%Refer to the appendix for more details on how this actually works.

\item The complete solution to the decoding problem is  {$\hat{e}_{\Lambda_0} = [\hat{e}_{[J]}, \hat{e}_{[\bar{J}]}]$}, where $J$ and $\bar{J}$ denote the least and most reliable information set of bits, respectively. Knowing that  {$\hat{e}_{[J]}$} satisfies  {$\Lambda_{[J]}\hat{e}_{[J]} = \bar{s}$} then it is easy to see that  {$\Lambda \hat{e}_{\Lambda_0} = \Lambda [\hat{e}_{[J]}, \mathbf{0}] = \bar{s}$}.

\item The last step is to take the solution $\mathbf{e}_{\Lambda_0} = [e_{[J]}, \mathbf{0}]$, and map it to the original bit-index order:  {$\hat{e}_{\Lambda_0} \rightarrow \hat{e}_\text{OSD-0}$}. We call the rearranged vector, $e_\text{OSD-0}$, the OSD-0 solution. 

\end{enumerate}

\begin{figure}
\centering
    \includegraphics[width = 0.25\textwidth]{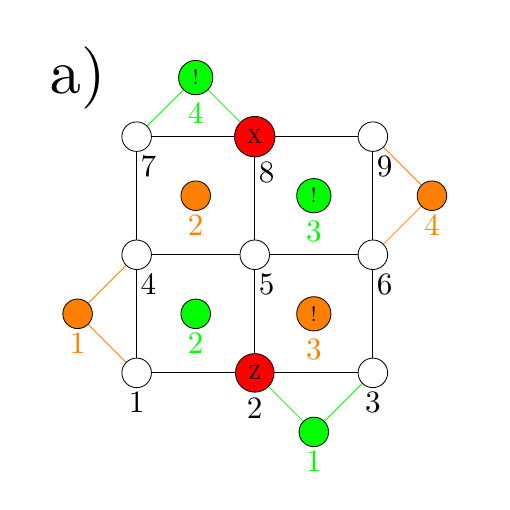} \\
    \centering
    \includegraphics[width = 0.3\textwidth]{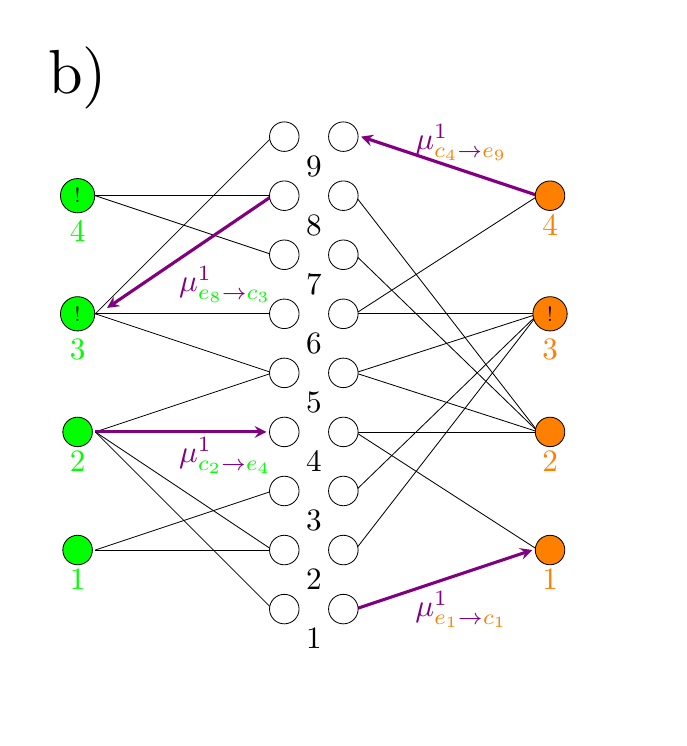} \\
    \centering
    \includegraphics[width = 0.2\textwidth]{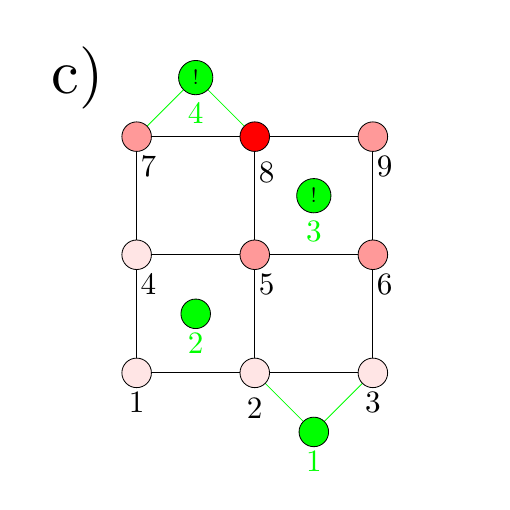} 
    \includegraphics[width = 0.2\textwidth]{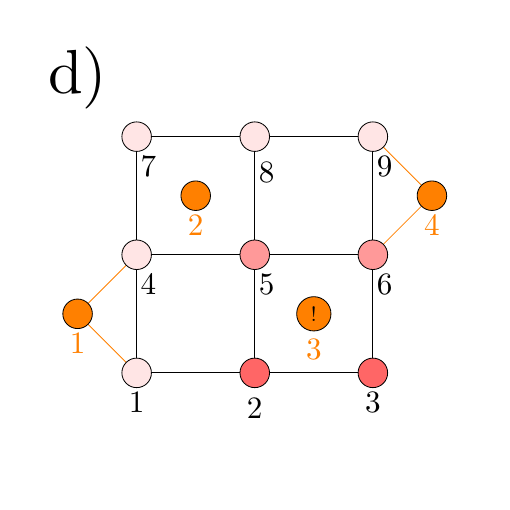} \\
    \centering
    \includegraphics[width = 0.3\textwidth]{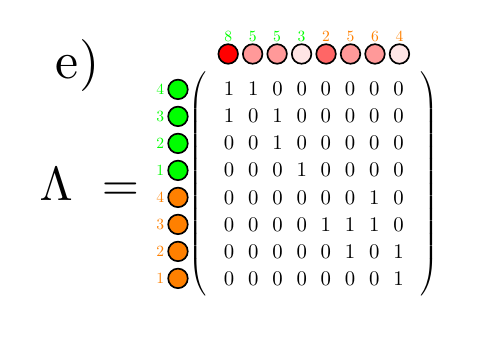} \\
    \centering
    \includegraphics[width = 0.25\textwidth]{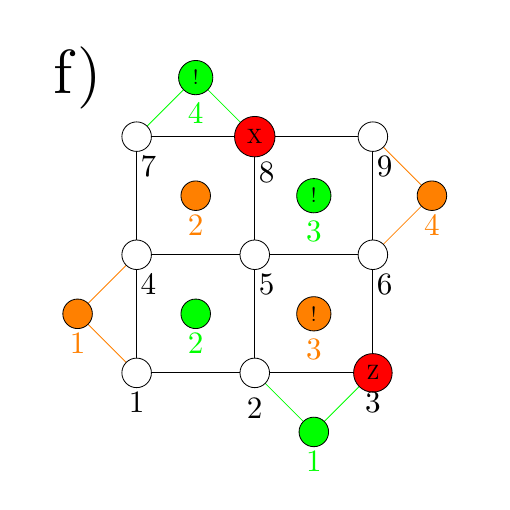}
    \caption{Graphical description of a BPOSD-0 process for a specific syndrome in a  {distance-3 rotated} code  {in a)}.  {The process begins by running BP over the Tanner graph in b) resulting in the marginal probabilities in c) and d) (lower llr implies redder color). The parity check matrix is ordered according to the llrs. The invertible matrix associated to the least reliable set is selected in e). Gaussian elimination is done with the obtained $\Lambda$ matrix to solve the syndrome problem, leading to a recovered error matching the syndrome in f). } }
    \label{BPOSD0 fig}
\end{figure}

In FIG. \ref{BPOSD0 fig}, an example is provided where a specific error produces a syndrome which is later decoded finally obtaining a recovered error.  {In FIG. \ref{BPOSD0 fig} a)}, the error and the syndrome within the  {rotated} code are introduced, data qubits, $X$ and $Z$-checks are labelled in black, green and orange colors respectively.  {In FIG. \ref{BPOSD0 fig} b)}, one can see a graphical representation of the BP method, where messages are sent from data qubits to checks and vice versa. Notice, how there are two white circles for each data qubit label, that is because the two BP graphs are independent, and so each data qubit will have two resulting marginal probabilities, one for the $X$ operators and another one for the $Z$ ones.  {In FIG. \ref{BPOSD0 fig} c) and d)}, the resulting marginal probabilities are represented within the data qubits of the  {rotated} code with $X$-check qubits and $Z$-check ones. { Data qubit being redder indicate its marginal probability of recovering a non-trivial operator being larger, and thus its llr being lower, and whiter otherwise.} {In FIG. \ref{BPOSD0 fig} e)}, the matrix $\Lambda$ is represented. The chosen data qubits are the ones which represent the independent columns with the higher marginal probabilities, the four columns on the left are extracted from the $X$-check subgraph and the remaining four from the $Z$-check subgraph, as indicated by the labels on top of the matrix. This matrix is inverted following the earlier described process, reaching a recovered error which is depicted   {In FIG. \ref{BPOSD0 fig} f)}. Notice how the error is not the same set of Pauli operators as the inputted one, nevertheless, it is within the same stabilising group, and thus, successfully corrects the error. 

\textbf{OSD-$\mathbf{w}$:} Higher order OSD is similar to OSD-$0$, the difference being that we now consider solutions  {$\hat{e}_{\Lambda_w} = [\hat{e}_{[J]}, \hat{e}_{[\bar{J}]}]$} where  {$\hat{e}_{[\bar{J}]} \neq \mathbf{0}$}. OSD-$w$ begins by running the first five steps of OSD-$0$ and computing  {$\hat{e}_{[J]}$}, which for the sake of notation we will now denote by  {$\hat{e}^{w=0}_{[J]}$}. Once this is done, different candidates  {$\mathbf{\hat{e}}_k$} can be found by making choices for  {$\hat{e}_{[\bar{J}]}$} and solving
 {
\begin{equation} \label{eq:osdw-sol}
    \hat{e}_{\Lambda_w} = [\hat{e}_{[J]}, \hat{e}_{[\bar{J}]}] = [\hat{e}^{w=0}_{[J]} + \Lambda^{-1}_{[J]}\Lambda_{[\bar{J}]}\hat{e}_{[\bar{J}]}, \ \hat{e}_{[\bar{J}]}],
\end{equation}}

where $\Lambda_{[\bar{J}]}$ is the submatrix obtained by taking the columns of $\Lambda$ indexed by $\bar{J}$. The solution  {$\hat{e}_w$} given in \eqref{eq:osdw-sol} satisfies the OSD syndrome equation  {\begin{equation}
    \label{eq:osdw}
 \Lambda \hat{e}_{\Lambda_w} = \bar{s}\end{equation}}
 
 for any choice of  {$\hat{e}_{[\bar{J}]}$}. The premise behind OSD-$w$ (considering  {$\hat{e}_{[\bar{J}]} \neq \mathbf{0}$}) is to find solutions  {$\mathbf{e}_{\Lambda_w}$} of lower weight than  {$\hat{e}_{\Lambda_0}$}. The  {$\hat{e}_{[\bar{J}]} \neq \mathbf{0}$} component has dimension $n-\text{rank}(H)$, implying that testing all possible configurations for  {$\hat{e}_{[\bar{J}]}$} is intractable beyond a small value of $n$. For this reason, it is important to design a strategy that makes good choices for the  {$\hat{e}_{[\bar{J}]}$} candidates. This is best approached\footnote{A different strategy was employed in the works that first introduced qOSD. However, slight performance improvements were shown in \cite{osd} when using the combination sweep strategy.} using the so called \textit{combination sweep strategy}. This greedy search method works as follows:

\begin{enumerate}

\item At the start of the search, the OSD-$w$ solution is equal to the unordered OSD-$0$ solution:  {$\hat{e}_{\Lambda_w} = \hat{e}_{\Lambda_0}$}.

\item Sort the bits in the  {$\hat{e}_{[\bar{J}]}$} subvector of the solution according to the soft-outputs of the original BP decoding attempt. Note that this step is already built into OSD-0 when re-arranging the parity check matrix according to the BP outputs. 

\item Test all possible weight-$1$ configurations of  {$\hat{e}_{[\bar{J}]}$}. There are a total of $n - \text{rank}(H)$ weight-$1$ candidates. If the weight of any of the candidates is lower than the weight of  {$\hat{e}_{\Lambda_0}$}, update the choice of  {$\hat{e}_{\Lambda_w}$}.

\item Try all possible weight-$2$ configurations in the first $w$ bits of  {$\hat{e}_{[\bar{J}]}$}. Obviously, the order\footnote{It is worth mentioning that in \cite{osd}, the order $w$ of the OSD algorithm is referred to as the search depth.} $w$ will be upper bounded by $n-\text{rank}(H)$ the dimension of  {$\hat{e}_{[\bar{J}]}$}. The number of candidates is given by the binomial coefficient $\binom{w}{2}$. If the weight of any of the candidates is lower than the weight of the current choice for the solution, update the choice of  {$\hat{e}_{\Lambda_w}$}.

\item The final step is analogous to that of the order $0$ version: map the solution to the original bit-index order accordingly:  {$\hat{e}_{\Lambda_w} \rightarrow \hat{e}_\text{OSD-w}$}.

\end{enumerate}

OSD-$w$ entails testing a total of $n-\text{rank}(H) + \binom{w}{2}$ candidates for  {$\hat{e}_{\Lambda_w}$}, out of which the minimum Hamming weight solution, or at least a better choice than  {$\hat{e}_{\Lambda_0}$}, will have been found. As the order $w$ of the combination sweep strategy is increased, the likelihood of finding the solution of minimum Hamming weight to \eqref{eq:osdw} increases. At the same time, this also implies additional computational demand, negatively impacting the complexity of the algorithm and its runtime performance.

\subsubsection{OSD Complexity}

 {The complexity of the BP process is $\mathcal{O}(nj)$ \cite{BP_comp1, BP_comp2}, where $n$ indicates the number of physical qubits and $j$ is the average number of non-trivial elements per column in the parity check matrix. Moreover,} in \cite{qldpc2}, the authors show that most of the performance improvements provided by qOSD postprocessing are achieved with OSD-0. While increasing the order $w$ of the algorithm yields benefit over setting $w=0$, for various QLDPC code families the improvement provided by running the algorithm with $w>0$ is marginal. It is important to mention that in \cite{qldpc2}, the authors use a different algorithm to the combination sweep strategy \cite{osd}. Instead, they apply an exhaustive approach where they test all possible permutations in the first $w$ bits of $e_{[\bar{J}]}$. Given that OSD-$0$ requires the solving of a linear system, its complexity will be at most $\mathcal{O}(n^3)$ (although there are matrix inversion algorithms with better complexity, they are quite impractical). The exhaustive approach to OSD-$w$ has a complexity in the general case of $\mathcal{O}(n^3 + n2^w)$. Although the combination sweep also has an edge in terms of complexity ($\approx \mathcal{O}(w^\alpha n^3)$), it is likely that OSD-$0$ is the only version of the algorithm that can be successfully implemented in a real time system \cite{osd, raven3}. Recently, an OSD-inspired reduced complexity approach known as \textit{stabilizer inactivation} has been proposed. In \cite{inactivation} the authors showed that this strategy, which has $\mathcal{O}(n^2\text{log}n)$ in the worst case, achieved a higher threshold for the family of generalized bicycle codes.

% la ultima complejidad?
% BELIEF MATCHING!

\subsubsection{Performance and threshold}

\begin{figure}
    \centering
    \includegraphics[width = 1\columnwidth]{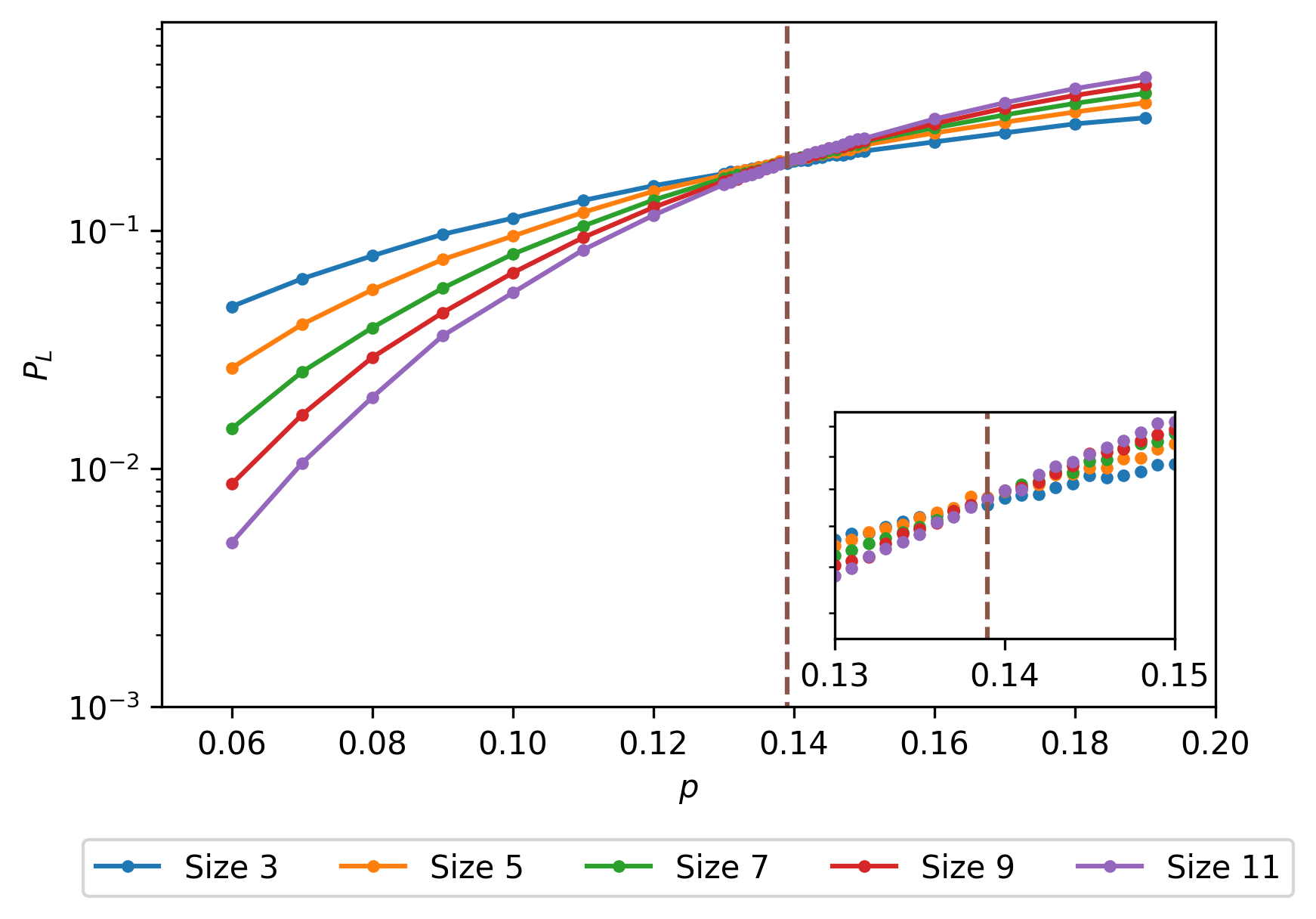}
    \caption{Logical error probability with dependence on the physical error probability under depolarizing noise with BPOSD-0 decoding.}
    \label{bposd-0}
\end{figure}

As has been previously seen in FIG. \ref{bpdepo}, the BP decoding method has no probability threshold by itself and so is not a usable decoding method for the decoding of the surface code. Nevertheless, considering BPOSD-0 yields an enhancement which produces a probability threshold of $0.139$ under depolarizing data qubit noise. This result can be seen in FIG.  \ref{bposd-0}. This threshold can be further expanded when considering higher BPOSD orders at the expense of a higher complexity. Moreover, the BPOSD method is also compromised when considering $Z$-noise bias, which can be explained by the $Z$-check subgraph becoming more dense than the $X$-check one and thus being more prone to failure. Table \ref{tablabposd} reviews several thresholds for three different ordered statistics decoding processes under different biases.

% \begin{table}
% \begin{center}
% \begin{tabular}{|c|c|c|c|}
%   \hline
%   \multicolumn{1}{|c|}{$\eta$} & \multicolumn{3}{c|}{$p_{th}O(10^{-3})$} \\
%   \cline{2-4}
%   \multicolumn{1}{|c|}{} & BPOSD-0 & BPOSD-2 & BPOSD-4 \\
%   \hline
%   1/2 & .139 & Data 2 & Data 3 \\
%   \hline
%   1 & Data 4 & Data 5 & Data 6 \\
%   \hline
%   10 & Data 1 & Data 2 & Data 3 \\
%   \hline
%   100 & Data 4 & Data 5 & Data 6 \\
%   \hline
%   1000 & Data 4 & Data 5 & Data 6 \\
%   \hline
% \end{tabular}
% \caption{Probability threshold values for different biases and orders in the rotated planar code under the BPOSD decoding scheme.}
% \label{tablabposd}
% \end{center}
% \end{table}

\begin{table}
\begin{center}
\begin{tabular}{|c | c|} 
 \hline
 $\eta$ &  $p_{th}$ \\ [0.5ex] 
 \hline\hline
 1/2 &  0.139 \\ 
 \hline
 1 &  0.138 \\
 \hline
 10 &  0.098 \\
 \hline
 100 &  0.094 \\
 \hline
 1000 & 0.092  \\ [1ex] 
 \hline
\end{tabular}
\caption{Probability threshold values for different biases and orders in the rotated planar code under the BPOSD-0 decoding scheme.}
\label{tablabposd}
\end{center}
\end{table}

\subsubsection{Measurement errors}
% belief matching
% measurement errors --> soft syndrome information
% single-shot decoders
% BPOSD for circuit level noise

There have been many proposals for handling circuit-level noise by using belief propagation. An important proposal for handling this type of errors in the surface code is the decoder deemed as belief-matching \cite{fragile}. This decoder is base on a combination between the BP and MWPM algorithms, idea that was previously explored in \cite{Crigerbeliefmatching} with perfect syndrome measurements. The original belief-matching algorithm achieved a $17.76\%$ threshold for the rotated planar code when depolarizing noise was considered over the data qubits. The decoding complexity of the algorithm was maintained when compared to MWPM when parallel processing was allowed \cite{Crigerbeliefmatching}.  {Given the impressive performance of this BP+MWPM approach, it was generalized to handle noisy gates and SPAM errors in \cite{fragile}, resulting in an excellent decoding strategy for such planar codes}. The basic idea behind the belief-matching algorithm is similar to the BPOSD decoder in the sense that the soft ouptuts of the sum-product algorithm are used for the posterior decoding algorithm to make use of them. In this sense, the weights of the the graph in the MWPM decoder are produced by using such a posteriori information. Regarding the generalization of the belief-matching for circuit-level noise, the authors discussed how to construct the circuit-level Tanner graph that takes the measurement circuits of the surface code into account so that belief propagation can be run over such graph and then use the soft output for obtaining the MWPM solution \cite{fragile}. Note that $\mathcal{O}(d)$ measurement rounds are required for this. The authors found a threshold of $0.94\%$ for the belief-matching algorithms with circuit level noise, which is comparable to the belief-find decoder also proposed in such article, which follows the same procedure but using UF instead of MWPM and achieves a threshold of $0.937\%$ \cite{fragile}. The authors discuss that their performances are similar due to the fact that most of the information needed for decoding is provided by the BP part of the algorithm. 

Another approach to deal with noisy syndromes is to consider soft syndrome information for the BP graph used for decoding as it was done for QLDPC codes in \cite{softRav}. In such work, the authors discuss the fact that when a measurement is noisy, the input information of the Tanner graph regarding the information of the syndrome can be soft, i.e. a probability distribution conditional to the obtained noisy measurement. Therefore, the fact that the measurement outcome might not be precise is also fed to the BP algorithm. However, this study only considered the fact that the noisy measurements are a result of SPAM errors, neglecting circuit level noise. The authors discuss that this is considered to be future work in this direction. Interestingly, such paper was based on the previous work by Pattison et al. that discussed the use of soft measurement information in the context of MWPM and UF for surface codes \cite{softmwpmuf}. The authors concluded that their modified decoders using soft information improved the threshold obtained by hard decision decoders for the circuit-level noise considered. In this sense, considering this approaches for BP decoders discussed for the surface code might be an interesting path to follow for dealing with circuit level noise. 

 {Moreover, for the latest articles of \cite{bravyildpc, QuEra}, an approach was taken where, via the software package \cite{stim}, one can compute the parity check matrix of the circuit-level noise, which contains all single error mechanisms in the syndrome extraction circuit as columns. Once this parity check matrix is obtained, given a syndrome, one can return an error applying BP+OSD. Applying BPOSD to circuit-level noise yields very good results although it is under constraint of considering a much larger parity check matrix to be inverted. Refer to Appendix \ref{appA} for a more detailed description of the multiple measurement-round measurement extraction and decoding.}

Finally, single-shot decoding using BPOSD was investigated for the 4D toric code in \cite{singleshot}. Single-shot decoding refers to estimating the error when noisy measurements are present in a single measurement round, i.e. without needing to measure the usual $\mathcal{O}(d)$ rounds. In such article, the authors propose to decode data qubit errors and noisy syndrome measurements altogether in a single stage via the BPOSD decoder. The authors discuss that they obtain better thresholds than using multiple-measurements and other single-shot decoder such as the cellular automaton decoder. However, they use a phenomenological noise model for faulty syndrome measurements as discussed before, i.e. they do not consider full circuit-level noise, which they consider to be future work \cite{singleshot}. Single-shot decoding is being actively investigated at the time of writing, specially from the point of view of QLDPC codes \cite{singleBomb,singleBreuck}. This approach was originally proposed for some families of topological codes, and due to its advantage in terms of measurement rounds and, presumably, performance, seems to be interesting to study more codes that may admit this kind of decoding to deal with circuit-level noise.

\subsection{The Tensor Network decoder}

Tensor Network (TN) decoders are decoding methods, proposed by Bravyi et al. for the  {planar} code, that aim to resolve the DQMLD problem.
% i.e. to estimate which is the most probable logical coset based on the syndrome information \cite{mps}. 
% Note that, up to this point, all decoders considered have followed a QMLD logic, that is, their aim has been to seek for the most probable Pauli error (QMLD) since, as has been seen earlier, the error recovery problem contemplating error logical classes (DQMLD) belongs to the $\#P$ complexity class \cite{PcompleteDec}. 
For this reason, the TN decoder is also usually referred as the maximum likelihood decoder (MLD)\footnote{Note that this nomenclature might be somewhat confusing as it does not explicitly represent the fact that it is a degenerated decoder.}\cite{mps}. Fortunately, surface codes admit a natural representation in terms of Tensor Networks (TN) \cite{TN}, feature that can be used for approximating the DQMLD problem targeting the most probable error coset. 
Decoding quantum error correcting codes with TNs was first considered in \cite{andy}
% , where the authors described for the first time the equivalence between decoding a quantum code and contracting a TN for quantum turbo, polar and branching multiscale entanglement renormalization ansatz (MERA) codes \cite{branchingmera,mera}
. Later, the TN decoding was particularized for the planar code in \cite{mps}, and, afterwards, in \cite{tuckett2}, the method was expanded to the rotated planar code yielding significant results. Lastly, in \cite{chubb2}, the method was generalized to any quantum 2D code. 

In broad terms, the idea behind TN decoding consists in recovering an error $E_{rec}$ that shares the observed error syndrome in a quick manner so that when combining it with the actual channel error the resulting element lays in the normalizer of the code. Once this is done, the probability of the logical error cosets of the normalizer are computed to select the most probable one. For the specific case of codes encoding a single logical qubit (such as the rotated planar code), there will be four logical cosets: $I_L, X_L, Y_L$ and $Z_L$. It is worth mentioning that $E_{rec}$ does not need to be the most probable error since its unique purpose is to transport the error to the zero syndrome coset \cite{logicalsparse}. After $p(E_{rec}I_L), p(E_{rec}X_L), p(E_{rec}Y_L)$ and $p(E_{rec}Z_L)$ are computed, the most probable error class, $\mathcal{L}$, determines the recovery operation $\hat{E}=E_{rec}\mathcal{L}$.

\subsubsection{Introduction to TNs}

In order to better understand the functioning of the TN decoder, we will first briefly introduce some generic terms about tensor networks. TNs are a class of variational wave functions used in the study of many-body quantum systems in which the quantum relation from tensor to tensor is known \cite{TN}. TNs are used in many research fields for studying complex and correlated systems with large configuration spaces. How they work is based on the fact that there are configurations much more significant than others, and so, given a certain threshold, one can omit the least reliable configurations in order to work with a reduced and manageable subspace.

\begin{figure}
    \centering
    \includegraphics[width = \columnwidth]{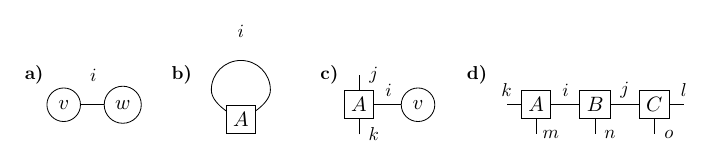}
    \caption{Some examples of tensor operations. \textbf{a)} indicates a scalar product of two vectors  {as shown in equation \eqref{TNeq1}}, \textbf{b)} indicates the trace of a matrix  {as shown in equation \eqref{TNeq2}}, \textbf{c)} indicates the product of a vector with a matrix  {as shown in equation \eqref{TNeq3}} and \textbf{d)} indicates a matrix product state  {as shown in equation \eqref{TNeq4}}.   }
    \label{Tnexamp}
\end{figure}

% Siguiente frase sacada de la wikipedia
TNs accept an intuitive notation. FIG. \ref{Tnexamp} depicts some examples of tensors represented through various plots, where vectors are represented as circles, and matrices and higher order tensors, as squares.  {Each plot represents the following expressions from left to right}:

    \begin{align}
        &\sum_iv_iw_i,  \label{TNeq1}\\
        &\operatorname{Tr}(A),\label{TNeq2}\\
        &\sum_iA_{i,j,k}v_i,\label{TNeq3}\\
        &\sum_{i,j}A_{k,m,i}B_{i,n,j}C_{j,o,l}.\label{TNeq4}
    \end{align}

In a TN, some, or all the indices can be contracted according to a specific pattern, as shown in FIG. \ref{Tnexamp}. If all the indices are contracted, the TN results in a scalar, as in the case of the two first examples of FIG. \ref{Tnexamp}. The lines connecting tensors from one to the other represent the indices such that there is a sum over all their possible values, whereas lines that are not connecting to anywhere represent free indices. Those indices that correspond to the original physical Hilbert spaces are called \emph{physical} indices, and the rest (connecting the different tensors with each other) are called \emph{bond} indices, and are responsible for the entanglement in the quantum many-body wavefunction. The range of values of a bond index is called \emph{bond dimension}, following common TN terminology \cite{mps,TN}.

\begin{figure}
    \centering
    \includegraphics[width = \columnwidth]{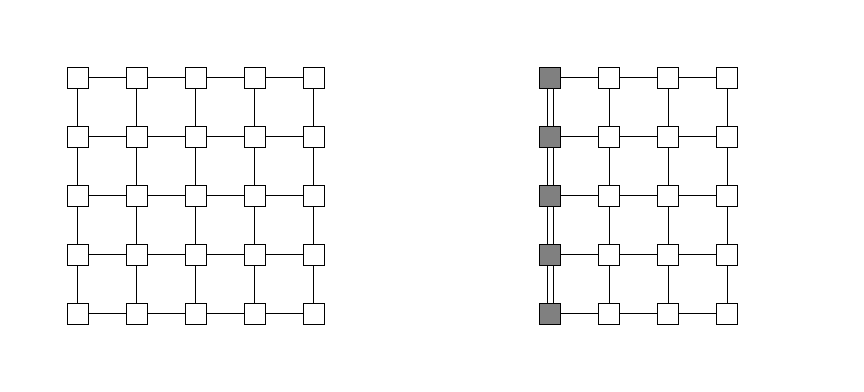}
    \caption{A tensor network which contracts the left most column with the second one.  {When doing the contraction, the vertical bond dimensions are truncated to avoid exponential increase of the complexity.}}
    \label{MPO}
\end{figure}

Consider now the following case: we have a 2D tensor network such as the one showed in FIG. \ref{MPO}, which is the one encountered when computing expectation values of a projected entanglement pair state (PEPS). The constituent tensors on the bulk have bond dimension 4, while the ones on the boundary have lower bond dimensions. One can consider the left-most (or right-most) column as a matrix product state (MPS) and columns within the bulk of the PEPS as matrix product operators (MPO). When contracting the left MPS with its nearest MPO, the overall 2D TN will be reduced by one column as shown in FIG. \ref{MPO}, and for the new column which is obtained, the vertical bond dimension will have increased in an exponential manner. Contracting the entire 2D TN  will most likely be an unfeasible challenge as a result of the exponentially increasing bond dimension. Fortunately, we can still obtain meaningful results for continuous column contraction while avoiding an exponential growth \cite{TN}. For the cases we consider, this is made by using the QR matrix decomposition altogether with the singular value decomposition (SVD), which is used to truncate the vertical bond-dimensions \cite{mps}. Truncating the bond dimension of the column contractions will yield an approximate result to the overall contraction as opposed to an analytical result which, nevertheless, allows for a defined run time since, when the contraction results in a last column, its vertical contraction can be carried at polynomial time \cite{chubb2}.

\subsubsection{TN decoding for the planar code}

\begin{figure}
    \centering
    \includegraphics[width = 1.1\columnwidth]{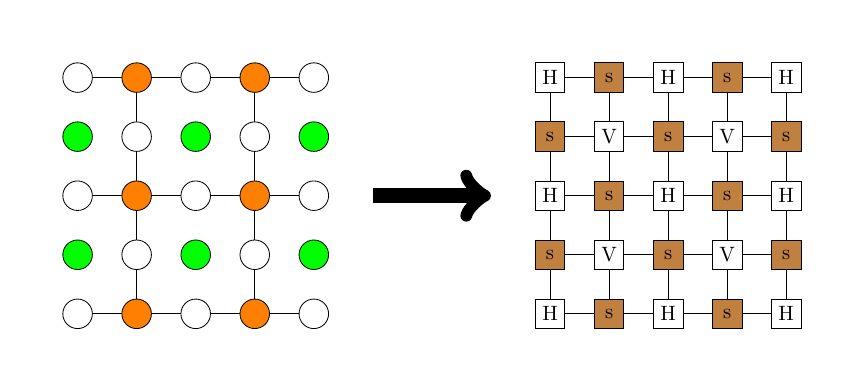}
    \caption{ On the left, a distance-3 planar code, on the right the tensor network which represents it. Stabilizer checks are denoted through brown squares and labelled as $s$, and horizontal and vertical data qubits are denoted as $H$ and $V$ respectively.}
    \label{planartn}
\end{figure}

As mentioned earlier, when facing a non-trivial syndrome in a code, a TN decoder will seek to find a Pauli error correspondent to the syndrome, $E_{rec}$, and the most probable logical coset given the chosen error $E_{rec}$. In this sense, the planar code can be associated to a PEPS tensor network \cite{mps} following:

\begin{equation}
    p(E_{rec}\mathcal{L}) = \sum_{\alpha, \beta} T(\alpha ; \beta),
\end{equation}

where $p(E_{rec}\mathcal{L})$ is the probability of the error being within the error class $\mathcal{L}$ and $\alpha_i, \beta_j \in \{ 0,1\}$ indicate the application of the stabilizer generator operators. An arbitrary application of a stabilizer element operator would be denoted as $g(\alpha, \beta) := \prod_i(A_i)^{\alpha_i}\prod_j(B_j)^{\alpha_j}$, where $\alpha_i$ indicate $X$-stabilizer operators and $\beta_j$ indicate $Z$-stabilizer operators. The resulting tensor network is graphically represented in FIG. \ref{planartn}. Note how now all stabilizer generators are considered as identical while the data qubits are labelled as horizontal (H) or vertical (V). This consideration is equivalent, now, horizontal data qubits are considered to be operated with $Z$-stabilizer operators from top and bottom and $X$-stabilizer operators from left and right, while the opposite happens for vertical data qubits.

\begin{figure}
    \centering
    \includegraphics[width = .3\columnwidth]{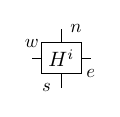}
    \includegraphics[width = .3\columnwidth]{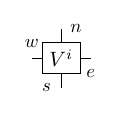}
    \includegraphics[width = .3\columnwidth]{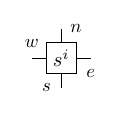}
    \caption{ On the left, graphical representation of the tensor nodes which constitute the planar code tensor network. On the right, element values of the tensors, the superindex indicates the labelling of the tensor node.}
    \label{nodes}
\end{figure}

Under these considerations we can define the tensor nodes which form the overall planar code PEPS tensor network in FIG. \ref{nodes},  {which will be defined by the following expressions corresponding to the tensor figures from top to bottom}:

    \begin{align}
        H^i_{n,e,s,w} &= p(E^i_{rec} + (nZ) + (eX) + (sZ) + (wX)),\\
        V^i_{n,e,s,w} &= p(E^i_{rec} + (nX) + (eZ) + (sX) + (wZ)),\\
        s^i_{n,e,s,w} &= \delta_{n,e,s,w}.
    \end{align}

The elements of the data qubit tensors are the probabilities of experiencing Pauli errors given by the indices. Moreover, the indices $n$, $e$, $s$ and $w$ indicate the tensors to their north, east, south and west; respectively. The indices are binary and indicate if either of the adjacent stabilizer generators operates non-trivially on them as can be seen in the right side of FIG. \ref{nodes}, where one can see that the state of a data qubit tensor is both dependent on the indices and the correspondent Pauli operator on the qubit from the error $E_{rec}$. On the other side, the stabilizer tensor nodes are defined by Kronecker deltas, i.e. they are either 1 when $n = e = s = w$  {or} $0$ otherwise. That is because the stabilizer generators either operate on all their nearest data qubits or do not operate on any of them, as was earlier shown in FIG. \ref{rotatedplanardemo}.

Since the probability of error of data qubits is independent from one to another, considering an error $E_{rec}$ associated to a syndrome and contracting the tensor network from FIG. \ref{planartn} results in the summation of products of tensor probabilities under all the combinations of the stabilizer generators, which itself is equivalent to finding the probability of the error class $p(\mathcal{L}E_{rec})$. Unfortunately, and as explained before, contracting columns increases the vertical bond dimension exponentially and, thus, the reasonable approach consists in truncating the vertical bond dimension after each contraction to a truncated value $\chi$. This is often done through the Schmidt decomposition \cite{mps}, which allows the resulting tensor after the column contraction to be represented as a sum of products of smaller tensors called "Schmidt tensors". The SVD is applied to said Schmidt tensors resulting in the Schmidt values, the truncation follows by only considering the "Schmidt tensors" correspondent to the highest $\chi$ Schmidt values. This allows for obtaining approximate values of the probabilities of the error classes, which can then be used for decoding at a higher precision than any of the aforementioned algorithms, if the truncation of the bond dimension is high enough.
\begin{figure}

\begin{subfigure}[b]{.45\textwidth}
\centering
    \includegraphics[width = 0.45\textwidth]{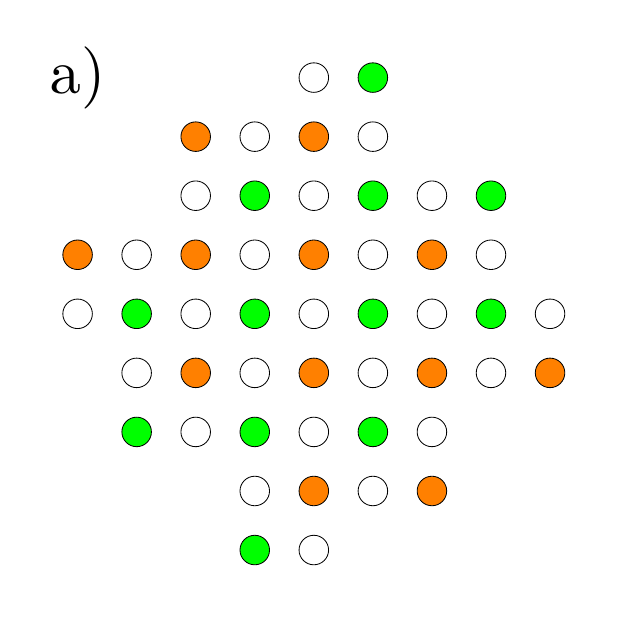}
    \includegraphics[width = 0.45\textwidth]{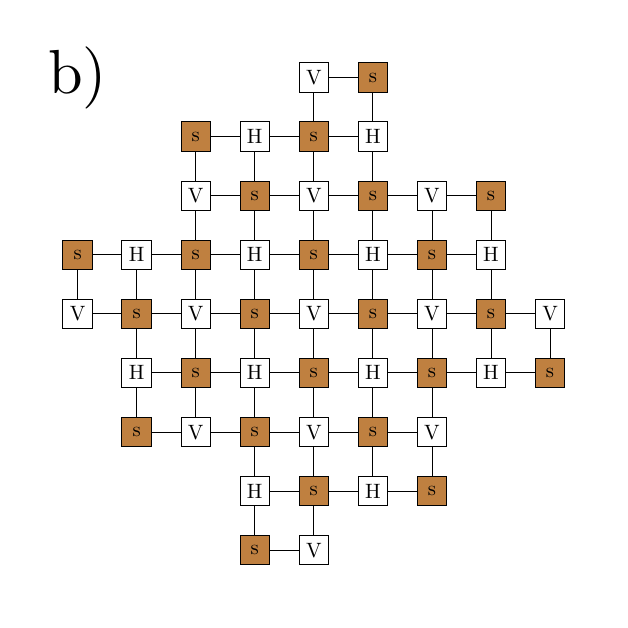} \\
    \includegraphics[width = 0.7\textwidth]{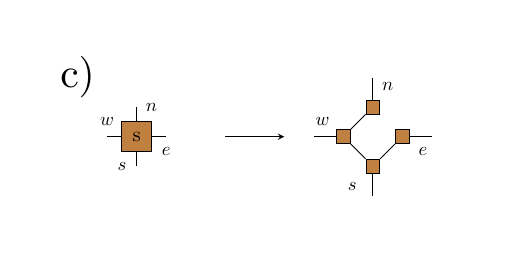} \\
    \includegraphics[width = 0.7\textwidth]{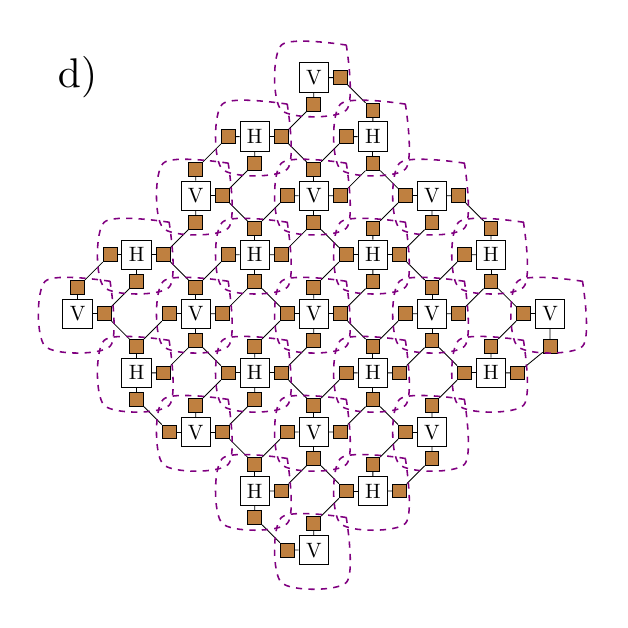} \\
    \includegraphics[width = 0.7\textwidth]{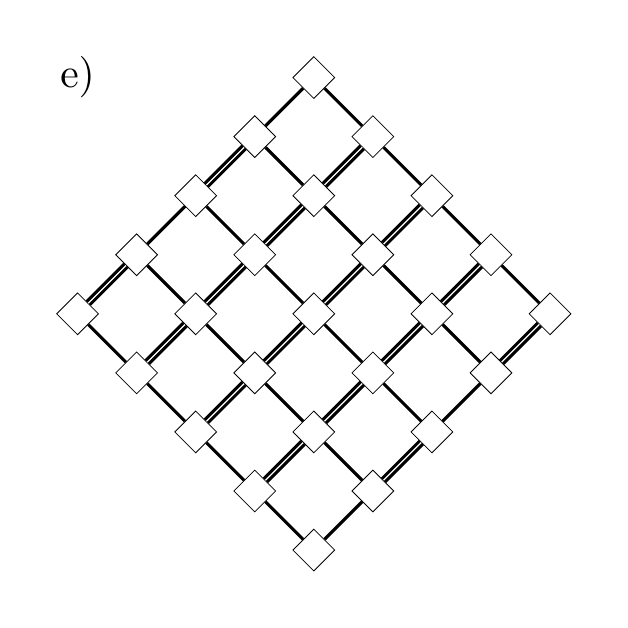}
\end{subfigure}
\caption{In a), a rotated planar code of distance-5. In b) the tensor network one could extract following \cite{mps}. In c), the tensor nodes correspondent to stabilizer generators are split in 4 tensor nodes connected with themselves. In d), a new tensor network arrangement is proposed for the rotated planar code. Notice the tensor network can be separated in sets of tensor nodes which are encircled in dashed violet lines. In e), the aforementioned encircled tensors are contracted and a new tensor network is obtained.}
\label{rpctn}
\end{figure}

\subsubsection{TN decoding for the rotated planar code}

For the rotated planar code, the tensor network representation of the system is not as straight forward as in the case of the planar code, since the code can no longer be mapped directly to a PEPS tensor network. An elaborate and efficient way of adapting the TN decoder in \cite{mps} for rotated planar codes, while greatly improving their performance was elaborated in \cite{tuckett2} as illustrated in FIG. \ref{rpctn}. As depicted in the figure, the rotated planar code is first adapted to the tensor network model proposed in \cite{mps}, which does not correspond to a PEPS state. In order to map such TN to the desired representation, the stabilizer nodes are split in 4 smaller tensor nodes which preserve the delta tensor structure themselves \cite{qecsim}, i.e. the values are given by the Kronecker delta. Afterwards, the data qubit nodes altogether with their adjacent stabilizer nodes are contracted producing the desired PEPS. Note how the resulting PEPS is no longer isotropic, i.e. if the PEPS is rotated 45º counter clock-wise, most of the vertical bonds will be of dimension 2 while the horizontal ones will be of dimension 1.

\begin{figure}

\begin{subfigure}[b]{.45\textwidth}
\centering
    \includegraphics[width = 0.7\textwidth]{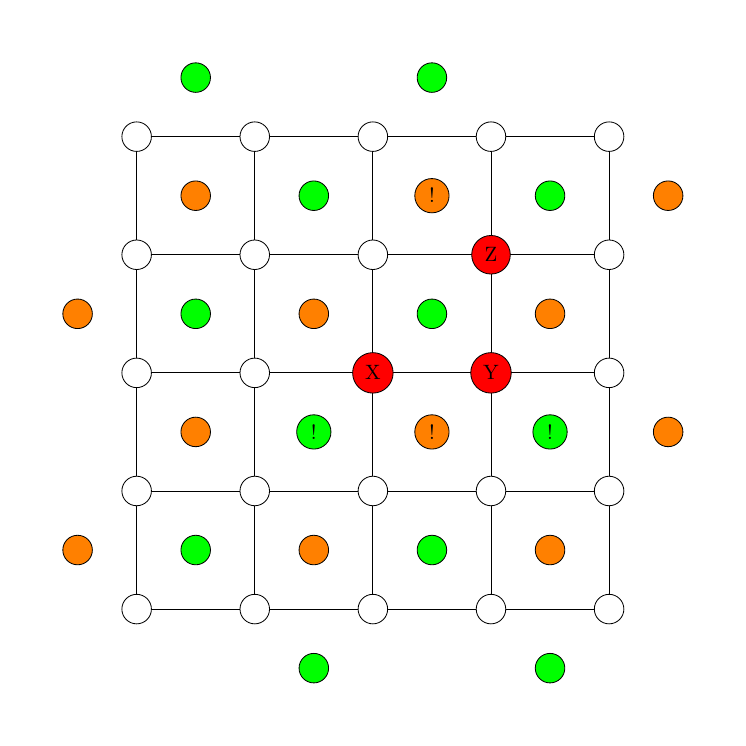} \\
    \includegraphics[width = 0.7\textwidth]{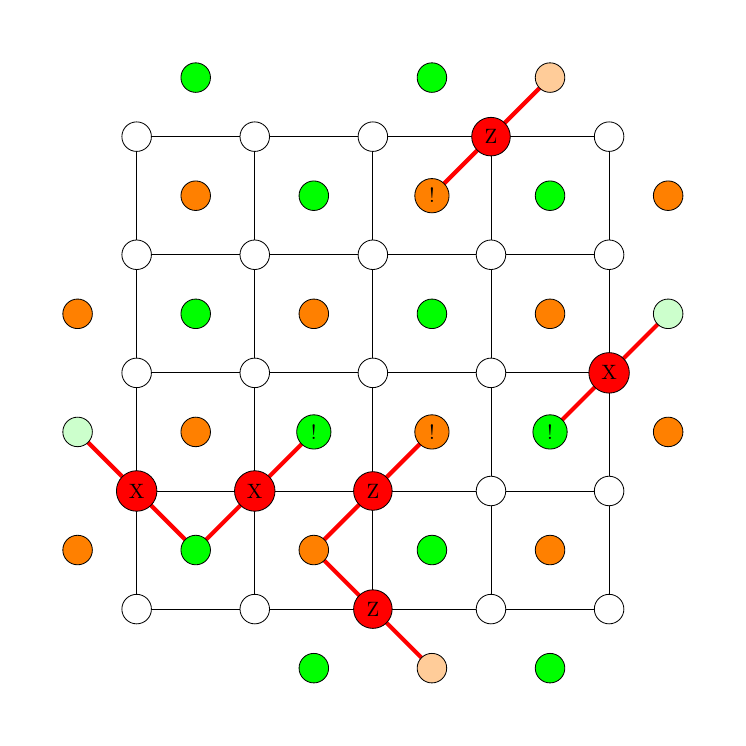} \\
    \includegraphics[width = 0.45\textwidth]{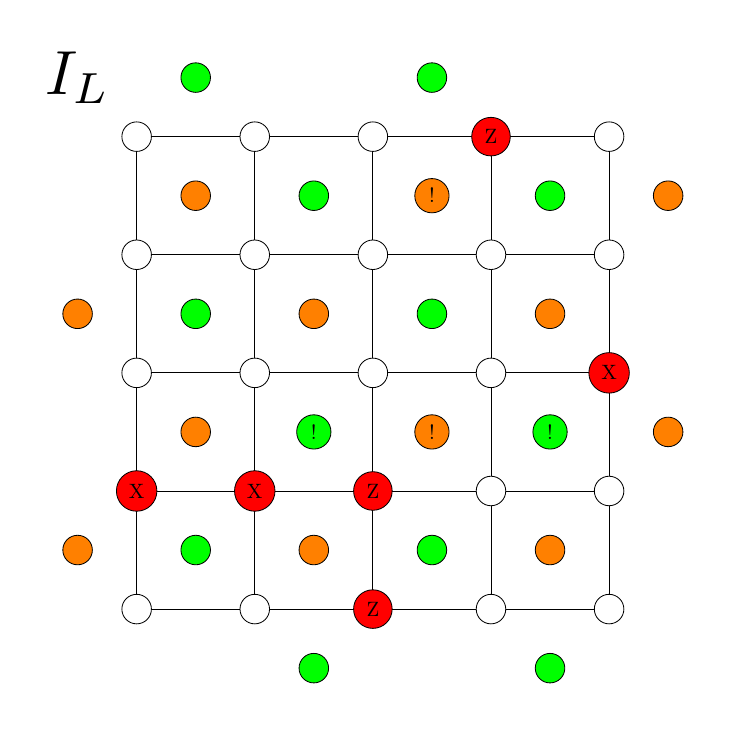}
    \includegraphics[width = 0.45\textwidth]{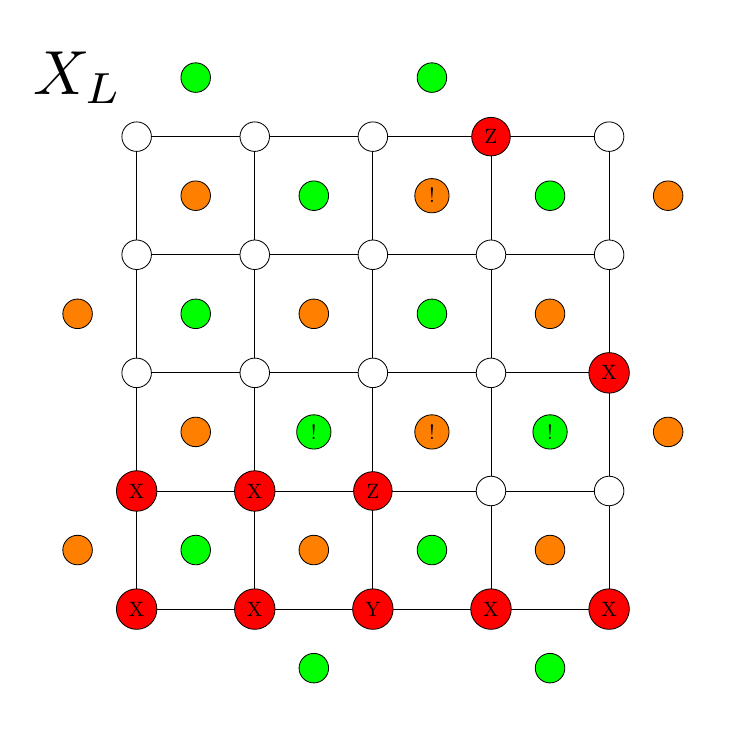} \\
    \includegraphics[width = 0.45\textwidth]{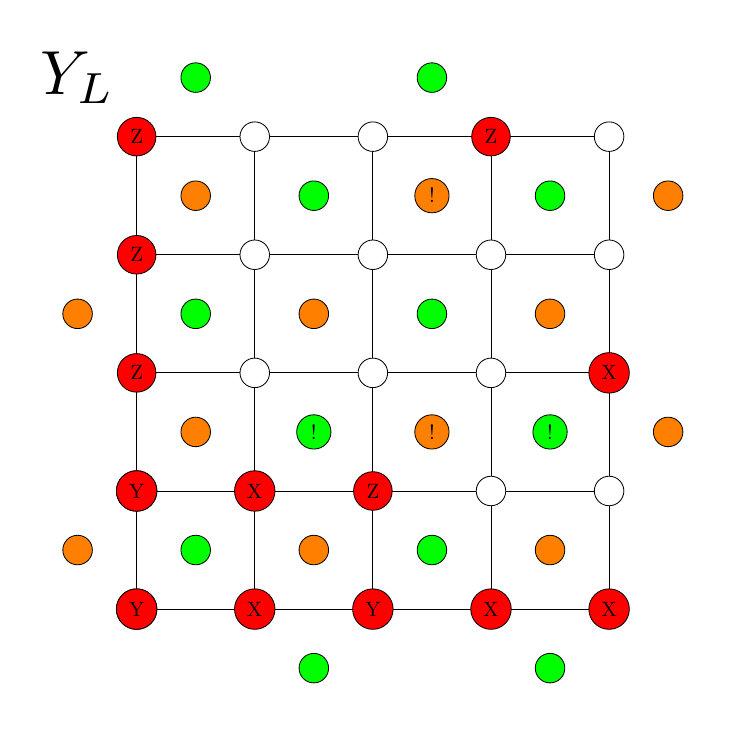}
    \includegraphics[width = 0.45\textwidth]{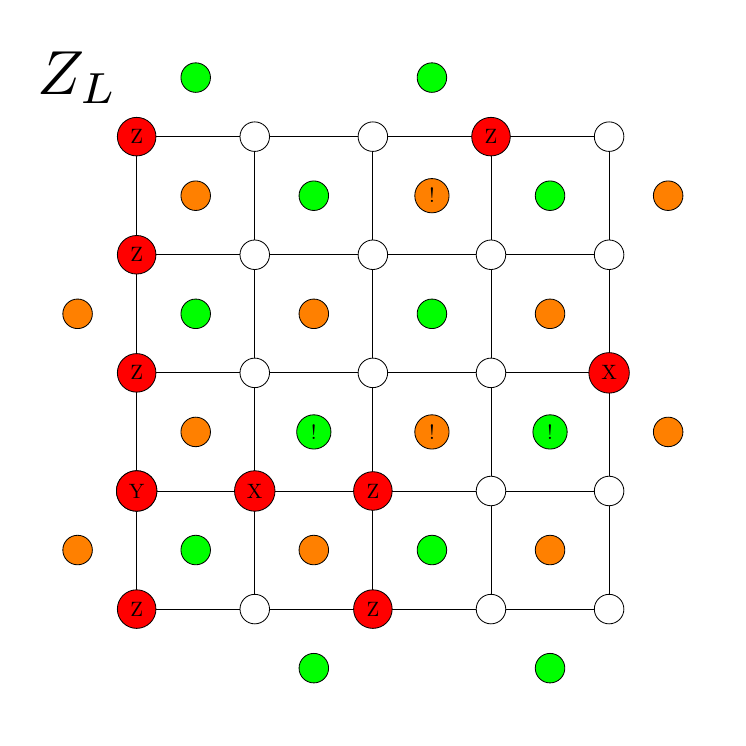} 
\end{subfigure}
\begin{subfigure}[b]{.45\textwidth}
\centering
    \includegraphics[width = 0.7\textwidth]{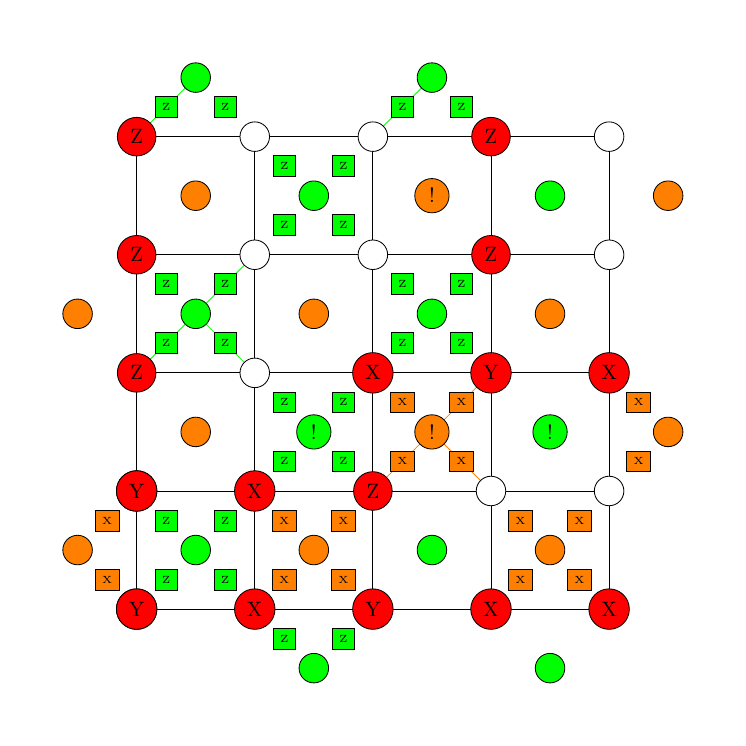}
\end{subfigure}
\caption{Graphical representation of a TN decoding process for a specific syndrome in a  {distance-5 rotated} code.}
\label{tensornetworkdecodingprocess}
\end{figure}

Given this new structure, the rotated planar code can be decoded through the TN decoder. In FIG. \ref{tensornetworkdecodingprocess}, we can see an example of the decoding process of an arbitrary syndrome through the TN decoder for a $5\times 5$ rotated planar code. In the top image, the error $E$ is presented altogether with the measured syndrome and in the second image an error $E_{rec}$ which returns the code to the codespace is presented. This error is found by generating chains from every non-trivial syndrome element to its nearest virtual check. Afterwards, the probability cosets of the four logical operators are computed by contracting the associated tensor network. The combination of each of the logical operators with $E_{rec}$ is also presented in the third and fourth row. Finally, the coset with the highest probability is $Y_L$ for this example and, thus, the recovered error by the TN decoder is $E_{rec}Y_L$. In the figure at the bottom, the resulting state of the code after recovery defined by $E_{rec}Y_LE$ can be seen, and, as shown by the highlighted stabilizer operators it belongs to the stabilizer set, implying that the correction has been successful.

\subsubsection{Performance and threshold}

\begin{figure}
    \centering
    \includegraphics[width = 1\columnwidth]{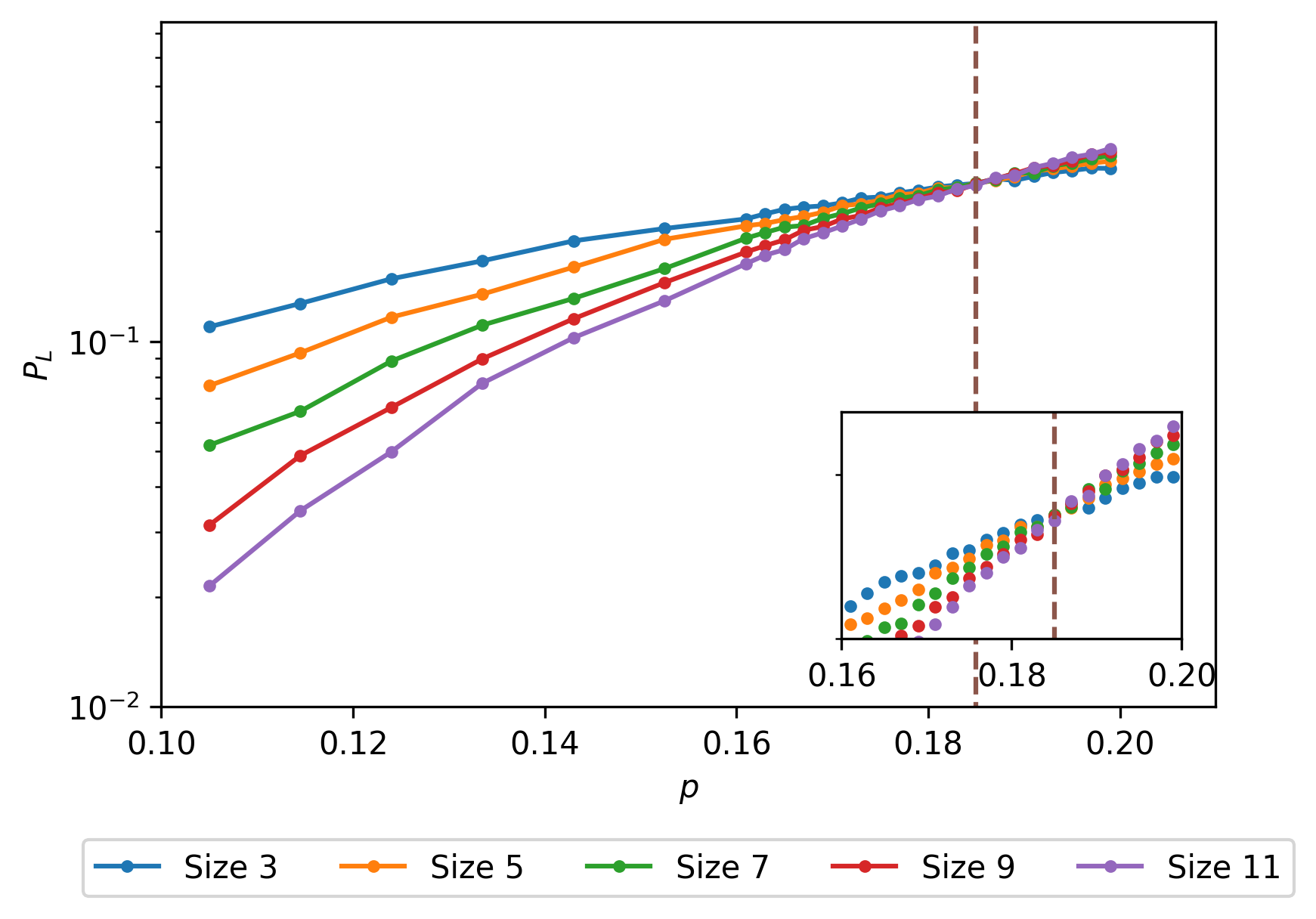}
    \caption{Logical error probability with dependence on the physical error probability of the rotated planar code under depolarizing noise decoded with the TN decoder with bond dimension $\chi = 16$.}
    \label{mpsthreshold}
\end{figure}

The tensor network decoder has the highest code threshold out of the ones considered in this work equal to $0.185$ under depolarizing noise with a bond dimension $\chi = 16$, as can be seen in FIG. \ref{mpsthreshold}.  {It is worth mentioning that this exceptional performance is highly related to the value of $\chi$. Recall that $\chi$ is the truncation is the maximum number of Schmidt tensors that we will be considered when contracting the code. A larger distance will increase the number of Schmidt tensor exponentially, thus, the maximum value of $\chi$ scales in an exponential manner as larger code distances are considered.} Recent work has been done studying $\chi$ values for achieving convergence of the tensor network decoding method under several noise model and code tailoring conditions \cite{tuckettxy, tuckett2}.

Moreover, as can be seen in Table \ref{tablatn}, tensor network decoding also suffers under biased noise in a significant manner. There have been recent studies which have investigated the effect of biased noise in surface codes and ways to enhance its performance reaching significant results which allowed for lower values of $\chi$ in order to obtain a convergence within the tensor network problem \cite{xzzx, tuckett2}. At the current time, the exceptional performance of the tensor network decoding scheme motivates researchers to seek for methods to accelerate its poor run time.

\begin{table}
\begin{center}
\begin{tabular}{|c | c|} 
 \hline
 $\eta$ &  $p_{th}$ \\ [0.5ex] 
 \hline\hline
 1/2 &  0.185 \\ 
 \hline
 1 &  0.175 \\
 \hline
 10 &  0.111 \\
 \hline
 100 &  0.097 \\
 \hline
 1000 & 0.096  \\ [1ex] 
 \hline
\end{tabular}
\caption{Probability threshold values for different biases of the rotated planar code under Pauli noise decoded with the TN decoder with bond dimension $\chi = 16$.}
\label{tablatn}
\end{center}
\end{table}

\subsubsection{Complexity}

The complexity of the tensor network decoding method is mainly given by two important tasks: the contraction of the MPS with its nearest MPO and the truncation of the resulting state. The contraction of the MPS with the MPO has a complexity of $\mathcal{O}(d\chi^2)$ while the truncation technique has a complexity $\mathcal{O}(d\chi^3)$ \cite{mps,NPTN}. Since for every matrix product operation there is a need for $d$ truncation techniques, one for each tensor node within the MPS, the resulting truncation complexity is $\mathcal{O}(d^2\chi^3) = \mathcal{O}(n\chi^3)$. Thus, the truncation complexity ends up defining the overall complexity of the decoding method.

The resulting complexity is a high price to pay for a DQMLD method, as larger code sizes are considered, the required $\chi$ should be also larger for taking into account the most probable configurations within the stabilizing cosets. The cubic growth with the bond dimension truncation harshly compromises the possible usability of the TN decoder in real-time decoding. Moreover, the Google Quantum AI team observed that this decoder is many orders of magnitude slower than the MWPM implementations used for their experiments \cite{googleSurf}. 

\subsubsection{Measurement errors}

Considering measurement errors in tensor network decoding is a cumbersome task, since it requires a space-time syndrome extraction which yields additional tensor nodes to take into account. The Google team in \cite{googleSurf} studied the tensor network decoding process in a distance-5 code for 25 syndrome extraction rounds by considering a Tanner graph between the syndrome elements and the circuit-level error mechanisms in a similar manner than the method used in \cite{fragile}. Afterwards, this Tanner graph can be rewritten as a tensor network that evaluates the probability of a logical coset outcome and, it can be planarized and contracted considering a specific truncation \cite{bentensor}.

As with the other decoding methods, considering measurement errors significantly increases the complexity of the already complex tensor network decoding method. Nevertheless, the high cost of performing TN decoding allowed for the best performance in the experimental rotated code from Google \cite{googleSurf}.

\subsection{Other decoders}
Many other decoders have been proposed through the literature for decoding topological codes other than the mainstream MWPM, UF, BPOSD and TN decoders. Many of them have been proposed to decode specific families of topological codes other than the rotated planar code such as the toric or the the color code. In this section we briefly introduce those decoders and discuss their capabilities.

\subsubsection{Cellular-automaton decoder}

 {Before the cellular-automaton decoder is introduced, we introduce the concept of cellular-automaton. A cellular automaton \cite{careview} can be described as a set of cells in a grid, each of which is described by one of a finite number of states. For each cell, a set of cells called the neighbourhood is defined relative to it. An initial state for all the cells is selected at time instance $t=0$. Then, a new configuration is generated according to a fixed rule dependent to the state of each cell and the state of the cells in its neighbourhood, advancing $t$ by 1. Thus, the overall system evolves in time following its initial state and the fixed rule.}

% Buscar cita
 {An example of a cellular automaton is the ``Game of Life" \cite{gameoflife}. In the Game of Life there is a square grid in which the neighbourhood of every cell is composed by its left, right, top, bottom and diagonal cells. For the game of life, each cell can take two values, "alive" or "dead". Moreover, given a time $t+1$, the updated state of every cell will be given by the following rules:}

 {\begin{itemize}
    \item Any alive cell with fewer than two alive neighbours dies.
    \item Any alive cell with two or three alive neighbours continues alive.
    \item Any alive cell with more than three alive neighbors dies.
    \item Any dead cell with exactly three alive neighbours becomes an alive cell.
\end{itemize}}

 {An instance of this cellular-automaton is illustrated in FIG. \ref{cellularautomatonexample}. The right alive cell in the top figure dies because it is only adjacent to a single alive cell and the top-left cell becomes alive in the second plot because it is adjacent to three alive cells. The remaining stay alive because they have 2 or 3 alive adjacent cells.} 

% Figura
\begin{figure}
    \centering
    \includegraphics[width = .5\columnwidth]{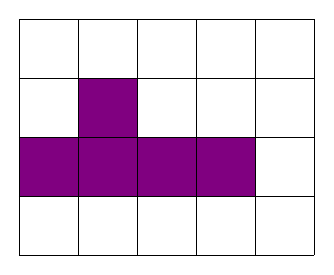}
    \includegraphics[width = .5\columnwidth]{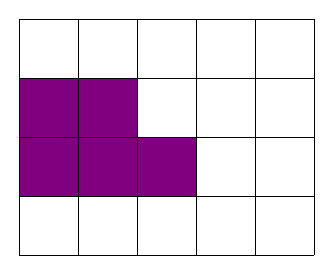}
    \caption{ {Illustration of two time instances in the Game of Life. On the top figure, an illustration of a configuration in time $t$, where violet squares are live cells and white ones are dead. On the bottom figure, the configuration of the cellular automaton at time instance $t+1$ is presented.}}
    \label{cellularautomatonexample}
\end{figure}

 {Cellular-automaton decoders were first proposed for decoding the toric code \cite{automata}. This class of decoders was constructed with the aim of reducing the decoding problem to simple local update rules. The authors justify that cellular automata do fit such requirements and, thus, propose the use them in order to decode toric codes.}

 {
The approach on the first cellular-automaton decoder consists in considering two cellular automata, a main one and an auxiliary one, where cells the states of the checks given an error \cite{automata}. Within the main cellular automaton each cell can be in one of two states, trivial or non-trivial. FIG. \ref{cellularautomatonexample} provides a portrayal of the mapping of a toric code under a syndrome to its cellular automaton representation.
}
\begin{figure}
    \centering
    \includegraphics[width = \columnwidth]{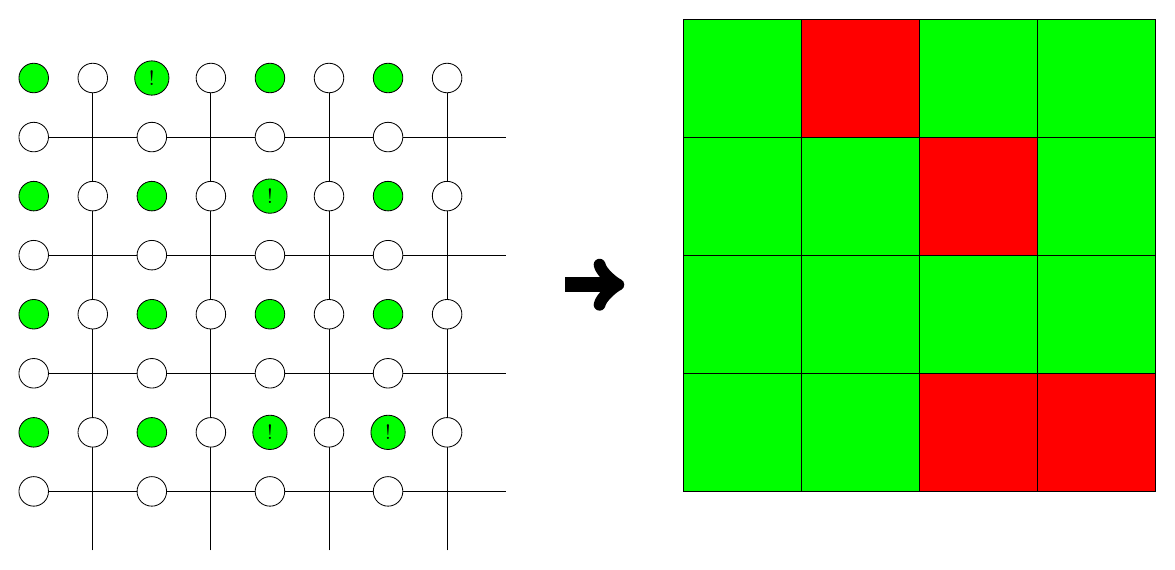}
    \caption{ {Illustration of the mapping of a syndrome in the toric code to a the check cellular automaton. In the left, a distance-4 toric code with a syndrome, where non-trivial syndrome elements are denoted with exclamation marks. To the right, the cellular-automaton to consider, where green cells indicate trivial checks and red ones indicate non-trivial checks.}}
    \label{cellularautomatonexample}
\end{figure}

 {The cellular-automaton decoder considers that every non-trivial cell $x$ corresponds to a charge $q(x)$. Moreover, an electrostatic field $\phi(x)$ is considered within every configuration of the cellular automaton. The fixed rule that characterizes the main cellular automaton system is that on every time instance, non-trivial cells consider their neighbourhood, and move with a probability of $1/2$ to the adjacent cell with highest field value, otherwise, they remain in the same spot as in the previous instance. Whenever two non-trivial cells collide they disintegrate and the decoder considers them as matched. The field is then changed following the next cellular automaton configuration. The cellular-automaton decoder will terminate when all non-trivial checks have been disintegrated and all cells are trivial. At the end, the recovered error will correspond to operators matching collided pairs.
}

 {The field utilized by the main cellular automaton in each iteration is provided by the auxiliary cellular automaton, which, for this general explanation, will take the same lattice form as the main cellular automaton. In the first time instance $t=0$, every cell within the auxiliary cellular automaton will have a value $q_E(x)$, which will be $q$ or $0$ if they correspond to a non-trivial or trivial cell respectively. Afterwards, for every time instance, the state of the auxiliary cellular automaton cells will evolve as:}

\begin{equation}
    \phi_{t+1}(x) = (1-\eta)\phi_t(x) + \frac{\eta}{2D}\sum_{\langle y,x\rangle}\phi_t(y) + q_E(x),
\end{equation}

 {where $\phi_t(x)$ indicates the field in the cell $x$ and the time instance $t$, $0<\eta<1/2$ is a smoothing parameter we can freely choose, $D$ is the dimension of the auxiliary cellular automaton lattice, which, for our case, would be 2 and the sum $\langle y,x\rangle$ is over all the cells in the neighbourhood of the $x$ cell. In order to set the field for the main cellular automaton, $c$ evolution instances are contemplated on the auxiliary cellular automaton. The $c$ parameter is tunable and known as the field velocity and needs to grow with logarithmic dependence with the size of the auxiliary matrix in order to establish meaningful fields for the main cellular automaton \cite{automata}. }

% \begin{figure}[h!]
%     \centering
%     \includegraphics[width = 0.75\columnwidth]{Figures/CAexample.png}
%     \caption{Visual representation of the 2D cellular automaton decoder for the toric code with periodic boundaries. The blue dots, represented by lattice $V$, are the data qubits while the green boxes, represented by lattice $\Lambda$, depict the cellular automaton decoder. Figure taken from \cite{automata}.}
%     \label{CAdecoder}
% \end{figure}

There have been other proposals for the update rule of the cellular automaton decoder such as the so called Toom's rule \cite{toom,toom2} or the sweep rule \cite{sweep,sweep2}  {altogether with other lattice structures for the auxiliary cellular automaton such as 3D toric codes \cite{automata}}. Most of the cellular automaton decoders have been studied for toric codes with boundaries in $d$ dimensions \cite{automata,toom,toom2,sweep,sweep2}, but some lattices without boundaries have examined \cite{sweep2}. Additionally, it was shown that the decoders for the toric code can be applied to decode color codes \cite{colortoric} and, thus, cellular automaton decoders can also be applied for this family of QECCs. Unfortunately, its application for a rotated planar code has not been investigated.

Regarding the threshold achieved by this class of decoders it is noteworthy to say that the noise model considered has been primarily an i.i.d. phase-flip model \cite{automata,toom,toom2,sweep,sweep2}. In this sense, a threshold of $8.2\%$ was claimed in \cite{automata} for a 2D toric code, while a MWPM decoder exhibits a $10.31\%$ threshold for such code \cite{mwpmtoricphase}. However, a threshold of $15.5\%$ was claimed in \cite{toom2} for the cubic toric code using the sweep rule. In terms of the complexity of this decoder, it can naturally be integrated in a parallel way by communicating the processing cores locally and, therefore, it is usually seemed as a fast decoder \cite{automata,sweep,sweep2}. In this sense, the 2D cellular automaton of \cite{automata} requires a number of updates in the order of $\log ^{5} (d)$ updates, while the 3D implementation requires a number of updates in the order of $\log^3 (d)$. As the authors discuss, this comes from the fact that the cellular-automaton decoder accepts a parallel implementation from construction \cite{automata}. The sweep decoder implementation in \cite{sweep} requires a runtime $\mathcal{O}(d)$, so linear in $d$.

An important feature of cellular-automaton decoders is that they are robust for the case that measurement errors are considered \cite{sweep,sweep2}. Specifically, this decoder is deemed to be a single-shot decoder, which implies that measurement errors can be dealt with a single decoding round (recall Appendix \ref{appA}). This is specially interesting for a decoder since as discussed before, the way in which decoders usually deal with noisy syndromes is by performing $d$ measurement rounds and then solving the space-time like obtained graph \cite{sweep2}.

\subsubsection{Renormalization group decoder}

The renormalization group (RG) decoder was proposed as a fast and efficient decoder for the Kitaev toric code \cite{renormalization}. The main idea behind this decoder is that the toric code can be seen as a concatenated code so that the decoding problem can be resolved from the smaller codes that form it.  {Concatenating two codes refers to using the logical qubits obtained by the first code as the physical qubits for the second one \cite{gottesman}. Thus, the obtained logical qubits have two layers of protection. Logically, we can add an arbitrary number of protection layers in exchange of a higher qubit overhead. The 9-qubit Shor code is an example of code concatenation since the qubits protected by means of a phase-flip repetition code are encoded with a bit-flip repetition code \cite{shorQEC}. Mathematically, the concatenation of encoding an arbitrary quantum state $\ket{\psi}$ can be expressed as
\begin{align}\label{conc}
    \ket{\psi}_L^{2} &= U_2(U_{1}(\ket{\psi}\otimes\ket{0}^{\otimes n_1 - k_1})\otimes\ket{0}^{n_2-k_2}) \\ & =U_2(\ket{\psi}_L^1\otimes\ket{0}^{n_2-k_2}),
\end{align}
where $U_1$ and $U_2$ refer to the encoding operations of the codes with parameters, $[[n_1,k_1,d_1]$ and $[[n_2,k_2,d_2]]$, respectively; and $\ket{\psi}_L$ refers to the fact that it is a logical state over some code. We restrict expression \eqref{conc} to 2 concatenation layers for simplicity.}

\begin{figure}[h!]
    \centering
    \includegraphics[width = 0.75\columnwidth]{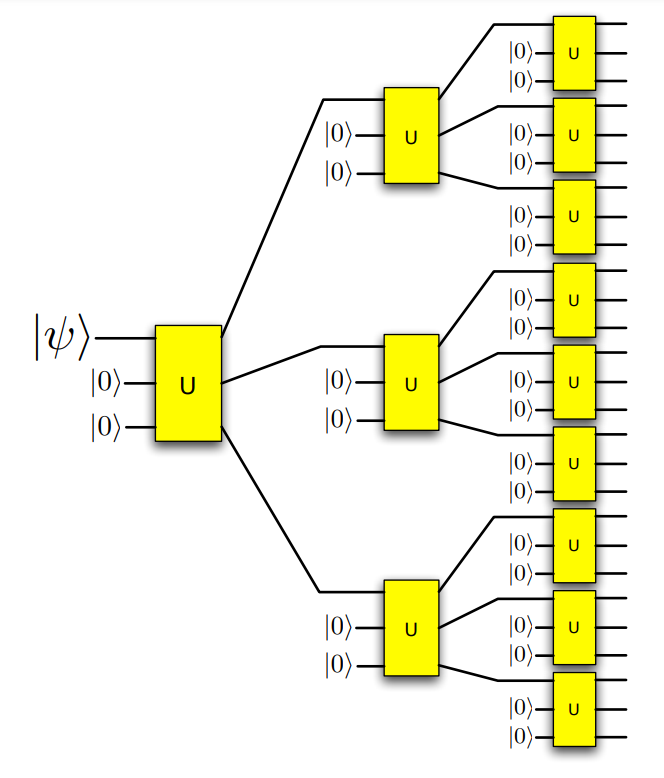}
    \caption{ {Graphical depiction of $n$ layers of code concatenation. For this example, the same code is repeated every layer. Such code encodes a single qubit within three qubits. Figure taken from a talk by Duclos-Cianci (Accessible in: \url{https://av.tib.eu/media/35296})}.}
    \label{figconc}
\end{figure}
\begin{figure}[h!]
    \centering
    \includegraphics[width = 0.75\columnwidth]{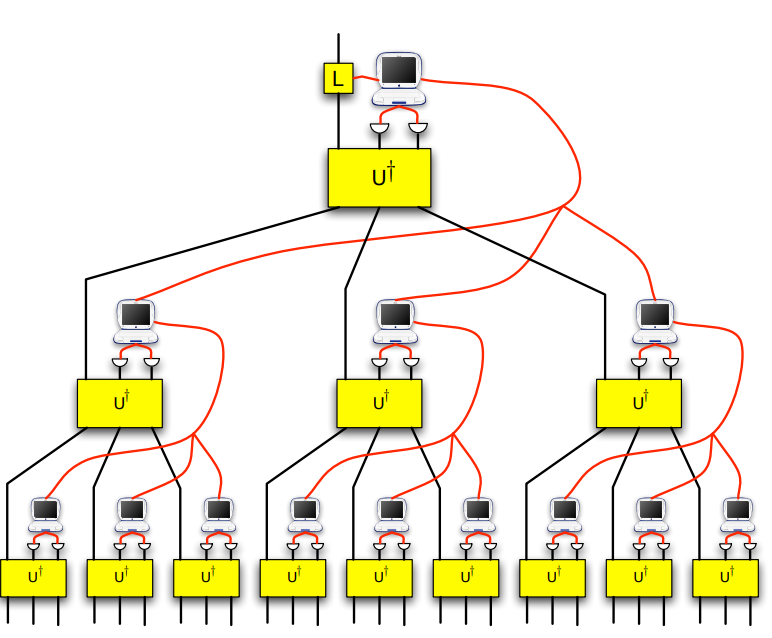}
    \caption{ {Graphical depiction of the decoding process for the concatenated code in FIG. \ref{figconc}. The process goes from the bottom to the top. The red lines represent the fact that the logical error probability is passed to the next layer as apriori information for decoding. The $L$ operator in the top layer represents the estimated logical error. Figure taken from a talk by Duclos-Cianci (Accessible in: \url{https://av.tib.eu/media/35296})}.}
    \label{figconcdec}
\end{figure}

 {FIG. \ref{figconc} shows a graphical representation of code concatenation. Decoding codes that present this concatenated structure can be done by decoding each of the layers, which consists of several smaller codes. Specifically, the process begins by decoding each of the small codes of the last layer of concatenation (considering the syndrome information of such layer) and other than taking a hard decision for the logical error, the soft information of the logical error probability is passed to the next layer as a priori information. This process is repeated until the last layer is reached, where the hard decision for the estimated logical error is done. For the first decoding step, the a priori information is the physical error probability. FIG. \ref{figconcdec} shows this process graphically. Importantly, recall that the optimal decoding problem, i.e. considering degeneracy, for stabilizer codes is not efficiently solvable when the code size grows. However, it can actually be dealt with efficiently for small codes by brute force. Therefore, codes with a concatenated structure as the one presented in FIG. \ref{figconc} can actually be optimally decoded by following the described process. }

\begin{figure}[h!]
    \centering
    \includegraphics[width = 0.9\columnwidth]{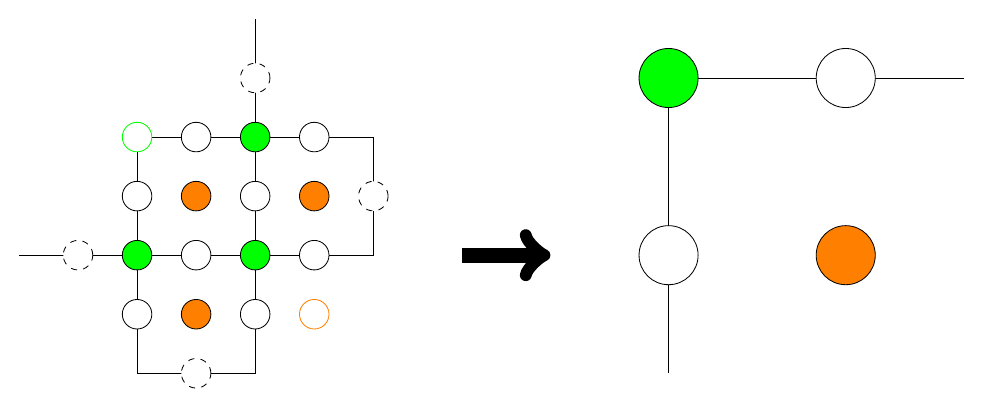}
    \caption{Example of a code block decomposition used for RG decoding.  {The unit cells consist of 12 data qubits. The data qubit and check color convention is the same as through the rest of the article. Dashed data qubits refer to qubits shared between unit cells. Filled checks are used in decoding each layer while empty ones are passed to the next concatenation layer}. Since this code block is an open boundary toric code, two logical qubits are encoded  {as depicted in the right part of the figure. Note also that here the information of the unused checks of the left is used}. Figure inspired from \cite{renormalization}.}
    \label{RGexample}
\end{figure}

\begin{figure}[h!]
    \centering
    \includegraphics[width = 0.9\columnwidth]{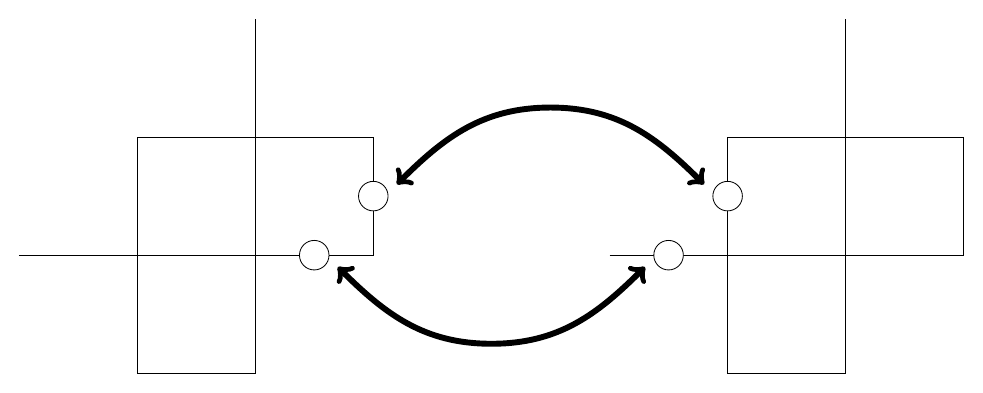}
    \caption{ {Example of two unit cells of the toric code decomposition where shared qubits exchange information by means of belief propagation to reach consistency.}}
    \label{BPRGexchange}
\end{figure}

 {The proposal of the RG decoder for the toric code was motivated by such ideas \cite{renormalization}.} The toric code is not exactly a concatenated code, but the authors propose that it can be  {approximated as} one when allowing the blocks of the concatenation to share qubits, so the  {the blocks} are actually overlapped \cite{renormalization}.  {Each of the small codes in this approximated concatenation is selected to be an small (12 qubit) open boundary toric code. This unit cell is shown in FIG. \ref{RGexample}. In the unit cell depicted there, the dashed data qubits represent shared qubits between neighbouring unit cells in the periodic structure of the toric code. In addition, the measured syndromes of the filled checks are used in the decoding of the unit cells, while the empty check information is used to decode the next concatenation step (as seen in the right figure filled checks).} Following this rationale, the RG decoder computes the logical error probabilities of the smallest codes in the concatenation, which then are used as the noise model for the second layer of codes on the concatenation.  {Since the concatenated structure is an approximation, the soft information obtained for the shared qubits is different in each of the blocks. This implies that the marginal probabilities of error obtained after decoding differ, i.e.}
 {\begin{equation}
    P_{L,i}(E) \neq P_{R,i}(E),
\end{equation}
where $i$ refers to one of the shared qubits, $L$ and $R$ refers to let and right cells and $E$ the error experienced. To solve this inconsistency, the authors proposed to use belief propagation to obtain the merged soft information from the independently obtained ones and, thus, force the marginal to be equal, i.e. $P_{L,i}(E) = P_{R,i}(E)$. FIG. \ref{BPRGexchange} shows a graphical depiction of two neighbouring unit cells for which shared qubits consistency is required, and obtained via BP. This process is done, as described for concatenated coding before, until the last layer of concatenation is reached and} the estimation of the error is obtained for correction. The authors discuss that since the small codes for the concatenation are small open-boundary topological codes, those can be decoded by brute force \cite{renormalization}. 

As stated, the renormalization decoder was originally proposed for the Kitaev toric code, so with periodic boundaries, but modifications for decoding color codes \cite{renormalization2}, the 3D cubic code \cite{renormalization3,hdrg,hdrg1} qudit topological codes \cite{hdrg,hdrg2,hdrg3} and the four-dimensional toric code (both for periodic and open boundary conditions) \cite{renormalization4} have also been proposed. Interestingly, the RG decoder was the first known algorithm to decode the color code. The RG decoder has not been investigated for the rotated planar code discussed through this paper, but it could be valid, in principle, by using the ideas for decoding open boundary four-dimensional toric codes \cite{renormalization4} since the rotated planar code is essentially a rotated toric code with boundaries.

Regarding the code threshold achieved by RG decoders, its standard version shown a threshold of $7.8\%$ for the depolarizing noise model, which falls short when compared to the $15.5\%$ threshold obtained by decoding it with the MWPM decoder. However, the application of belief propagation decoder before the RG decoder was proposed in \cite{renormalization}. This was done so that the smallest codes of the concatenation can start the decoding with the probability distribution given by BP implying that the RG is more accurate globally. By doing so, the BP+RG decoder achieved a $16.5\%$ threshold, which exceeds that of the MWPM decoder by itself \cite{renormalization}. This, however, was comparing the performance with the MPWM alone, i.e. belief-matching would also increase the threshold \cite{fragile, Crigerbeliefmatching}. In addition, it does not exceed the threshold achieved by the tensor network decoder, which stands between $17\%$ and $18.5\%$ \cite{mps}.

Similar to the cellular automata decoder, the RG decoder admits a parallelization of the decoding problem since each of the codes forming a layer of the concatenation can be decoded in at the same time \cite{renormalization}. This implies that the complexity of the algorithm is a function of the number of concatenation layers, which scales logarithmically with the distance and, thus, number of qubits of the code. Therefore, the decoding complexity of the standard RG decoder is in the order of $\mathcal{O}(d)$ if the complete parallelization is done, while in the order of $\mathcal{O}(d^2\log{d})$ if the decoding is done serially \cite{renormalization}. Subsequently, this decoder is considered to be a fast decoder. The BP+RG approach for increasing the threshold will suffer a toll in the complexity coming from  {also} having to execute BP. It is noteworthy to say that whenever faulty syndrome measurements coming from circuit level noise are considered, the RG decoder needs $d$ rounds of measurements for dealing with them, implying that the complexity of the decoder will also increase \cite{sweep2}.

\begin{figure*}
    \centering
    \includegraphics[width=\textwidth]{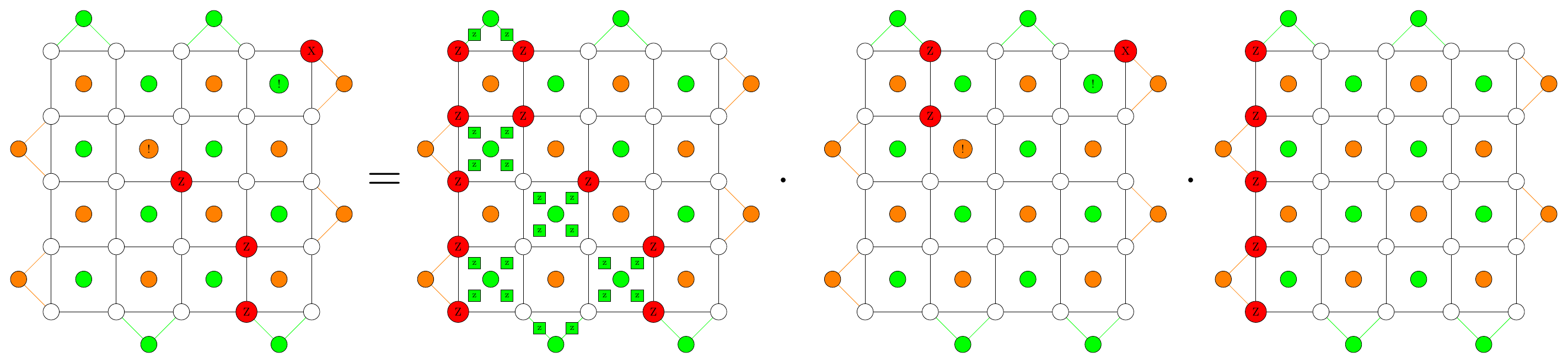}
    \caption{ {Example of decomposing a Pauli error (left), $E$, into its stabilizer (middle, left), $S$; pure error (middle,right), $T$; and logical error (left), $L$; components as in equation \eqref{dec}. The green squares are depicted to show that such Pauli chain forms an stabilizer for the $S$ component. Note that the pure error is the only Pauli string triggering syndromes. Figure motivation taken from \cite{neural}.}}
    \label{decompNN}
\end{figure*}

\subsubsection{Neural-network decoders}
Neural-network (NN) decoders are a result of applying machine learning (ML) techniques for decoding quantum error correction codes \cite{neural}. In order to do so, the authors of the NN decoder proposed reducing the decoding problem to the well-studied ML problem of classification, which generally consists on optimizing the assignment of known labels (generally low-dimensional) to known inputs (generally high-dimensional) with the scope of afterwards using the optimized assignment to label inputs that are unknown. A toy example of a classification problem in ML is to label photos of cats and dogs, i.e. to say which photos depict cats and which dogs \cite{catdog}. 
The stage in which the classifier is optimized with known data is named training. Hence, NN decoders need a training stage before being able to decode. In the context of the analogy with machine learning, the authors propose to decompose an error into three multi-qubit Pauli operators as
 {\begin{equation}\label{dec}
    E = STL,
\end{equation}}
where $S$ is an element of the stabilizer, $T$ is any fixed Pauli that produces the error syndrome (usually referred as pure error {, and it is the part of the error that produces the syndrome \cite{reviewpat}}) and $L$ is a logical Pauli operator \cite{neural}. Note that this decomposition is standard in the field of quantum error correction \cite{reviewpat}.  {FIG. \ref{decompNN} shows an example of the decomposition in equation \eqref{dec}, note that the triggered syndrome is associated to the $T$ component of the error.} Due to the degenerate nature of QECCs  {(Recall that errors up to an stabilizer act equivalently on the logical qubits)}, any recovery operator $E' = S'TL$ will successfully correct the corrupted codeword implying that the stabilizer element of the recovery operator can be assigned arbitrarily with no impact in the logical error rate of the correcting method. In addition, the authors discuss that the pure error can be produced using a parallel table look-up since each element of the error syndrome depends on a unique pure error independently of the others \cite{neural} {\footnote{ {This has been previously seen in the TN decoder, where $E_{rec}$ solves the syndrome by matching every non-trivial syndrome element to the nearest virtual check.}}. In this sense, the authors are able to remove the pure error part.} The algorithms performing such two assignments  {(selection of $S$ and $T$ components)} are named as the simple decoder. Hence, the NN decoder is based on the fact that since the rotated planar code encodes one logical qubit, the  {possible} logical operators associated are $\mathrm{\hat{I},\hat{X},\hat{Y},\hat{Z}}$ (the hat refers to the fact that those operators are defined over the logical qubit) and, therefore, decoding can be done  {by means of} a classification problem with those four labels.

The way in which the classification problem was dealt with in \cite{neural} was by using feed-forward neural networks.  {Neural networks are machine learning models that aim to mimic the neuron structures of biological brains. In this sense, neural networks are defined by a graph consisted of many artificial neuron layers (nodes) which are connected among them (edges).} The neurons are defined by a so called-activation function  {(that defines the interactions between the artificial networks)} which depends on an array of weights, $\bar{w}$, of length $m$ and a bias $b$.  {Since the NN is of the feed-forward type, the neurons calculate the nonlinear activation function 
\begin{equation}\label{activation}
    \bar{y}=(1+\mathrm{exp}(-(\bar{w} * \bar{x} + b)))^{-1},
\end{equation}
where $\bar{x}$ is the input  {(the measured syndrome in the first layer)} and $*$ refers to the inner product, and pass it to the subsequent layer until the output layer is reached\footnote{ {Since the considered NN was feed-forward the data exchanged among layers is done in just one direction, from the input to the output. There are instances of NNs that consider both directions for the information flow. Those are referred to as recurrent neural networks.}}, where the result of the activation function is rounded to $\{0,1\}$\cite{neural}. The combinations of the outputs define which logical error is estimated and combined with the results of the simple decoder for estimating the complete recovery operator. Note that the estimated error will match the syndrome, but it is not guaranteed to be of minimum-weight due to the heuristic nature of the neural network method.} The layers between the input and output layers are named as hidden layers in the terminology of ML. As explained before, a neural network requires a training stage before it can operate over a set of unknown data. Therefore, for the NN decoder, a training stage where the weights and biases of the neurons are trained is needed. The authors of \cite{neural} select the average cross-entropy as the cost function to optimize the activation function and produce a training set directly sampling errors at an error probability where the MWPM decoder has a $25\%$ logical error rate. The cost-function was minimized by stochastic gradient descent, while the number of elements of the training set was limited to at most $10^6$ samples.

\begin{figure}[h!]
    \centering
    \includegraphics[width = 0.75\columnwidth]{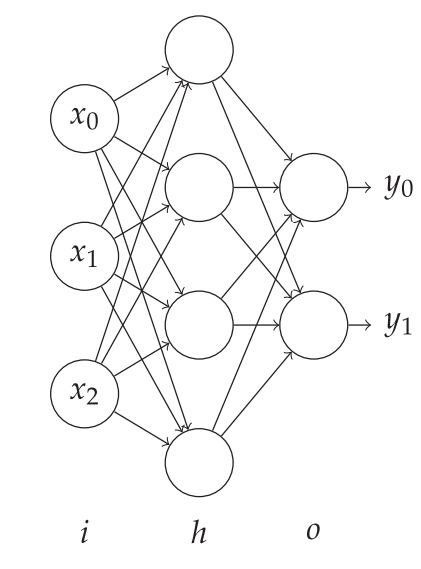}
    \caption{Graphical depiction of the feed-forward neural network used for the NN decoder. The example consists of an input layer (i), a hidden layer (h) and an output layer (o). The inputs represent the syndrome data, while the outputs refer to the estimated logical error. Each of the neurons in every layer compute the activation function  {(for this example it is given in equation \eqref{activation})} as a function of their received information and pass the information to the next layer. Figure taken from \cite{neural}.}
    \label{NNexample}
\end{figure}

The NN decoder in \cite{neural} proved to be an efficient method for decoding small distance, $d\in\{3,5,7\}$, rotated planar codes achieving similar performance as the MWPM decoder while maintaining a similar complexity. As a result of the interesting performance of this decoder, several studies discussing ML decoders for the rotated planar code have been proposed such as the deep neural network based decoder in \cite{NNdeep,compareNN} or the reinforcement learning based decoder in \cite{NNreinforce}. The deep neural network decoder proved a better threshold than MWPM at the cost of a higher complexity, while the reinforcement learning decoder proved an almost linear complexity at the cost of a worse code threshold \cite{compareNN}. Combinations of NN decoders with RG decoders have also been proposed \cite{NNRG}. Recently, NN decoding methods for surface codes that do not rely on an specific error model, i.e. that are purely based on hardware obtained data, have been proposed \cite{nnData1,nnData2}. Machine learning based decoding methods have also been proposed for other topological codes such as the toric code \cite{NNTC1,NNTC2,NNTC3,NNTC4,NNTC5,NNTC6,NNTC7,NNTC8,NNTC9}, the color code \cite{NNCC1} or the semionic code \cite{NNSM1}. Each of the proposed ML decoders for each of the codes provide either a faster decoding or a better threshold than other methods, showing the natural trade-off between accuracy and speed.

Circuit level noise has also been considered for neural network decoders, i.e. faulty syndrome data arising from noisy gates and measurements \cite{neural,NNdeep,NNCC1,NNcham}. In a similar fashion as other decoders, dealing with noisy measurements requires approximately $d$ syndrome measurement rounds, implying that the runtime of the decoders increases. Importantly, NN decoder should be trained using this error model so that they can cope with it, i.e. training is specific to the noise model in consideration \cite{neural}. The performance of the NN decoder of \cite{neural} achieves the same performance as MWPM for this scenario.

To sum up, NN decoders are promising in the quest of obtaining accurate and fast decoders for decoding not only rotated planar codes but also other families of error correcting codes. Some of the problematic behind this family of decoders is the training cost, which may scale exponentially with the code distance \cite{sweep2}. This is one of the main reasons behind the fact that most of the NN-based decoders have only been tested for small distances $d<10$ \cite{NNTC9}. Interestingly, promising results have been obtained for decoding the toric code with an NN+UF decoder that achieves high threshold with almost linear complexity, while having been tested up to $d=255$ \cite{NNTC9}. However, the need of a training stage before operation implies that NN and ML decoders might not be the best choice when the noise fluctuates over time, as it occurs for superconducting qubit platforms \cite{TVQC,fastfading}. Additionally, it is important to state that mapping the decoding problem to a ML classification problem works well for the rotated planar code or other topological codes since they encode a low number of logical qubits, implying that the possible logical errors is not too high. However, since the number of logical errors scales as $2^{2k}$ with the number of logical qubits \cite{logicalsparse}, $k$, these methods may not be implementable for other families of codes with higher coding rates such as QLDPC codes.

\subsubsection{MaxSAT decoder}
The MaxSAT decoder has been recently proposed as an efficient algorithm to decode color codes \cite{maxsat}. This decoder is based on the analogy between the decoding problem of the color code and the LightsOut puzzle when the error model in consideration is the bit-flip noise. The LightsOut puzzle refers to a problem where a lattice whose  {vertices} are associated with switches and  {faces with }lights  {that }can be either on or off. In such lattice, toggling a switch on the lattice implies that all its neighbouring lights change their previous state, i.e. if they were on, they are turned off and vice versa. The LightsOut problem then tries to find out a sequence of switch actions so that all lights are turned off. The puzzle has two important properties: toggling a switch twice is the same as not doing anything and the state of a light only depends on how often its neighbour switch have been toggled (independence on the order). The authors argue that this problem is similar to the decoding process of a color code by doing the following analogy:
\begin{itemize}
    \item Data qubits $\leftrightarrow$ switches.
    \item Checks $\leftrightarrow$ lights.
    \item Syndrome $\leftrightarrow$ initial light configuration.
    \item Decoding estimate $\leftrightarrow$ switch set that is a solution.
    \item Minimum-weight decoding estimate $\leftrightarrow$ solution with minimum switch operations.
\end{itemize}

\begin{figure}[h!]
    \centering
    \includegraphics[width = 0.75\columnwidth]{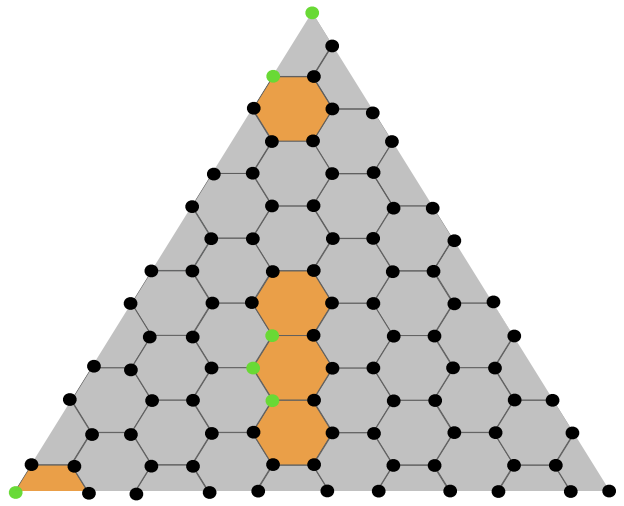}
    \caption{LightsOut and color code ($d=11$) decoding analogy. The marked faces represent the initial configuration (syndrome), while the green data qubits represent a possible solution of the problem which is not necessarily minimal. Figure taken from \cite{maxsat}.}
    \label{MaxSATexample}
\end{figure}

Therefore, the authors exploit such an analogy to propose a decoding methods based on solving the LightsOut puzzle. The main idea is to find the minimal solution set for the puzzle, which is equivalent to a decoding estimate of minimum weight matching the syndrome measurement. In order to obtain the minimal solution of the LightsOut puzzle, the authors formulate such problem as a maximum satisfiability (MaxSAT) problem, which generally refers to the problem of determining the maximum number of clauses of a Boolean function that can be made true by assigning true values to the variables \cite{maxSATgen}. Following this rationale, the authors formulate the decoding problem as the MaxSAT problem 
\begin{equation}\label{eq:maxsatClauses}
    \begin{split}
     \forall f\in F :& \bigoplus_{v\in\mathcal{F}_\mathrm{switches}(f)}  \mathrm{switch}_v = \mathcal{F}_\mathrm{init}(f), \\
    & \forall f\in F: \mathrm{not}(\mathrm{switch}_f),
    \end{split}
\end{equation}
where $f\in F$ refer to the lights or faces, the $\mathrm{switch}_v$ are Boolean variables representing if a switch is toggled, $\mathcal{F}_\mathrm{switches}(f)$ is a discrete function that takes a face $f$ and returns the set of switches surrounding it; and  $\mathcal{F}_\mathrm{init}$ is a Boolean function that describes the initial configuration of the system (associated with the measured syndrome). $\bigoplus$ denotes the exclusive-or (XOR) Boolean operation. In equation \eqref{eq:maxsatClauses}, the constraints in the first line represent the satisfiability problem, and the second are the soft constraints so that such problem is maximum.  {Thus, as a result of the analogy presented before, solving such problem leads to an error of minimum weight (from the soft constraints) that matches the syndrome (from the satisfiability problem).} Hence, color code decoding can be done by solving such MaxSAT problem \cite{maxsat}. The authors discuss many MaxSAT solvers for decoding the color code.

% performance and runtime
The MaxSAT decoder for bit-flip error noise in color codes was proven to have a very high threshold of $10.1\%$ \cite{maxsat}. This is near the optimal $10.9\%$ threshold obtained by the MPS or TN decoder, and substantially exceed the $\approx 9\%$ of the MWPM, $8\%$ of the UF and $7.8\%$ of the RG decoders. This excellent threshold does not come free of charge since it implies a higher decoding runtime. The authors do not discuss the complexity of the MaxSAT decoder, but they do explain that the runtime is slower than the ones of MWPM, UF and RG decoders, while it is faster than the TN decoder. This is consistent with the typical performances versus complexity trade-off discussed throughout this tutorial. The toll in runtime comes from the fact that the MaxSAT problem is an NP-hard problem \cite{maxsat}. Importantly, the authors argue that the MaxSAT decoder is faster whenever the physical error rate is lower, which is a desired feature as QECCs are expected to be working in sub-threshold noise levels.

In conclusion, the MaxSAT decoder is an efficient decoding method for the color code that exhibits a very high code threshold for bit-flip noise with a more reasonable complexity when compared with the tensor network decoder. However, at the time of writing, the MaxSAT decode is yet a limited approach, mainly due to the fact that only bit-flip noise is considered. This is very important since the LightsOut analogy proposed by the authors in \cite{maxsat} depends on such error model assumption. Additionally, no faulty measurements have been considered and it is rather unclear if the decoder will be able to operate whenever circuit noise level is considered. Nevertheless, this decoding method has been very recently proposed and generalizations of it for noises with other structures (importantly the depolarizing channel) and other families of topological codes can be expected as future work.

\subsection{Software Packages}
% decir que paquetes de software hay en internet y que hace cada uno. Tabla de resumane tambien

\begin{figure*}
\centering
    \includegraphics[width = 0.7\textwidth]{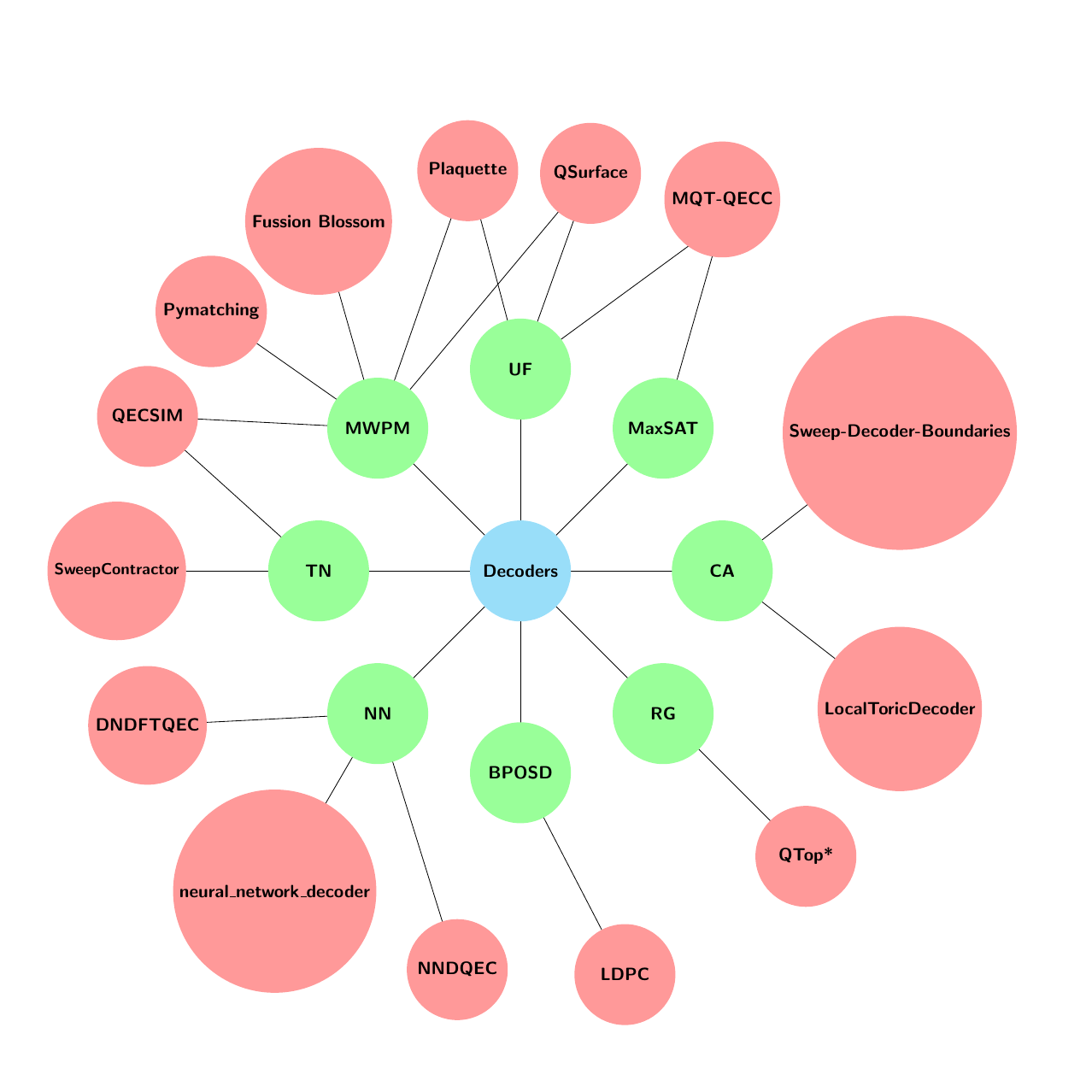}
    \caption{Sketch representing the available software packages for each mentioned decoder. Each edge connects each decoder (green circle), with the software packages which support it (pink circles). The decoders mentioned in the figure are the Cellular Automata decoder (CA), the MaxSAT decoder, the Union Find decoder (UF), the Minimum Weight Perfect Matching decoder (MWPM), the Tensor Network decoder (TN) the Neural Network decoder (NN), the Belief Propagation + Order Statistics Decoder (BPOSD) and the Renormalization Group decoder (RG).}
    \label{softwareDecs}
\end{figure*}

In this section an overview regarding the existing software packages for surface code decoding simulation will be covered. Fortunately, the scientific community has provided a large number of open source repositories for performing simulations of several decoders of  {surface codes} and other topological codes and, thus, we will cover several of them (the most popular and used ones) as a reference for people on the community of quantum error correction. We provide a graphical overview in FIG. \ref{softwareDecs}.

% Ruido
Whenever considering numerical simulations of a surface code, it is important to be able to generate the samples of the noise following the distribution of interest for a certain scenario so that the decoder operates over such noise model. In this sense, generating samples for data qubit noise is pretty straightforward by considering the specific distribution of the Pauli channel for each of the qubits (See section \ref{secnoisemod}). Nevertheless, whenever circuit-level noise for faulty check measurements is considered, sampling the noise is not a trivial task \cite{stim}. Therefore, the \textit{Stim} repository by Craig Gidney was developed for sampling circuit-level noise for the sake of obtaining samples of noise that accurately resemble the errors that a surface code experiences over such scenario \cite{stim}. Stim samples circuit noise to all qubits within the code, including measurement qubits, at a fast speed and is the simulator of choice when considering stabilizing circuit noise. This repository can analyze a $d=100$ surface code circuits in approximately 15 seconds for sampling circuit shots at a rate of 1000 samples per second. 

% MWPM
Regarding software implementations of the {MWPM} decoder there are several open source repositories which successfully simulate realizations of the decoder at a sufficient speed. The most popular and quickest one considering serial computation is Oscar Higgot's the \textit{Pymatching} repository \cite{sparse, pymatching}, the latest version of which, {uses} the  {sparse blossom} implementation of the MWPM decoder. The \textit{Fusion Blossom} algorithm \cite{fussionBlossom} is also available at the GitHub repository\footnote{\url{https://github.com/yuewuo/fusion-blossom}} by Yue Wu which, as explained before, albeit having a worse serial performance than Pymatching, allows parallel decoding improving the complexity at each additional node and eventually surpassing the performance of Pymatching. Additionally, the \textit{QECSIM} repository\footnote{\url{https://github.com/qecsim/qecsim}} by David Tuckett also stands as a fair simulator for the minimum weight perfect matching which also allows other decoding methods. Another recent software package that incorporates Pymatching and Fusion Blossom (by calling those repositories) was posted in the \emph{Plaquette} repository\footnote{\url{https://github.com/qc-design/plaquette}} by the QC Design team. Interestingly, such repository incorporates Stim too, allowing the combination of all those repositories in a single one.

% UF
An implementation of the UF decoder can be found in the \emph{Qsurface}\footnote{\url{https://github.com/watermarkhu/qsurface}}. The package allows the simulation of the standard union-find decoder \cite{UF} as well as a modification of it by the authors of Qsurface which they name as Union-find Node-Suspension decoder. The authors claim that an improved threshold can be obtained by using such modified decoder while maintaining complexity. Moreover, such package includes an implementation of the MWPM algorithm. The previously discussed Plaquette package also includes an implementation of the UF decoder by Delfosse and Nickerson. The inclusion of this decoder is specially interesting for the Plaquette package since the circuit-level noise obtained from Stim can be directly used for testing the UF decoder.

% BPOSD
The BPOSD algorithm was originally thought as a general decoder for QLDPC codes \cite{qldpc2}. However, a surface code can be seen as a sparse code as the distance of the code increases (due to the fact that the weight of the stabilizer is constant independently of the size). In this sense, the  {\emph{LDPC}} package\footnote{\url{https://github.com/quantumgizmos/ldpc}} by Joschka Roffe implements the BPOSD decoder for general QLDPC codes. For simulating surface codes, the parity check matrix \cite{qldpc5} of such code should be used whenever defining the QLDPC code to be decoded by the BPOSD implementation.  {This package is intended as a module for building and benchmarking LDPC codes. At this moment, the package includes the BPOSD decoder, but a new version of it is expected with implementations of UF and modified BP decoders, among others. We do not include the dependencies to those other decoders in FIG. \ref{softwareDecs} since the release is not out yet.}

% MPS
The aforementioned QECSIM repository \cite{qecsim} does also include an implementation of the tensor network decoder. The package include the possibility of tuning the truncation parameter that is used for the TN decoder. Another implementation of the tensor network decoder can be found in the \emph{SweepContractor} repository\footnote{\url{https://github.com/chubbc/SweepContractor.jl}} by Cristopher Chubb which is based in the sweep line algorithm of \cite{chubb2}.

% Otros
% cellular automaton
In the case of cellular automaton decoders, the repository\footnote{\url{https://github.com/MikeVasmer/Sweep-Decoder-Boundaries}} \emph{Sweep-Decoder-Boundaries} by Michael Vasmer implements the version of the decoder using the sweep rule as presented in \cite{sweep,sweep2}. As discussed in the papers, the implementation of the decoder admits simulations using circuit-level noise. Additionally, the repository \emph{LocalToricDecoder} by Kasper Duivenvoorden implements the version of the cellular automaton using Toom's rule as proposed in \cite{toom,toom2}. We have been unable to find software packages implementing the other cellular automaton decoders discussed.

% renormalized group 
Concerning the RG decoder, the \emph{QTop} repository\footnote{\url{https://github.com/jacobmarks/QTop}} by Jacob Marks presents an implementation of it. However, it is noteworthy to say that the author stated in the repository that such software package is not being actively maintained and, thus, it may be unreliable. {In any case}, the code is available there for anyone that may be interested in testing it or using it to code its own version of RG decoders.

% neural network
With respect to neural network decoders, the repository\footnote{\url{https://github.com/Krastanov/neural-decoder}} \emph{Neural Network Decoders for Quantum Error Correcting Codes} by Stefan Krastanov implements the neural network decoder proposed in \cite{NNTC2} for the toric code. Moreover, Pooya Ronagh developed {another} repository\footnote{\url{https://github.com/pooya-git/DeepNeuralDecoder}} \emph{Deep Neural Decoders for Fault-Tolerant Quantum Error Correction} for the deep neural network decoder proposed in \cite{NNdeep} for decoding rotated planar codes. Interestingly, the authors of both repositories programmed the decoders so that other neural networks than the ones in \cite{NNdeep, NNTC2} {could} be incorporated. Hence, those repositories provide freedom to integrate other neural network decoders of the literature. Finally, the repository\footnote{\url{https://github.com/baireuther/neural_network_decoder}} \emph{neural\_network\_decoder} by Paul Baireuther contains the implementation of the neural network decoder for the color code in \cite{NNCC1}.

% MaxSAT
To finish with this section, there is a repository with the implementation of the MaxSAT decoder presented in \cite{maxsat}. The repository\footnote{\url{https://github.com/cda-tum/mqt-qecc}}, developed at the Technical University of Munich, is named \emph{QECC: An MQT tool for Quantum Error Correcting Codes written in C++} and is intended to be a more general tool for QEC than just the MaxSAT decoder for color codes. At the time of writing, the authors of the repository state that the project is still in early development and, thus, will contain other QEC software tools. The main ideas used for the software package were presented in \cite{TUMqecPaper}. Either way, at the moment of writing, the MaxSAT decoder and a decoder for QLDPCs based on the UF decoder \cite{ldpcuf} are present in the software package.

\section{Discussion} \label{secdiscussion}
\begin{figure*}[t]
    \centering
    \includegraphics[width = 0.8 \textwidth]{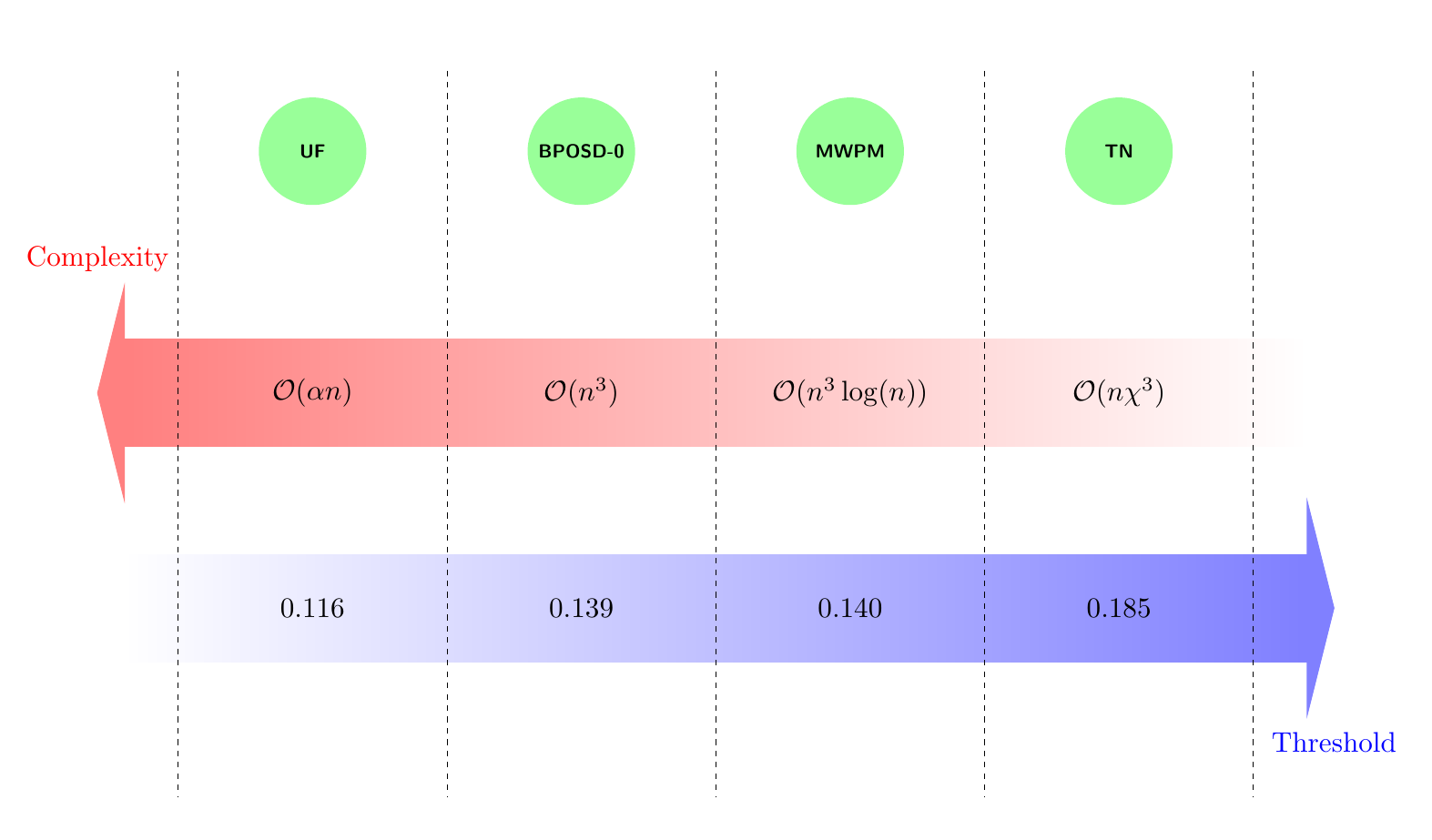}
    \caption{Graphical illustration showcasing the complexity and probability threshold of the decoders studied in this work. This comparative figure is inspired by a figure in \cite{maxsat}.}
    \label{pthcomp}
\end{figure*}

% mini intro

The construction of a fault-tolerant quantum computer remains as the Holy Grail of quantum computing so that the groundbreaking theoretical potential of such technology can be achieved. The key element to solve this quandary is quantum error correction. Quantum information has proven to be so frail that the scientific community has accepted that quantum computing will be unrealizable without QEC. One of the principal elements of a QECC is the decoder, or the classical algorithm that is used to estimate which error has corrupted the quantum information. Due to the urgent necessity of obtaining fast and accurate enough decoders that can operate in real-time for experimentally implemented QECCs, the number of papers related to this is increasing in a very fast pace. In this sense, the QEC community is strongly pushing the state-of-the-art of the topic so that fault-tolerance can be a reality as fast as possible.

% Hablar de resultados en Google, como belief-matching y mps han resultado en buenas contribuciones. (Incluir gráfica de Google?)

% Como se ajustan los métodos de decodificación estudiados a lo mencionado anteriormente, incluir imagen
As presented in the main text, the main decoders for the rotated planar code family are the MWPM, UF, BPOSD and TN decoders. FIG. \ref{pthcomp} is an illustrative representation of the accuracy versus runtime trade-off discussed through the text for those decoders. At the time of writing, it seems that the MWPM is an strong candidate due to its high threshold and the fact that it always returns the  $X$ and $Z$-errors of minimum weight matching the observed syndrome\footnote{Note that this is not the same as returning the overall minimum weight errors since $X$ and $Z$-errors are evaluated independently and, as such, $Y$-errors are considered as events of double weight as opposed to Pauli errors of weight 1.}. In addition, the MWPM decoder presents a considerable worst-case complexity when compared to other decoding methods, but recent implementations of the algorithm have proven to present an almost linear expected complexity \cite{sparse,fussionBlossom}, making this method even more powerful yet. On the other hand, UF also stands as a fair contestant due to its decoding speed which comes from a linear worst-case complexity. Nevertheless, as discussed before, the MWPM implementations with an average complexity lower than $\mathcal{O}(n^2)$ \cite{sparse}, even reaching linearity given that enough nodes for parallelization are available \cite{fussionBlossom}, make the candidacy of the UF decoder to significantly lose strength. {In any case}, it is noteworthy to state that many of those fast MWPM implementations were constructed based on many ideas introduced by UF decoding. The BPOSD-0 decoder lays in a middle ground between the MWPM and UF decoders whenever the rotated planar code is considered. As discussed before, increasing the search depth of the OSD algoirthm has a considerable impact on the worst-case complexity of the decoder and, thus, the benefits of this decoder will probably be lost due to the fact that the threshold of the code would not increase as much. In this sense, it seems that the MWPM decoder is a better candidate for the specific instance of surface codes. Nonetheless, the BPOSD algorithm was proposed as a general decoder for QLDPC codes \cite{qldpc2} and, therefore, it is the best candidate for decoding such family of codes which is being studied thoroughly at the time of writing. As seen in the earlier section, its implementation allows the circumnavigation of split beliefs in conventional belief propagation decoding, which have posed a great hurdle to conventional BP decoding for QLDPC codes \cite{qldpc2, raven1}. The TN decoder achieves the highest code threshold for the rotated planar code, albeit at the expense of the largest worst-case computational complexity. Both the benefit and drawback of this algorithm is a result of the fact that it is an effort to solve the maximum-likelihood (ML) problem explicitly, and this is why it is usually referred as the ML decoder\footnote{Note that in the terminology of Section \ref{secstabcodes} the TN actually solves the DQMLD problem, but the literature on the TN decoder refers to this just as ML.}. In this sense, the TN decoder is an almost brute force approach where such brute force search is limited by the bond dimension, $\chi$, which is the approximation parameter that is needed for this method to be actually implementable for codes of considerable size. Hence, the bond dimension is the parameter that will be in charge of the accuracy and speed of the algorithm. Due to the enormous growth in bond dimension in the surface code tensor network contraction, the tensor network is not really considered for real-time decoding but is studied cautiously as an effective degenerate quantum maximum likelihood decoder. Also this decoder presents the possibility of the experimental implementation of codes beyond break-even with poorer hardware capabilities due to its high threshold. The conclusion to this is that the choice of the decoder for a future real-time implementation of surface codes remains as an open question. This quandary is even more complex taking into account the existence of other decoding approaches such as the CA, RG, NN or MaxSAT decoders discussed in this tutorial. Each of those may provide interesting features that may be beneficial for specific families of codes or even adaptations of them may be interesting for the context of rotated planar codes. For example, machine learning  {based} decoders are gaining popularity on the recent times, probably due to the huge interest in machine learning in general.

In this line, recent breakthroughs have been achieved in the experimental implementations of surface codes over superconducting qubit  {and neutral atom} platforms \cite{wallraffSurf,googleSurf,neutralQEC}. Those experimental implementations were based on the rotated planar code discussed throughout this tutorial. In \cite{wallraffSurf}, Andreas Wallraff's group at ETH Zurich led an experiment based on the rotated planar code of distance-3 by means of a processor consisted of 17 transmon qubits. Krinner et al. used the obtained data after multiple measurement rounds in order to run the MWPM decoder for estimating the errors corrupting the quantum information. In \cite{googleSurf}, the Google Quantum AI realized an experiment based on the rotated planar codes of distances 3 and 5 by means of their expanded Sycamore device consisted of 72 transmon qubits. The authors actually run experiments for the distance-5 rotated planar codes since the actual data for the distance-3 code can be obtained from such code. Acharya et al. used the obatined data to decode the errors by means of the TN and belief-matching decoders\footnote{Recall that the belief-matching decoder combines the output of a BP decoder in order to reweight the edges of the graph in the MWPM problem \cite{fragile}.}. In this way, the authors proved that the logical error rate obtained by the rotated planar code improved as the distance of the code increased for both decoders. This result is really important since the code was operating in a sub-threshold noise level and it represents an experimental proof that the performance of the codes can be increased by increasing the distance (recall the threshold theorem).  {Furthermore, Mikhail Lukin's group at Harvard, with collaborators of QuEra computing, conducted QEC experiments in a neutral atom processor \cite{neutralQEC}. Among other relevant QEC experiments, Bluvstein et al. implemented a transversal CNOT gate between the logical qubits of two rotated planar codes for distances $d=\{3,5,7\}$. Importantly, the authors used the transversal CNOT to do a Bell-pair experiment and showed that the experiment error decreases with code distance. It is important to say that the authors discussed the application of a correlated decoder, that considers decoding both logical qubits jointly rather than independently, in order to obtain such result. Two correlated decoding algorithms are proposed, based on belief-find and TN decoders. The experiments used a single round of syndrome measurements, leaving multiple-round extraction as future work. }Importantly, the experiments performed the execution of the decoding algorithm by dealing with the measured data in a post-processing stage, i.e. no real-time decoding was actually performed. However, these experiments represent the state-of-the-art of experimental implementations of surface codes and they are an effort to progress towards the ultimate goal of fault-tolerant quantum computes.

% Faltan citas
Following the present discussion, and as it has been seen throughout the text, the decoding stage stands as a pivotal operation within the successful functioning of, not only  {surface codes}, but any correcting code in general. One should be aware of the fact that the named code threshold depends on three aspects: the code family in consideration, the decoding method for the code and the noise model in consideration \cite{thresh1,thresh2}. In this sense, it is straightforward to see that designing decoding algorithms is crucial in terms of the ability of a family of codes to correct errors. The intrinsic locality and degeneracy of  {surface codes} allows for several decoding processes to present themselves as valid candidates for future real device implementation. Nevertheless, in addition to the code threshold for a decoding method, two other critical aspects stand as limiting factors for the implementation of surface codes decoding in a real-time fashion when implemented on experimental hardware: run time and circuit-level noise.

% Discutir con Josu runtime, meter gráfica
Decoding runtime refers to the actual time that a decoder requires to ouptut an estimate of the error that has corrupted the logical qubit encoded by a code. Hence, this inherently imposes a delay in the error correcting system that is hazardous for its actual operation. This comes from the fact that, once the syndrome measurement has been done, the data qubits will still continue to suffer from errors as they will not stop to decohere. Thus, if the delay between measuring the checks and applying the recovery operation is too high, then the algorithm will fail, with high probability, the actual error configuration when recovery is executed (See appendix \ref{appBacklog} for a discussion of why real-time decoding is required). This is especially important by taking into account that the decoherence times of state-of-the-art qubits are short and the generation of syndrome data is done a fast rate\footnote{Note that this is not true for all qubit technologies such as ion traps, but those present additional problems such as the fact that their operation times, quantum gates, are very slow.} \cite{approximatingDec}. As seen in through this tutorial, it is pretty fair to state that maintaining code threshold while reducing the actual complexity of a decoder is a hard problem. This is a result of the trade-off of being fast against considering more error configurations, which actually relates to code performance. This logic comes from the fact that the traditional comparison of accurate and fast is done in terms of code threshold and worst case complexity. However, it is a recent trend to benchmark runtime by expected complexity, i.e. an average of the number of operation required to decode different error patterns. The rationale behind this is that worst-case events are generally exponentially rare, implying that their impact will not be as high in code performance as other more frequent cases. This effect should be more important whenever hardware noise improves lower to sub-threshold physical error rates and, thus, expected runtime might be a better benchmark for decoding algorithms. In this sense, designing faster decoding algorithms that are able to maintain the capabilities of the code in terms of decoding accuracy is a critical issue for the QEC field. It is important to state that significant advances in this direction are being obtained with proposals such as sparse blossom \cite{sparse} or Fusion Blossom \cite{fussionBlossom}. Furthermore, recently, sliding-window decoding schemes have been proposed that allow parallel decoding of the syndrome data by means of matching like decoding algorithms (MWPM and UF), thus, dealing with the exponential backlog problem of real-time decoding while maintaining logical fidelity \cite{slidingBrowne,slidingTan}.

% Discutir measurement errors
On the other side, considering circuit-level noise is fundamental so that surface codes are successful when implemented in real hardware. As explained before, the consequence of such noise associated to imperfect gates, SPAM and error propagation results in measurement errors that make the decoder not to be able to correctly estimate the actual error that has corrupted the data qubits of code (refer to Section \ref{secnoisemod}). In this sense, considering circuit-level noise is sometimes referred as fault-tolerant error models as it is considering all the errors of the elements of the system \cite{neural}. Circuit-level noise is usually alleviated by means of multiple measurement rounds, in the order of $\mathcal{O}(d)$, so that the additional errors that led to measurement errors can be taken into account for the decoding problem (refer to Appendix \ref{appA} for a more detailed description). Nevertheless, such multi-round measurement results in a toll for complexity and, hence, runtime. This directly hinders the previously discussed necessity of being fast in decoding so that error correction is successful. Additionally, even when the multiple measurement protocol is applied, the code threshold significantly decreases (usually an order of magnitude) in comparison to thresholds that only consider data qubit noise \cite{fowlerReview,fragile}. The direct result of this is that hardware requirements for sub-threshold code implementation become even more stringent. Therefore, circuit-level noise adds another layer of complexity to the already difficult decoding implementation in real hardware. In this sense, improving gate fidelities and decoherence parameters, as well a other noise sources, of state-of-the-art quantum processors will be a very important for obtaining fault-tolerant quantum computers. 

% No sé si quedaba rara esta parte teniendo en cuenta que estábamos hablando de surface codes
%Additionally, we have discussed that some decoders have the property of being single-shot decoders (recall the Cellular Automaton decoder). Decoders of such characteristics have the property that they do not require many measurement rounds in order to deal with circuit-level noise, making them interesting candidates \cite{automata,singleshot}. Hence, studying new decoders that may exhibit this properties for topological codes while maintaining performance and complexity is especially interesting for the field. Importantly, not all families of codes admit this type of decoding, e.g. (rotated) planar codes or 2D toric codes \cite{singleshot}.

% Que Pat mire la parte de single shot decoding porque creo que no se puede hacer en el surface code.

% Hablar de biases hacia Z
Moreover, tailoring both codes and decoders to the specific structure of the noise that the qubits experience is being investigated so that the actual correcting performance for such noise. A widespread example of this is the bias towards $Z$-errors that many state of the art qubit technologies such as silicon spin qubits or ion traps exhibit \cite{approximatingDec,bias1}. As described in section \ref{secnoisemod}, this comes from the fact that the dephasing time, $T_2$, of such qubits is much smaller than the energy relaxation time, $T_1$. As previously discussed, running the aforementioned decoders over biased noise models usually contributes to a worse threshold than for the depolarizing channel. This, however, occurs due to the fact that the code/decoder is not tailored to work over noise exhibiting such tendency. Specifically, since $X$ and $Z$-errors are decoded independently, then one of the decoding problems results to be more dense if the noise is biased, i.e. it will experience errors with higher weight more probably, and, thus, the actual threshold will be lower since such decoding sub-problem will fail more often. In this sense, the community has actually studied how to deal with some of those noise structures so that performance is maintained or even enhanced. Some examples include:

% Añadir una imagen para cada uno

\begin{itemize}
    \item \textbf{The rectangular  {rotated} code:} using a  {rotated} code with rectangular shape makes the minimal weight logical error $Z_L$ to require more physical $Z$ operators, increasing the $Z$-distance $d_Z$ and, consequently, increasing the number of $Z$-errors it can correct: $(d_Z-1)/2$ \cite{rectangularbias}. On the negative side, this is at the expense of increasing the number of necessary data qubits to $d_Xd_Z$. In FIG. \ref{rectangularcode}, a rectangular code of distances $d_X = 5, d_Z = 7$ is shown, notice how  the logical operator $Z_L$ requires additional Pauli operators than for the $X_L$ operator. This code construction may use the reviewed decoders in order to estimate the channel errors.

\begin{figure}
    \centering
    \includegraphics[width = 0.8 \columnwidth]{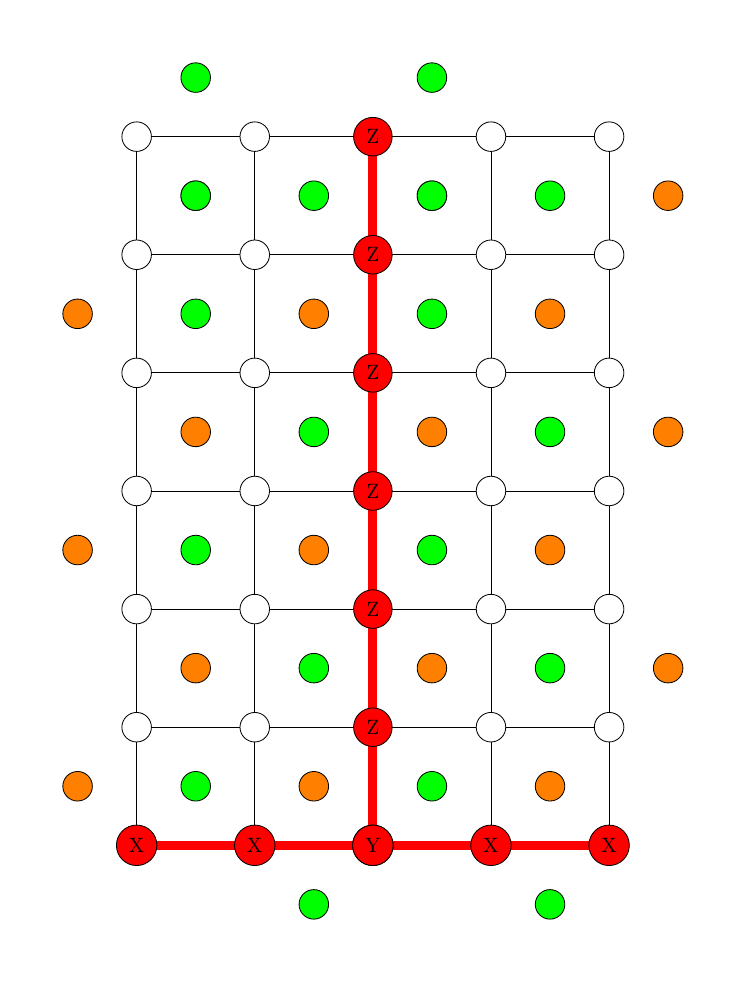}
    \caption{Graphical representation of a $d_X = 5, d_Z = 7$ rectangular rotated planar code. $Y_L=X_LZ_L$  {operator is} also represented.}
    \label{rectangularcode}
\end{figure}
 
    \item \textbf{Clifford deformation:} in Clifford deformed surface codes, the check qubits are conjugated respect to a unitary operator from the Clifford group \cite{cliffGroup} so that other valid stabilizers are obtained \cite{roffeinid}. By means of this stabilizer modifications, the new stabilizer codes can be constructed so that they are more susceptible to $Z$-noise. For example, the XY-code is a surface code which, instead of having $X$ and $Z$-checks, has $Y$ and $X$-checks,  both of which are susceptible to $Z$-noise. A graphical depiction of the XY-code is given in the top figure of FIG. \ref{clifforddeformation}. As a result of such susceptibility to $Z$-errors, the associated decoder can make use of the correlation between $X$ and $Y$-errors, i.e. $XY = Z$, in order to be more accurate in detecting $Z$-noise \cite{fragile, recMWPM,tuckett2}. Another popular Clifford deformed surface code is the XZZX surface code \cite{xzzx,tailoredXZZX}. For this code, all check qubits act equally on their surrounding data qubits operating two $X$ and two $Z$ operators as shown in the bottom figure of FIG. \ref{clifforddeformation}. One can see that the commutation between the checks is maintained, since adjacent checks either anti-commute twice or directly commute. An interesting characteristic of the XZZX code is that there is only one set of $Z$ physical operators which produce a logical operator. As is shown in FIG. \ref{clifforddeformation}, all the products of this logical operator with the stabilizer set produce physical operators which are composed of $X$, $Y$ and $Z$ operators. This feature {lowers} the probability of a logical phase-flip, $Z_L$, error under a highly $Z$-biased noise model. Moreover, for the case in which the $Z$-bias is infinite, e.g. a pure dephasing channel, the XZZX surface code operates as a series of disjoint repetition codes that can be decoded independently. The data qubits of such repetition codes are the top-left to bottom-right diagonals of the overall XZZX code, as can be seen in the bottom of FIG. \ref{clifforddeformation}, where the diagonal consisting of two $Z$ physical errors produces syndromes in their top-left and bottom-right sides. Given this condition, the threshold of the XZZX code reaches an outstanding $50\%$ \cite{xzzx,tailoredXZZX}, while reducing the amount of required qubits significantly. Therefore, it can be seen that a bias towards $Z$-errors is not only not problematic, but also beneficial if the code and decoder are tailored for such model.

\begin{figure}
    \centering
    \includegraphics[width = 0.9 \columnwidth]{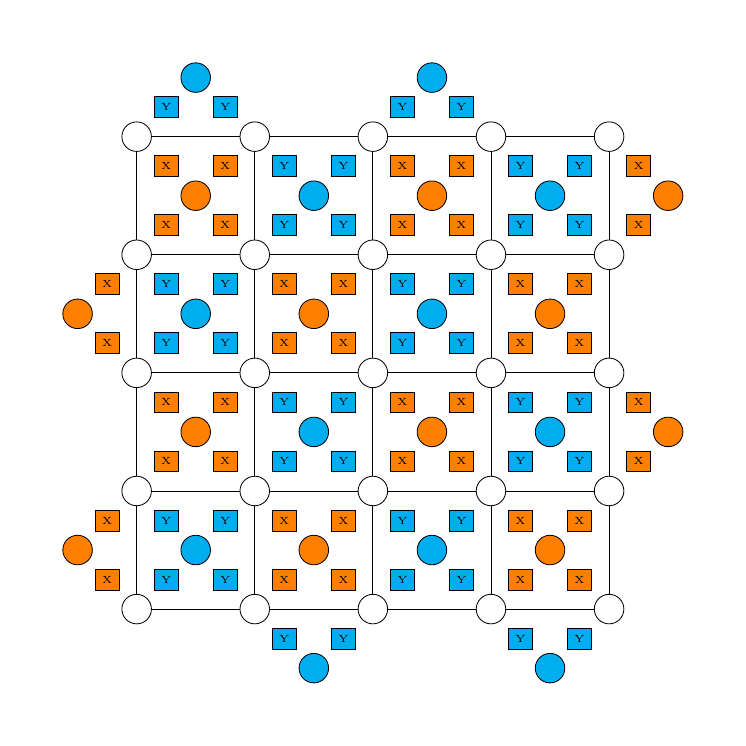} \\
    \includegraphics[width = 0.9 \columnwidth]{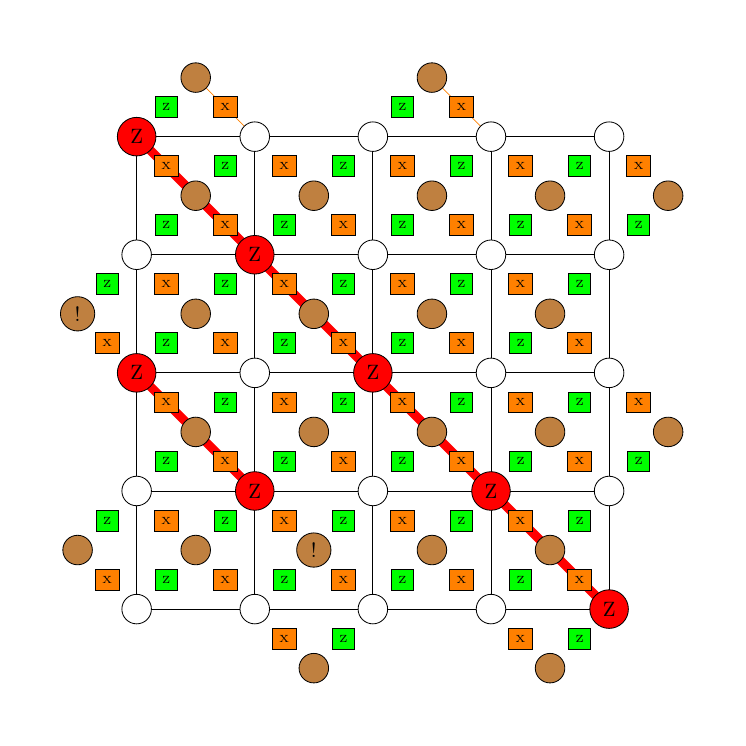} 
    \caption{On the top, a graphical representation of a rotated planar XY code with the operators of the stabilizer generators.  {We represent $Y$-checks with blue to distinguish them from $Z$-checks (green troughout the text).} On the bottom, the same representation but for the rotated planar XZZX code altogether with a $Z_L$ operator.  {We represent all checks with the brown color since all of them are equal now consisting on both $X$ and $Z$ stabilizers.}}
    \label{clifforddeformation}
\end{figure}
    
    \item \textbf{$X/Z$ correlation: } another important approach to improve the performance of surface codes is to make decoders aware of the fact that $Y$-errors represent a correlation between the $X$ and $Z$-errors. One downside of the independent $X$ and $Z$ decoding of CSS codes such as  {CSS surface codes} is that it makes the decoding blind of the correlation among those errors, i.e. it is usually considered that $p_Y = p_Xp_Z$. However, as seen in Section \ref{secnoisemod}, the Pauli noise models have $p_X \approx p_Y$ (both in biased and depolarizing) and, thus, by not taking this into account the decoder underestimates the frequency in which $Y$-errors corrupt the data qubits of the code. A method in order to circumvent this problem (without explicitly changing the structure of the code) consists in first decoding for one of the errors and, afterwards, using the partial recovered error in order to decode the other error. E.g., given an error in a CSS surface code, first decode the $X$-error, and then consider that, for recovering for the $Z$-error, every qubit has an updated probability of undergoing a $Z$-error defined by: 
    \begin{equation}
    \begin{split}
        &P(Z|X=1) = \frac{p_Y}{p_X + p_Y} \\
        &P(Z|X=0) = p_Z,
    \end{split}
    \label{graphcorrelation}
    \end{equation}
    where, $P(Z|X=1)$ is the probability that a specific data qubit undergoes a phase-flip $Z$ given that it has been considered to undergo a bit-flip $X$, and $P(Z|X=0)$ is the probability that a qubit which is considered to not have undergone a bit-flip has undergone a phase-flip $Z$. This can be done once \cite{fragile, xzcorrelation, fowlercorr} or in a recursive manner \cite{recMWPM, oldrec}. Moreover, it can be combined with the XY-code in order to enhance the susceptibility of the code towards $Z$-noise \cite{fragile, recMWPM}.
\end{itemize}

Thus, it is important from the decoding (and code construction point of view) to consider the actual noise that the qubits of the surface code experience. As seen for the XZZX code, this tailoring does actually even significantly improve the code performance without needing to lose resources. Even if we have discussed noise bias here, there are many other important subtleties in the nature of the noise for each qubits technology that should be considered for integration in real hardware. For example, multi-qubit error correlation and time-varying noise are examples of this. The fact that the actual distribution of the errors in a real quantum processor is independent seems to be generally false. In this sense, studying the correlated nature of noise is an incipient sub-field in the topic and, thus, tailoring decoding to such correlated nature should improve the performance of the codes over such models \cite{corr1}. Note that this quandary is not new for classical error correction \cite{tse}, implying that many ideas of dealing with such noise can be extrapolated to the quantum domain. Modifications of decoders to take into account correlated noise have been already proposed for quantum turbo codes over channels with memory, resulting in a considerably improved performance of the code when compared to the decoder that is blind to the correlation \cite{qtccorr}. Additionally, the noise experienced by superconducting qubits has been proven to be time-varying \cite{TVQC,fastfading}, which implies that the performance of the codes experiences a degradation. In this sense, studying adaptive decoders that can estimate the noise level \cite{estimatePauli} at a certain time in order to follow such fluctuating nature of the noise should considerably improve the performance \cite{TVestimAdapt}.

To conclude, quantum error correction is still in a primitive stage before the potential of quantum computing can be unleashed by fault-tolerant machines. In this sense, surface codes represent the most promising family of codes to be implemented in the early post-NISQ era, principally due to their locality feature (and, in the case of planar instances, two-dimensional qubit placing) and their high tolerance to quantum noise. As extensively reviewed in this article, decoders represent a central part of QEC methods as they are key elements in posing a threshold for a code. Moreover, we have discussed the importance of the runtime of this algorithms due to the accumulation of other errors if the estimation of the channel error results to be too slow. As a result of this, there exists an important trade-off between accuracy and speed of decoders which implies that a selection of a decoder for real-time decoding depends on many factors that go from the hardware noise level to the pace at which more errors accumulate. Hence, the selection of the best decoder for an experimental implementation of a surface code is still an open question that many research teams, both in academia and industry, are trying to resolve. Due to the extense zoo of possible qubit technologies being investigated nowadays, it is possible that many of the exisitng candidates, or new ones, are the best fit as a function of the specifics of each of the quantum computing platforms. There is much work left to do, and each of us on the field should contribute our share in this Herculean quest. We live in exciting times.

% Future work? Improvement in complexity for measurement errors, adaptation of decoder to real devices (i.ni.d. , architectural structures...), tailoring of codes to better resist noise (XZZX code, repetition code for high biases), multi-path summation...

%\section{Conclusion} \label{sec:conclusion}

% Mention reweighting of edges

\section*{Data availability}
The data that supports the findings of this study is available from the corresponding authors upon reasonable request.

\section*{Competing Interests}
The authors declare no competing interests.

\section*{Acknowledgements}
We want to acknowledge Nicolas Delfosse, Joschka Roffe, Pavel Panteleev, Christopher Chubb, David Tuckett, Michael Newman, Manabu Hagiwara, I\~nigo Barasoain and Javier Oliva for fruitful discussions. 

This work was supported by the Spanish Ministry of Economy and Competitiveness through the ADELE (Grant No. PID2019-104958RB-C44) and MADDIE projects (Grant No. PID2022-137099NB-C44), by the Spanish Ministry of Science and Innovation through the proyect Few-qubit quantum hardware, algorithms and codes, on photonic and solid-state systems (PLEC2021-008251), by the Ministry of Economic Affairs and Digital Transformation of the Spanish Government through the QUANTUM ENIA project call - QUANTUM SPAIN project, and by the European Union through the Recovery, Transformation and Resilience Plan - NextGenerationEU within the framework of the Digital Spain 2026 Agenda.

\bibliographystyle{quantum}
\bibliography{bibliography}

\section*{Appendices}

\appendix
\section{ {The backlog problem and why QEC requires real-time decoding}}\label{appBacklog}
\begin{figure}
    \centering
    \includegraphics[width = \columnwidth]{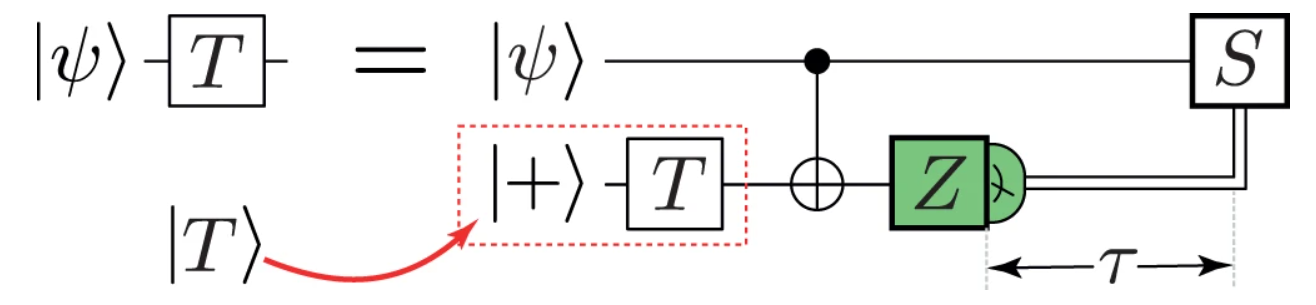}
    \caption{ {Application of a logical $T$-gate through the injection of a magic state by teleporting. The magic state should be distilled beforehand. Figure taken from \cite{slidingBrowne}.}}
    \label{magiState}
\end{figure}

 {As discussed in the main text, the decoding algorithms for  {surface codes} need to be fast enough to operate in real time. This is necessary because running fault-tolerant algorithms outside the reach of classical simulation requires the use of quantum algorithms that apply $T=\text{diag}(1,e^{i\frac{\pi}{4}})$ gates over the logical space of the code. The use of $T$ gates guarantees that the code supports the universal gate set Clifford+T that can be used to compile any quantum computation, a result known from the Solovay-Kitaev theorem \cite{solovayKitaev}. However, the Eastin-Knill theorem states that no quantum error correction code can implement a universal gate set in a transversal manner \footnote{ {A transversal logical gate is one that is implemented by performing physical gates over the physical qubits of the code in a manner that avoids the propagation of physical errors throughout the code. For example, the  {surface codes allows} the application of a logical $X$-gate by applying $X$ operators on the physical qubits horizontally from boundary to boundary (Recall FIG. \ref{rotateddegeneracy}).}}. In this regard,  {surface codes} support transversal implementation of logical Clifford gates but lack a transversal version of the logical $T$-gate. Fortunately, there are ways to circumvent the Eastin-Knill theorem with the purpose of achieving universal fault-tolerant computation. The most popular method is known as magic state injection \cite{slidingBrowne}. A magic state injection protocol can be seen in FIG. \ref{magiState}, where the protocol requires the distillation of a magic state \cite{fastDec} and the injection of such a state to the logical qubit by means of a teleportation circuit.}

 {Coming back to the topic of decoding algorithms, note that if one restricts the operation of the QEC to the set of Clifford gates, then due to the fact that the Clifford group is the normalizer of the Pauli group, it is not necessary to know the logical state for applying the circuit \cite{fastDec}. Since Clifford gates map Paulis into Paulis, the only thing that changes is the interpretation of the final measurements after the computation. Therefore, when running such circuits, one can record the syndrome history during the computation and post-process the errors using the recorded syndrome information. In this sense, one can construct a quantum memory without requiring real-time decoding \cite{fastDec}. However, the scenario drastically changes whenever useful 
computations are taken into account\footnote{ {By useful we mean quantum computations that cannot be efficiently simulated by classical means and, hence, require $T$-gates.}}. That is because when injecting the magic state, one must be certain if the $S$-gate after the measurement should be applied or not. Such a decision can only be reliably performed if the syndrome data before the application of said gate is decoded \cite{slidingBrowne}. If the correction of such a Clifford error is ignored, the resulting error might propagate into a much more complicated error that will corrupt the computation \cite{fastDec}. Therefore, in a fault-tolerant quantum computer, the classical processing of the syndrome data, that is, the decoding algorithm; should be fast enough so that syndrome data does not accumulate into an undecodable degree. This accumulation of syndrome data due to the latency of the decoding algorithm is usually referred as the backlog problem \cite{fastDec,slidingBrowne}. $r_\mathrm{proc}$ is defined as the rate at which the decoder processes the syndrome data while $r_\mathrm{gen}$ is the rate at which the syndrome data is generated (both quantities in bauds). As discussed by Terhal in \cite{fastDec}, if the ratio
\begin{equation}
    r_\mathrm{gen}/r_\mathrm{proc} = f  > 1,
\end{equation}
then a small initial backlog in the processing of the syndrome data will result in an exponential slow down as a function of the $T$-gate depth of the algorithm. More concretely, Terhal lower bounded the running time of $k$ $T$-depth quantum algorithms as $cf^k$ when $f >1$, for some constant $c$ \cite{fastDec, slidingBrowne}. Thus, the straightforward conclusion is that decoding algorithms must be fast enough to ensure that the injections of the magic states are delayed by only a constant amount of time (instead of an exponentially increasing amount of time).}

\section{Multiple-round decoding for measurement errors}\label{appA}

% Mencionar que hay otros métodos de decodificación como single-shot decoding para otros códigos (no surface codes)

Throughout this article, surface code decoders have been described as correcting errors arising from data qubit errors, i.e. considering that check measurements are noiseless. However, this simplified perspective is overly optimistic because circuit-level noise arising from gate and SPAM errors make syndrome measurement to be noisy too. As mentioned in the main article section on error models and decoding, circuit level noise will lead to incorrect syndromes. If we attempt to decode an erroneous syndrome, the recovered error will likely result in a logical error. Therefore, a new approach is needed to account this type of errors in the code. The common technique consists in performing multiple syndrome extracting rounds. When a check qubit undergoes a change in its measurement value, it implies a non-trivial syndrome element for such measurement round. Subsequently, decoding algorithms can be employed by connecting check qubits within a particular time instance to themselves in the following time, creating a graph containing space and time information of the errors. Recall that not every decoding algorithm requires this multiple-round measurements as, for example, the cellular automaton decoders \cite{automata,sweep2}. These type of decoders are usually known as single-shot decoders \cite{singleshot}. It is important to say that not every QEC family can admit a single-shot decoding algorithm \cite{singleshot}.

\begin{figure}
    \centering
    \includegraphics[width = 0.7\columnwidth]{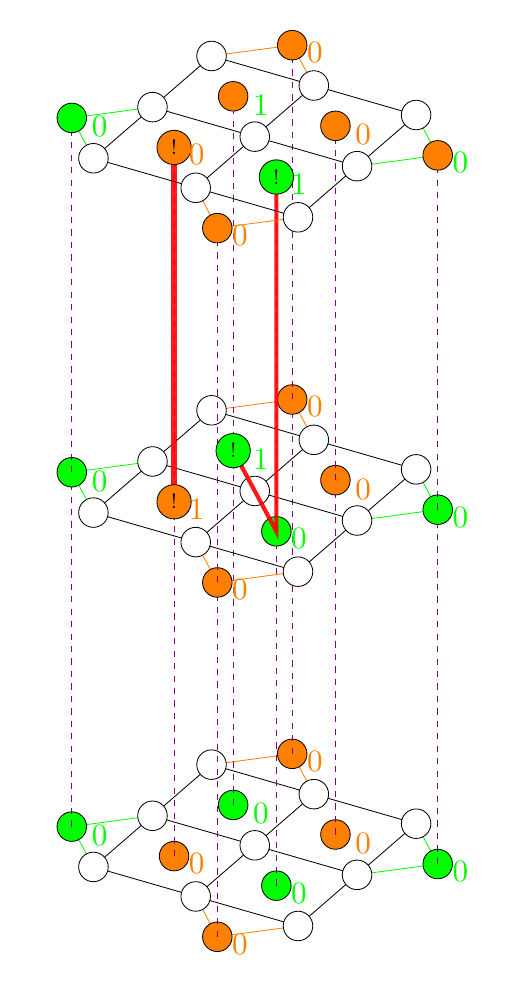}
    \caption{Graphical representation of an error in a space-time  {rotated} code.  {In the figure, each of the depicted rotated planar codes represent one measurement round ($d=3$ in this example). The solid red lines represent the recovered space-time error.}}
    \label{errormeas}
\end{figure}

FIG. \ref{errormeas} illustrates a graphical depiction of three decoding rounds. In this example, three rounds of syndrome extraction have been computed, where the temporal direction progresses upwards. During the initial syndrome extraction, all checks yield a measurement of 0, assuming an absence of errors. This set of measurements serves as the starting point. In the subsequent syndrome extraction, two checks undergo a measurement change: a green $X$-check transitions into 1, as does an orange $Z$-check. These altered check measurements are classified as non-trivial syndrome elements. In the third round, the measurement values of two checks change again, designating them as non-trivial syndrome elements. One of these checks is the orange check from the second decoding round, which reverts back to 0, while the other one is an $X$-check. It is worth noting that although the $X$-check from the second syndrome extraction retains a measurement value of 1, it is not considered a non-trivial syndrome element since its measurement value did not change.

% A partir de aquí es nuevo
The common approach for considering measurement error noise channels consists in introducing an error probability to the state initialization of the measurement qubit, noisy Haddamard and CNOT gates as described in the error model section and noisy measurement probabilities. Often, this is done by considering all gates within the syndrome measurement circuits from FIG. \ref{stabcircuit}, with probability $p$ \cite{fowlerReview}.  {For example, a CNOT is a two-qubit gate which,  under probability $p/15$, will produce non-trivial two-fold Pauli operator ($I\otimes X, I\otimes Y, \dots ,Z\otimes Z$)}. Often, the noise channel considered for gate errors is the depolarizing channel, but there have also been studies which have considered biased channels so as to reproduce results which may reflect the bias suffered by several experimental qubits.

To implement perfect matching methods like MWPM or UF, the dashed purple lines in FIG. \ref{errormeas} are taken into account. These lines represent edges that connect different time instances separated by one syndrome extraction. The resulting perfect matching will consist of space edges and time edges. Space edges correspond to Pauli errors within the data qubits, while time edges represent faulty measurements that do not require actual error recovery within the code. In FIG. \ref{errormeas}, a potential perfect matching is indicated by red lines, which consist of two time edges and one space edge.

% Esto también es nuevo

\begin{figure}
    \centering
    \includegraphics[width = 0.7\columnwidth]{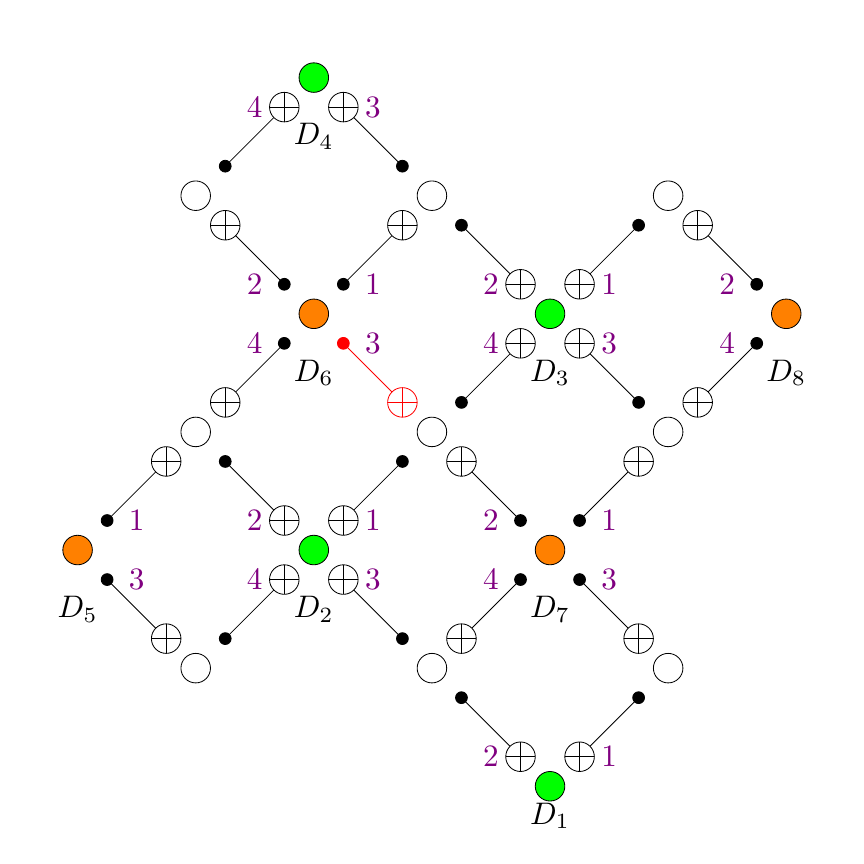}
    \centering
    \includegraphics[width = 1\columnwidth]{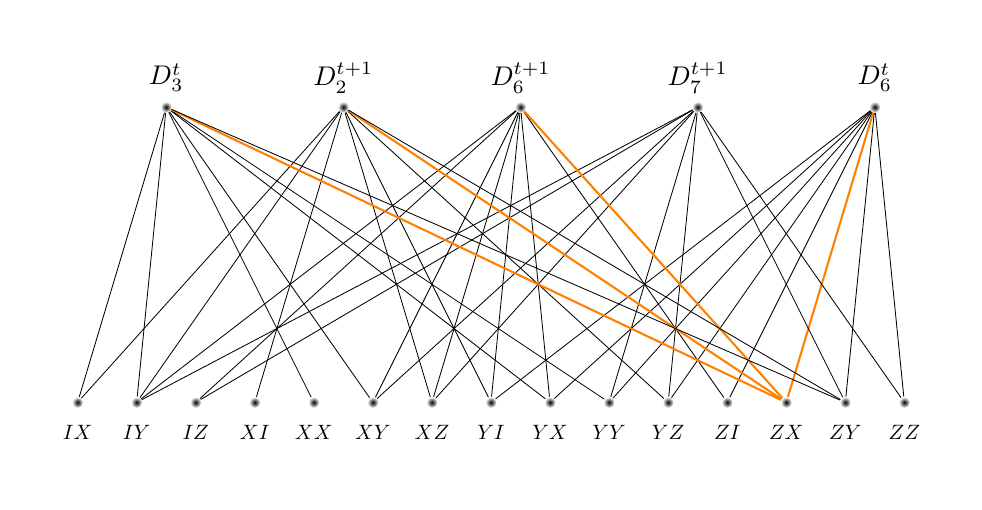}
    \caption{In the top, a distance-3  {rotated} code with the CNOT gates from the stabilizing circuit shown. The violet numbers indicate the order at which the CNOT gates are applied within the circuit of every measurement qubit, which themselves are labelled as $D_i$. Graphical representation of the circuit-level Tanner graph from a single CNOT, which is denoted in red. The figure is inspired from \cite{fragile}.}
    \label{tannerhiggott}
\end{figure}

An interesting approach to the decoding of surface codes under circuit-level noise and several decoding rounds was given in \cite{fragile}, where they consider the errors that may arise from the stabilizing circuit. When a set of errors is indistinguishable (they trigger the same measurement qubits), they are considered equal events under the combination of their probabilities. With these considerations in mind, a Tanner graph can be developed considering all syndrome measurement rounds. A graphic illustration is provided in FIG. \ref{tannerhiggott}, in the top figure one can see a planar representation of the stabilizing circuit of a distance-3  {rotated} code. Note that by following the CNOT order given by the violet numbers one can compute all the CNOT gates of the  {rotated} code stabilizing circuit with only 4 steps avoiding at all time that one qubit experiences two CNOT gates at the same time. The measurement qubits, which are also defined as detectors in the literature, are denoted with $D_i$ labels. After several stabilizing measurements we can describe the overall syndrome through the change in parities of the detectors as shown in FIG. \ref{errormeas}, i.e. the syndrome element $D_i^t$ will be 1 if the detector $D_i$ has changed its measurement outcome between the measurement time instances $t-1$ and $t$. Within the top figure, one of the CNOT gates are written in red, in the bottom figure, a Tanner graph relates the possible Pauli errors \footnote{Pauli operator to the left operates on the detector $D_6$, while the one on the right operates on the data qubit.} which may occur in the noisy CNOT gate with the detector events they would trigger. Consider the case of a $ZX$-error, which is highlighted in orange lines. The $X$ operator on the data qubit which is applied on the circuit time step $3$ will propagate to $D_3$ in the following circuit time step following the logic of FIG. \ref{stabcircuit} and producing a non-trivial $D_3^t$ due to the parity change in its measurement. In the following round of measurements the $X$-error from the data qubit will propagate to $D_2$ changing its parity and producing a non-trivial $D_2^{t+1}$. On the other hand, the $Z$-error correspondent to the detector will produce a measurement error, changing its parity and resulting in a non-trivial $D_6^t$. After another stabilizing round, the parity of the detector changes again since there is no measurement error and so it returns to the previous state, making $D_6^{t+1}$ also be non-trivial. FIG. \ref{tannerhiggott} only contains the Tanner graph correspondent to a CNOT error, in order to consider the entire  {rotated} code circuit level noise, one should consider all CNOT error Tanner graphs omitting operators which trigger the same set of operators, doing so allows to compute BP, which aids other decoding methods such as MWPM or UF in their task of decoding a syndrome \cite{fragile}. This method has also been considered for the BPOSD method in \cite{bravyildpc}, where the authors decode a low connection QLDPC code by considering the gate and idling Pauli errors.

% Hasta aquí

Usually, it is considered that a distance-$d$ surface code requires $\mathcal{O}(d)$ syndrome extraction rounds per decoding round \cite{lattice, fragile, chamberland} in order to preserve the distance property\footnote{A measurement error within a check which repeats itself over half the times of syndrome extraction rounds will produce a logical error.}. Considering several syndrome extractions further complicates the decoding process, since it significantly increases the number of nodes within decoding graphs and the connectivity of the graph. Specifically, a  {rotated} code of arbitrary distance-$d$, which has $d^2-1$ checks, results to have $d(d^2-1)$ checks. Moreover, even by considering multiple decoding rounds for dealing with faulty measurements, the threshold of the codes decrease dramatically due to such effect caused by circuit-level noise when compared to just data qubit noise \cite{fowlerReview, fragile, chamberland}.

In conclusion,  measurement errors pose a significant challenge to the successful implementation of  {surface codes}. The accurate extraction of syndromes is crucial for effective error correction. Faulty-measurements lead to lower code thresholds, implying that the hardware requirements for error correction are more stringent. Therefore, dealing with circuit-level noise and noisy measurements is critical to reach the fault-tolerant quantum computing paradigm.

\section{QECC numerical simulations}\label{appB}
Monte Carlo computer simulations of the $d\times d$ rotated planar codes with $d\in\{3,5,7,9,11\}$ have been performed with the objective of obtaining the performance curves and thresholds of the code when decoded with MWPM, UF, BPOSD and TN decoders.

Each round of the numerical simulation is performed by generating an $N$-qubit Pauli operator, calculating its associated syndrome, and finally running the decoding algorithm using the associated syndrome as its input (note that for the simulations performed in this paper, no SPAM or gate errors are considered). Once the error is estimated by the decoder, it is used to determine if a logical error has occurred on the codestate by using the channel error \cite{logicalsparse}. The operational figure of merit we use to evaluate the performance of these quantum error correction schemes is the Logical Error Rate ($P_L$), i.e. the probability that a logical error has occurred after the recovery operation. 

Regarding the software implementations of the decoders used to obtain the simulations in Section \ref{secdec} have been: QECSIM for the MWPM and tensor network decoder, the {LDPC} implementation of Joschka Roffe for the BPOSD decoder and an implementation coded by ourselves for the UF decoder by Delfosse and Nickerson.

For the numerical Monte Carlo methods employed to estimate the $P_L$ of the rotated planar codes, we have applied the following rule of thumb to select the number of simulation runs, $N_{\mathrm{runs}}$ \cite{montecarlo}, as

\begin{equation}
N_{\mathrm{runs}} = \frac{100}{P_L}.
\end{equation}

As explained in \cite{montecarlo}, under the assumption that the observed error events are independent, this results in a $95\%$ confidence interval of about $(0.8\hat{P_L} , 1.25\hat{P_L})$, where $\hat{P_L}$ refers to the empirically estimated value for the logical error rate.

\end{document}